\documentclass[aps,prd,nofootinbib]{revtex4-2}
\usepackage{amsmath,amssymb}
\usepackage{graphicx}
\usepackage{slashed}
\usepackage{verbatim}
\usepackage{dsfont,latexsym} 
\usepackage{color}
\usepackage{hyperref} 

\hypersetup{
    colorlinks,
    citecolor=black,
    filecolor=black,
    linkcolor=black,
    urlcolor=black
}

\newcommand{\be}{\begin{equation}}
\newcommand{\ee}{\end{equation}}
\newcommand{\ba}{\begin{eqnarray}}
\newcommand{\ea}{\end{eqnarray}}
\newcommand{\ban}{\begin{eqnarray*}} 
\newcommand{\ean}{\end{eqnarray*}}
\newcommand \nn {\nonumber}
\def\ck{{\check{k}}}

\def\d{\partial} 

\def\p{{\mathbf p}}
\def\q{{\mathbf q}}
\def\k{{\mathbf k}}
\def\x{{\mathbf x}}  
\def\y{{\mathbf y}}
\def\r{{\mathbf r}}
\def\z{{\mathbf z}}
\def\w{{\mathbf w}}
\def\v{{\mathbf v}}

\def\K{{\mathbf K}}
\def\ux{{\underline{x}}}
\def\uy{{\underline{y}}}

\def\uk{{\underline{k}}}
\def\up{{\underline{p}}}
\def\uq{{\underline{q}}}
\def\uw{{\underline{w}}}
\def\uv{{\underline{v}}}  
\def\P{{\mathbf P}}

\def\B{{\mathbf B}}

\begin{document}

\title{DIS dijet production at next-to-eikonal accuracy in the CGC}

\author{Tolga Altinoluk$^{a}$, Guillaume Beuf$\,^{a}$, Alina Czajka$^{a}$ and Arantxa Tymowska$^{a}$}
\affiliation{ Theoretical Physics Division, National Centre for Nuclear Research, Pasteura 7, Warsaw 02-093, Poland
}

\date{\today}

\begin{abstract}
We compute dijet production in Deep Inelastic Scattering at low $x$ in the dipole formalism at next-to-eikonal accuracy. We calculate the contributions induced by single photon exchange of either longitudinal or transverse  polarization.
 We include all types of corrections to the eikonal approximation in the gluon background field: 
 (i) finite longitudinal width of the target, (ii) interaction of the quark-antiquark pair with the subleading (transverse) component of the background field and (iii) dynamics of the target which is encoded in the $z^-$-coordinate dependence of the background field.  The final expressions for the dijet cross section are written as sum of a "generalized eikonal" contribution (where longitudinal momentum $p^+$ exchange between the target and the incoming  quark-antiquark pair is allowed since the average $z^-$ dependence of the background field is kept) and explicit next-to-eikonal corrections that involve decorated dipole and quadrupole operators. 
\end{abstract}

\maketitle

\vspace{-0.5cm}

\tableofcontents

\section{Introduction}
In the high-energy limit hadronic collisions are conveniently described by the effective theory called Color Glass Condensate (CGC) (see \cite{Gelis:2010nm, Albacete:2014fwa, Blaizot:2016qgz} for recent reviews and references therein). The CGC effective theory relies on the gluon saturation phenomena that is reached at sufficiently high energies in the Regge-Gribov limit, where the increase in the energy is provided by the decrease of the longitudinal momentum fraction (Bjorken $x$) carried by the interacting partons. In this limit, the gluon density of the interacting hadrons increases rapidly with the increasing energy. The rapid increase in the gluon density of the scattering hadrons slows down due to the non-linear interactions of the emitted gluons at sufficiently high energies, causing the aforementioned gluon saturation phenomenon which is characterized by a dynamical scale referred to as saturation momentum $Q_s$. Gluon saturation ideas were initially studied through nonlinearities of the classical Yang-Mills field theory in \cite{McLerran:1993ni, McLerran:1993ka, McLerran:1994vd} and the nonlinear functional evolution equation, Balitsky-Kovchegov/ Jalilian-Marian-Iancu-McLerran-Wiegert-Leonidov-Kovner (BK-JIMWLK) equation, derived in \cite{ Balitsky:1995ub, Kovchegov:1999yj, Kovchegov:1999ua, Jalilian-Marian:1996mkd, Jalilian-Marian:1997qno, Jalilian-Marian:1997jhx, Jalilian-Marian:1997ubg, Kovner:2000pt, Weigert:2000gi, Iancu:2000hn, Iancu:2001ad, Ferreiro:2001qy}. 

Within the CGC effective theory, one of the most frequently used observables to study the gluon saturation effects is deep inelastic scattering (DIS) on a dense target since it provides a clean environment to probe gluon saturation. DIS related observables are routinely computed in the dipole factorization \cite{Bjorken:1970ah,Nikolaev:1990ja} where the the incoming lepton emits a virtual photon which splits into a quark-antiquark pair. Then, the quark-antiquark pair is scattered on the dense target. The interaction between the quark-antiquark pair is described within the CGC framework and the rescatterings off the dense target are encoded in the Wilson lines. On the other hand, the splitting of the photon to the quark-antiquark pair is computed perturbatively. 

Even though the gluon saturation phenomenon has shown its hints in experimental data from Relativistic Heavy Ion Collider (RHIC) in the USA, the Large Hadron Collider (LHC) at CERN and HERA at DESY, no consensus has yet been reached concerning the discovery of the gluon saturation. One of the main features of the upcoming Electron-Ion Collider (EIC) in the USA is to provide a cleaner environment  to study the saturation effects than at the LHC, but at lower colliding energies. Moreover, it will provide a much higher luminosity than HERA, allowing the study of less inclusive and more discriminating observables. Despite the lower colliding energy, saturation effects should be enhanced at EIC compared to HERA thanks to the use of nuclear targets.
 Therefore, for saturation sensitive observables, a more precise theoretical framework is needed in order to fully benefit from the EIC. This precision can be provided either by performing the calculation of the observables at higher orders in coupling constant $\alpha_s$ or by improving the kinematical approximations adopted in the standard CGC calculations. 

Over the last ten years, we have witnessed a great effort to compute the next-to-leading order (NLO) corrections to the DIS related observables in the CGC. Computations of inclusive DIS for massless quarks \cite{Balitsky:2010ze, Balitsky:2012bs, Beuf:2011xd, Beuf:2016wdz, Beuf:2017bpd, Ducloue:2017ftk, Hanninen:2017ddy} and its fit to HERA data \cite{Beuf:2020dxl}, as well as inclusive DIS for massive quarks \cite{Beuf:2021qqa, Beuf:2021srj, Beuf:2022ndu} and its fit to HERA data \cite{Hanninen:2022gje}, diffractive structure functions \cite{Beuf:2022kyp}, diffractive dijet production in DIS \cite{Boussarie:2016ogo,Iancu:2021rup,Iancu:2022lcw}, exclusive light \cite{Boussarie:2016bkq, Mantysaari:2022bsp} and heavy \cite{Mantysaari:2021ryb, Mantysaari:2022kdm} vector meson production in DIS, inclusive dijet \cite{Caucal:2021ent, Taels:2022tza, Caucal:2022ulg}  and photon + dijet  production in DIS \cite{Roy:2019hwr}, inclusive \cite{Bergabo:2022tcu,Iancu:2022gpw} and diffractive \cite{Fucilla:2022wcg} dihadron, single inclusive hadron \cite{Bergabo:2022zhe} production in DIS have been performed at NLO accuracy (or partial NLO in some of the cases). 
 
On the other hand, as mentioned previously, another way of increasing the precision of the calculations in the saturation framework is to improve the adopted kinematical approximations. The key approximation used in the CGC framework is the eikonal one and in general it amounts to accounting for the contributions that are leading in collision energy and neglecting all the energy suppressed terms in the calculation of the observables. More precisely, from the point of view of the highly boosted target that is described by the background field ${\cal A}_a^{\mu}(x)$, eikonal approximation amounts to adopting the following three assumptions: (i) the background field is localized in the longitudinal directions (around $x^+=0$), (ii) only the leading component of the background field (which in our setup corresponds to ``$-$'' component) is taken into account during the interaction of the projectile parton with the target and other components of background field of the target (transverse and ``$+$'' components that are suppressed by the Lorentz boost factor) are neglected, and (iii) the dynamics of the target is neglected which amounts to assuming that the background target field is independent of $x^-$ coordinate due to Lorentz time dilation. These three assumptions together give the well-known shockwave approximation and in this case the background field of the target has the form
\begin{equation}
{\cal A}^{\mu}_a(x^-,x^+,\x)\approx \delta^{\mu -}\delta(x^+){\cal A}^-_a(\x)\, . 
\end{equation}
While eikonal approximation is reliable for the LHC energies and the computations adopting this approximation are quite successful to describe the experimental data, for energies at RHIC and the future EIC, energy suppressed corrections can become sizable and thus should be included in order to increase the precision of phenomenological studies. This idea motivated a lot of studies that aim to go beyond the eikonal approximation in the CGC framework by relaxing all three assumptions listed above. Initially, in Refs. \cite{Altinoluk:2014oxa, Altinoluk:2015gia} subeikonal corrections that stem from considering a finite width target are computed for the gluon propagator and its application to single inclusive gluon production and various spin asymmetries in central $pA$ collisions were studied at next-to-next-to-eikonal accuracy.  In Refs. \cite{Altinoluk:2015xuy,Agostini:2019avp}, it was shown that these corrections can be attributed to the modifications of the Lipatov vertex in $pp$ collisions. The effects of subeikonal corrections on the azimuthal harmonics for $pp$ \cite{Agostini:2019hkj} and for $pA$ \cite{Agostini:2022ctk,Agostini:2022oge} collisions were also investigated. Recently,  next-to-eikonal (NEik) corrections that are related with the transverse component of the background field \cite{Altinoluk:2020oyd} and the dynamics of the target  \cite{Altinoluk:2021lvu} have been computed for scalar and quark propagators. Apart from the aforementioned works that focus on the derivation of the subeikonal corrections to the parton propagators and their applications to observables, in \cite{Kovchegov:2015pbl, Kovchegov:2016weo, Kovchegov:2016zex, Kovchegov:2017jxc, Kovchegov:2017lsr, Kovchegov:2018znm, Kovchegov:2020hgb, Adamiak:2021ppq, Kovchegov:2021lvz, Kovchegov:2021iyc, Cougoulic:2022gbk, Kovchegov:2022kyy} quark and gluon helicity evolutions have been computed at next-to-eikonal accuracy. In \cite{Cougoulic:2019aja, Cougoulic:2020tbc} helicity dependent extensions of the CGC at next-to-eikonal accuracy have been studied.  In \cite{Chirilli:2018kkw, Chirilli:2021lif}  subeikonal corrections to both quark and gluon propagators have been calculated in the high-energy operator product expansion (OPE) formalism, and applied to study the polarized structure function $g_1$ at low $x$. Moreover, rapidity evolution of transverse momentum dependent parton distributions (TMDs) that interpolates between the low and moderate energies are studied in \cite{Balitsky:2015qba, Balitsky:2016dgz, Balitsky:2017flc, Balitsky:2017gis, Balitsky:2019ayf}. A similar idea is pursued in \cite{Boussarie:2020fpb, Boussarie:2021wkn} to study the interpolation between the low and moderate values of $x$ for the unintegrated gluon distributions. Finally, an approach based on longitudinal momentum exchange between the projectile and the target during the interaction have been followed in \cite{Jalilian-Marian:2017ttv, Jalilian-Marian:2018iui, Jalilian-Marian:2019kaf} to study the subeikonal effects. The effects of subeikonal corrections are also studied in the context of orbital angular momentum in \cite{Hatta:2016aoc, Boussarie:2019icw}. 

In the single photon exchange approximation, DIS process can be expressed as the product of a leptonic tensor, encoding the virtual photon emission by the incoming lepton, and a hadronic tensor, encoding the interaction of the virtual photon with the target. Integrating over the azimuthal angle of the scattered lepton, the hadronic tensor is projected into two scalar functions, as  
\begin{align}
\frac{d\sigma^{l+\textrm{target}\rightarrow l'+ \textrm{dijet}+X}}{d x_{Bj}\, d Q^2\, d {\rm P.S.}}
=&\, 
\frac{\alpha_{\textrm{em}}}{\pi\, x_{Bj}\, Q^2}
\left[
\left(1-y+\frac{y^2}{2}\right) \frac{d\sigma_{\gamma^{*}_T\rightarrow \textrm{dijet}}}{d {\rm P.S.}}(x_{Bj},Q^2)
+\left(1-y\right) \frac{d\sigma_{\gamma^{*}_L\rightarrow \textrm{dijet}}}{d {\rm P.S.}}(x_{Bj},Q^2)
\right]
\, .
 \label{lepton_to_photon_DIS} 
\end{align}
Let us denote $k_l^{\mu}$ the momentum of the incoming lepton, $q^{\mu}$ the momentum of the exchanged virtual photon (so that $k_l^{\mu}-q^{\mu}$ is the momentum of the scattered lepton), and $P_{\textrm{tar}}^{\mu}$  the momentum of the target. Then, the usual Lorentz invariant variables for DIS are defined as follows. The Mandelstam $s$ variable for the lepton-target collision is $s=(k_l+P_{\textrm{tar}})^2$, the photon virtuality is $Q^2=-q^2>0$, and the Bjorken variable is $x_{Bj}=Q^2/(2P_{\textrm{tar}}\!\cdot\! q)$. Finally, the inelasticity variable $y$ is defined as $y=(2P_{\textrm{tar}}\!\cdot\! q)/s=Q^2/(s x_{Bj})$. In Eq.~\eqref{lepton_to_photon_DIS}, $d {\rm P.S.}$ is the Lorentz invariant phase space measure for the produced hadronic final state (the dijet in our example). The two scalar functions appearing in Eq.~\eqref{lepton_to_photon_DIS} can be interpreted as cross sections for the virtual photon - target subprocess, in which the photon has either a transverse or longitudinal polarization. The aim of this paper is to calculate these two cross sections at NEik accuracy and at leading order in the QCD coupling $\alpha_s$, in which the produced quark and antiquark are identified as jets.

The outline of this paper is as follows. In Sec.~\ref{sec:LSZ}, we write the S-matrix elements for the virtual photon to dijet processes in terms of propagators in the background field of the target. At NEik accuracy, there are two contributions to these S-matrix elements, with the photon splitting to quark-antiquark pair occuring either before or inside the target.  
In Sec.~\ref{sec:quark_prop}, we recall our earlier results on the NEik corrections to the quark propagator through the whole target, and calculate the quark propagator from inside to after the target at eikonal accuracy. 
Sec.~\ref{sec:in} and \ref{sec:bef} are devoted to the calculation of the contribution to the S-matrix with photon splitting inside the target and before the target respectively.
Then, Sec.~\ref{sec:gammaL_cross_sec} and \ref{sec:gammaT_cross_sec} are devoted to the calculation of dijet production cross sections from longitudinal and transverse virtual photon respectively, appearing in Eq.~\eqref{lepton_to_photon_DIS}. Finally, the summary and outlook are provided in Sec.~\ref{sec:conclu}.
The detailed Dirac algebra calculations required to obtain the cross section are presented in Appendix~\ref{sec:Dirac_alg}.  
In Appendix~\ref{sec:zminus_dep_cross_sec}, we explain how to calculate the cross section for scattering processes off a dynamical background field (with dependence on the $z^-$ light-cone coordinate), generalizing the standard formula for the static background field case.


\section{Reduction formula for the S-matrix and integrated propagators\label{sec:LSZ}}

\subsection{LSZ-type reduction formula}

Let us consider the process in which a virtual photon of momentum $q$ and polarization $\lambda$ splits into a quark of momentum $k_1$ and an antiquark of momentum $k_2$, in the presence of a gluon background field ${\cal A}^{\mu}(x)$ representing the target. The S-matrix element for that process can be obtained following the LSZ approach.
The first non-zero contribution to the S-matrix is at order $e$ in QED, due to the photon splitting vertex. Moreover, that process does not require QCD interactions beyond the scattering with the background field. We are interested in the lowest order contribution in perturbation theory in a possibly strong background field, which is then of order $e\, g^0$ at the S-matrix level, with $g\, {\cal A}^{\mu}(x)$ resummed to all orders. At this order, the S-matrix can be written as
\footnote{
We use the metric signature $(+,-,-,-)$. We use $x^\mu$ for a Minkowski 4-vector. In a light-cone basis we have
$
x^\mu=(x^+,{\bf x},x^-)
$
where $x^\pm=(x^0\pm x^3)/\sqrt{2}$ and ${\bf x}$ denotes a transverse vector with components $x^i$. 
We will also use the notations 
$
\ux = (x^+,{\bf x})
$
and 
$\uk = (k^+,{\bf k})$. Finally we introduce the on-shell momentum 
$
\ck^\mu = (k^+,{\bf k},\ck^-)
$
constructed from $\uk =(k^+,{\bf k})$, with by definition $\ck^-=({\bf k}^2+m^2)/(2k^+)$ for a quark of mass $m$. We also use the condensed notations $u(1)\equiv u(k_1,h_1)$ and $v(2)\equiv v(k_2,h_2)$ for the Dirac spinors, where $h_1$ or $h_2$ is the light-front helicity.}
\begin{align}
S_{q_1 \bar q_2 \leftarrow \gamma^*} 
=&\,   
\int d^4z \,
 \epsilon^\lambda_\mu (q)\,  e^{-i q \cdot z} \;
 \langle 0|
 d_{\textrm{out}}(2) b_{\textrm{out}}(1)
 :\!\bar{\Psi}(z)
 \,(-i e e_f\gamma^\mu)\,
 {\Psi}(z)\!:\!
 |0\rangle
 \, .
 \label{LSZ_S_1} 
\end{align}
In Eq.~\eqref{LSZ_S_1}, $b_{\textrm{out}}(1)$ and $d_{\textrm{out}}(2)$ are the annihilation operators for the outgoing quark and antiquark in the asymptotic free Fock space, whereas $\epsilon^\lambda_\mu (q)\,  e^{-i q \cdot z}$ accounts for the incoming virtual photon, and the normal-ordered current operator comes from the photon splitting vertex. The quark field ${\Psi}(z)$ in Eq.~\eqref{LSZ_S_1} is a quantum field in a modified interaction picture, in such a way that the evolution of ${\Psi}(z)$ is generated by a Hamiltonian quadratic in the quantum fields, but with terms of any order in the background field. Hence, not only the free limit of the theory but also the interactions of quantum particles with the background field contributes to the evolution of ${\Psi}(z)$. Only the interactions between quantum particles are removed from the evolution of ${\Psi}(z)$ in this picture with respect to the Heisenberg picture. Moreover, we assume that the background field alone cannot lead to pair creation or pair annihilation of quantum particles. Then, the only possible contribution to the expectation value in Eq.~\eqref{LSZ_S_1} factorizes as
\begin{align}
 \langle 0|
 d_{\textrm{out}}(2) b_{\textrm{out}}(1)
 :\!\bar{\Psi}(z)
 \,(-i e e_f\gamma^\mu)\,
 {\Psi}(z)\!:\!
 |0\rangle
=&\, 
 \langle 0|
  b_{\textrm{out}}(1)
 \bar{\Psi}(z)
 |0\rangle\,
 (-i e e_f\gamma^\mu)\,
\langle 0|
 d_{\textrm{out}}(2) 
 {\Psi}(z)
 |0\rangle
 \, .
 \label{LSZ_S_2} 
\end{align}
Assuming that the background field decays fast enough at large $x^+$, the annihilation operators $b_{\textrm{out}}(1)$ and $d_{\textrm{out}}(2)$ can be expressed in terms of the quark field  ${\Psi}$ in the same modified interaction picture as~\cite{Altinoluk:2020oyd}
\begin{align}
b_{\textrm{out}}(1)
=&\, 
{\rm lim}_{x^+\to +\infty} \int d^2\x \int dx^-\,
e^{i\ck_1 \cdot x}\, 
\bar u(1) \gamma^+ {\Psi}(x)
\label{LSZ_b}
\\
d_{\textrm{out}}(2) 
=&\, 
{\rm lim}_{y^+ \to +\infty} 
\int d^2\y \int dy^- \,  e^{i \ck_2 \cdot y}\,
 \bar{\Psi}(y) \gamma^+ v(2)
\label{LSZ_d}
\, .
\end{align}
Finally, the Feynman quark propagator in the gluon background field is defined in terms of the field ${\Psi}$ in that picture as
\begin{align}
S_F(x,y) 
=&\,
 \langle 0|T\left(
  {\Psi}(x)\, 
 \bar{\Psi}(y)
 \right)
 |0\rangle\,
=
\theta(x^0\!-\!y^0)\,
 \langle 0|
  {\Psi}(x)\, 
 \bar{\Psi}(y)
 |0\rangle\,
 -
 \theta(y^0\!-\!x^0)\,
 \langle 0|
 \bar{\Psi}(y)\,
  {\Psi}(x)
 |0\rangle
 \, .
 \label{Feynman_quark_prop_def}
\end{align}
where the implicit spinor indices and color indices of the fields are not contracted, and where the minus sign in the second term is due to the Fermi-Dirac statistics for the quarks.
Hence, one has
\begin{align}
 \langle 0|
   {\Psi}(x)
 \bar{\Psi}(z)
 |0\rangle
=&\, 
S_F(x,z) \quad\textrm{for }x^+\to +\infty
\nn\\
 \langle 0|
 \bar{\Psi}(y)\,
  {\Psi}(z)
 |0\rangle
=&\, 
-S_F(z,y) \quad\textrm{for }y^+\to +\infty
\, .
\end{align}
All in all, one arrives at the expression
\begin{align}
&S_{q_1 \bar q_2 \leftarrow \gamma^*} =   
{\rm lim}_{x^+,y^+ \to +\infty} \int d^2\x \int dx^- \int d^2\y \int dy^- e^{i\ck_1 \cdot x}\,  e^{i \ck_2 \cdot y}\nn \\
& \times 
 \epsilon^\lambda_\mu (q) \int d^4z \, e^{-i q \cdot z} \;
 \bar u(1) \gamma^+ S_F(x,z)
 \,(-i e e_f\gamma^\mu)\,
\left(- S_F(z,y)
\right)
 \gamma^+ v(2)
 \, ,
 \label{LSZ_S} 
\end{align}
for the S-matrix element at lowest order $e\, g^0$ but with the interactions with the background field resummed to all orders.



For convenience, we define the reduced quark and antiquark propagators 
\begin{align}
&\tilde S^{q}_F(z)
={\rm lim}_{x^+ \to +\infty} \int d^2\x \int dx^- 
e^{i\ck_1 \cdot x} 
\bar u(1) \gamma^+ S_F(x,z)
\label{integrated_q_prop}
\\
&\tilde S^{\bar q}_F(z)
= {\rm lim}_{y^+ \to +\infty} \int d^2\y \int dy^-  
e^{i \ck_2 \cdot y} \; (-1)\; S_F(z,y)
\gamma^+ v(2)
\label{integrated_qbar_prop}
\, .
\end{align}
so that the S-matrix element can be written as
\begin{align}
&S_{q_1 \bar q_2 \leftarrow \gamma^*} = -i e e_f \, \epsilon^\lambda_\mu (q) 
 \int d^4z 
e^{-i q \cdot z}  \;
\tilde S^{q}_F(z)
\gamma^\mu\,
\tilde S^{\bar q}_F(z)
\label{S_red} 
\, .
\end{align}


\subsection{Vacuum case}

As a preliminary, let us consider the vacuum limit of Eq.~\eqref{LSZ_S}, corresponding to the absence of background field. No particle production is expected in that case, if the photon is spacelike, but it is instructive to check why this is the case in our setup. In the absence of background field, the quark propagators involved in Eq.~\eqref{LSZ_S} reduce to the standard Feynman propagator in vacuum
\begin{align}
&{S}_{0,F}(x,y) = 
\int \frac{d^4 p}{(2\pi)^4} \, 
e^{-i (x-y)\cdot {p}}\,
\frac{i (\slashed{{p}}+m)}{(p^2\!-\!m^2\!+\!i \epsilon)}
\label{vac_q}
\, .
\end{align}
In the vacuum, the reduced quark propagator \eqref{integrated_q_prop} is then
\begin{align}
\tilde S^{q}_{0,F}(z)
= &\,
{\rm lim}_{x^+ \to +\infty} \int d^2\x \int dx^- 
e^{ik_1 \cdot x} \int \frac{d^4 p}{(2\pi)^4} \, 
e^{-i (x-z)\cdot {p}}\,
\bar u(1) \gamma^+ \frac{i (\slashed{{p}}+m)}{(p^2\!-\!m^2\!+\!i \epsilon)}
\nn\\
= &\,
{\rm lim}_{x^+ \to +\infty} e^{ik_1^+ z^-}\, e^{-i\k_1 \cdot \z}\, e^{i\ck_1^-  x^+}
 \int \frac{d^4 p}{(2\pi)^4} \, 
(2\pi)^3 \delta^{(3)}(\up\!-\!\uk_1)
e^{-i (x^+-z^+) {p}^-}\,
\bar u(1) \gamma^+ \frac{i ({\slashed{\check{p}}}+m)}{(p^2\!-\!m^2\!+\!i \epsilon)}
\label{integrated_q_prop_vac_1}
\, .
\end{align}
Indeed, due to the presence of the $\gamma^+$, the term in $p^- \gamma^+$ drops from $\slashed{{p}}$. We can then replace $\slashed{{p}}$ by $\slashed{\check{p}}$ to emphasize that the numerator is independant of $p^-$. The integration over $p^-$ thus only receives a simple pole contribution, and one obtains  
\begin{align}
\tilde S^{q}_{0,F}(z)
= &\,
{\rm lim}_{x^+ \to +\infty} e^{ik_1^+ z^-}\, e^{-i\k_1 \cdot \z}\, e^{i\ck_1^- x^+}
e^{-i (x^+-z^+) {\ck}_1^-}\, 
\Big[\theta(x^+\!-\!z^+)\theta(k_1^+)-\theta(z^+\!-\!x^+)\theta(-k_1^+)\Big]
\bar u(1) \gamma^+ \frac{ ({\slashed{\check{k}}}_1+m)}{2k_1^+}
\nn\\
= &\,
 e^{i\ck_1 \cdot z}\: 
\frac{\theta(k_1^+)}{2k_1^+}\:
\bar u(1)\bigg[\{ \gamma^+,  \slashed{\check{k}}\}
 -  ({\slashed{\check{k}}}_1-m)\gamma^+
\bigg]
\nn\\
= &\,
 e^{i\ck_1 \cdot z}\: 
\bar u(1)
\label{integrated_q_prop_vac_2}
\, ,
\end{align}
using the identity $\bar u(1)({\slashed{\check{k}}}_1-m)=0$, and dropping $\theta(k_1^+)$ because the condition $k_1^+>0$ for the produced quark is already implicitly present in the definition of the observable. 

Similarly, for the antiquark reduced propagator \eqref{integrated_qbar_prop}, one finds in the vacuum case
\begin{align}
\tilde S^{\bar q}_{0,F}(z)
=&\, 
{\rm lim}_{y^+ \to +\infty} \int d^2\y \int dy^-  
e^{i \ck_2 \cdot y} \; 
\int \frac{d^4 p'}{(2\pi)^4} \, 
e^{-i (z-y)\cdot {p'}}\:
(-1)\:
\frac{i (\slashed{{p}}'+m)}{({p'}^2\!-\!m^2\!+\!i \epsilon)}
\gamma^+ v(2)
\nn\\
=&\, 
 e^{i\ck_2 \cdot z}\: 
v(2)
\label{integrated_qbar_prop_vac}
\, .
\end{align}

Now we can compute the vacuum contribution to the S-matrix element, as
\begin{align}
S_{q_1 \bar q_2 \leftarrow \gamma*}\bigg|_{\textrm{vac.}}
 =&\,
 -i e e_f \, \epsilon^\lambda_\mu (q)  \int d^4z 
e^{-i q \cdot z}  \tilde S^{q}_{F0}(z)
\gamma^\mu
 \tilde S^{\bar q}_{F0}(z)
 \nn\\
 =&\,
 -i e e_f \, \epsilon^\lambda_\mu (q)\;
 (2\pi)^4 \delta^{(4)}(q-\check{k}_1 - \check{k}_2)\;  \bar u(1) \gamma^\mu  v(2)
 \label{S-vac} 
 \, .
\end{align}

Hence, we obtain the 4-momentum conservation $\ck_1^\mu + \ck_2^\mu = q^\mu$ in addition to the on-shell conditions $\ck_1^2=m^2$ and $\ck_2^2=m^2$, and to the requirements $k_1^+>0$ and $k_1^+>0$ for produced particles. From all of these relations, it is possible to calculate the virtuality of the incoming photon as
\begin{align}
q^2
 =&\,
2q^+(\ck_1^-+\ck_2^-) -\q^2 
= 
2q^+ \left( \frac{(\k_{1}^2+m^2)}{2k_1^+} +\frac{(\k_{2}^2+m^2)}{2k_2^+} \right) -\q^2
 \label{virtuality_vacuum_1} 
 \, .
\end{align}
Introducing the notations
\begin{align}
z \equiv&\, \frac{k_1^+}{q^+} = 1- \frac{k_2^+}{q^+}
\nn\\
\K\equiv &\, \k_1 - z\q = -\k_2+(1\!-\!z)\q
\, ,
\end{align}
with the positivity of $k_1^+$ and $k_2^+$ implying $0<z<1$, one finds from Eq.~\eqref{virtuality_vacuum_1} 
\begin{align}
q^2
 =&\,
 \frac{(\K^2+m^2)}{z(1\!-\!z)} >0
 \label{virtuality_vacuum_2} 
 \, ,
\end{align}
meaning that the photon is timelike. However, the photon is always spacelike ($q^2<0$) in a DIS process. Hence, the condition $q^2<0$ is not compatible with the 4-momentum conservation $\ck_1^\mu + \ck_2^\mu = q^\mu$ and the other requirements, so that the vacuum contribution \eqref{S-vac} to the S-matrix element vanishes in that case. 
 We have simply checked the obvious statement that, there cannot be a DIS process (including particle production) without interaction with the target.


\subsection{General structure of the S-matrix element\label{sec:gen_struct}}

\begin{figure}
\setbox1\hbox to 10cm{
\includegraphics{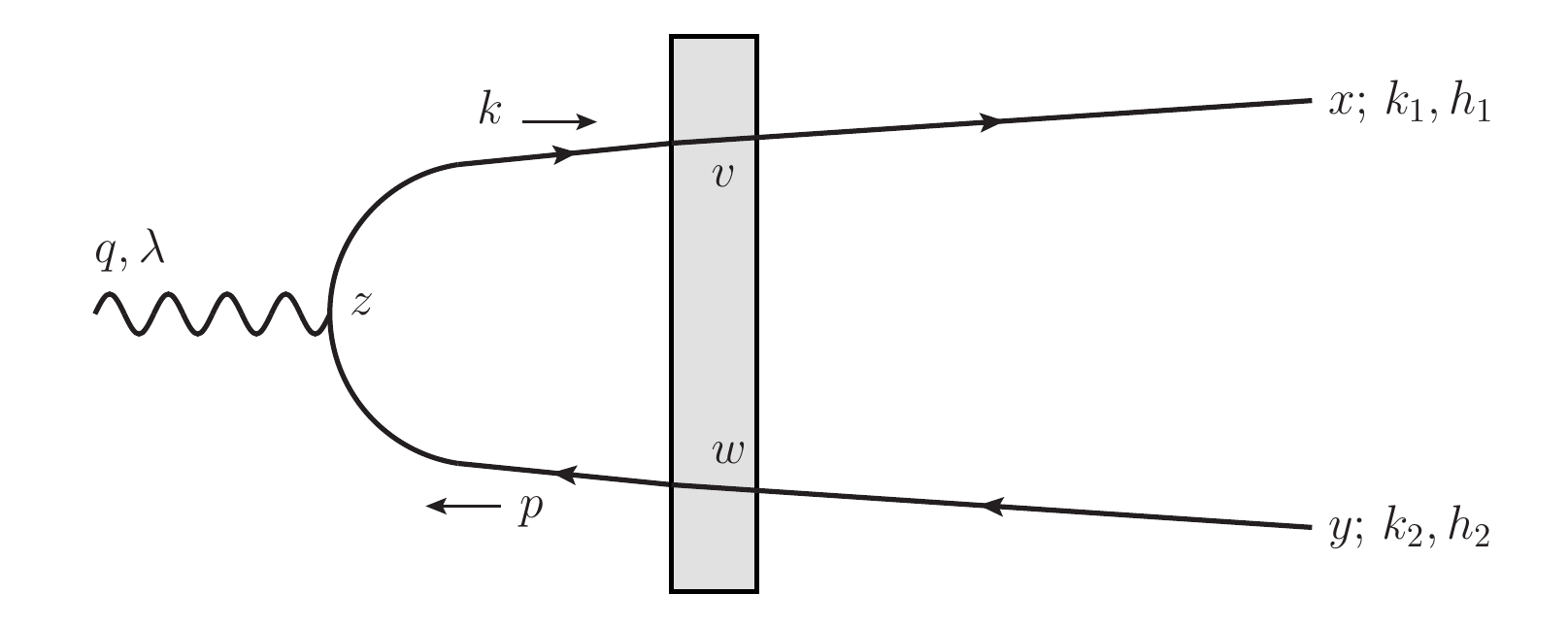}
}
\setbox2\hbox to 10cm{
\includegraphics{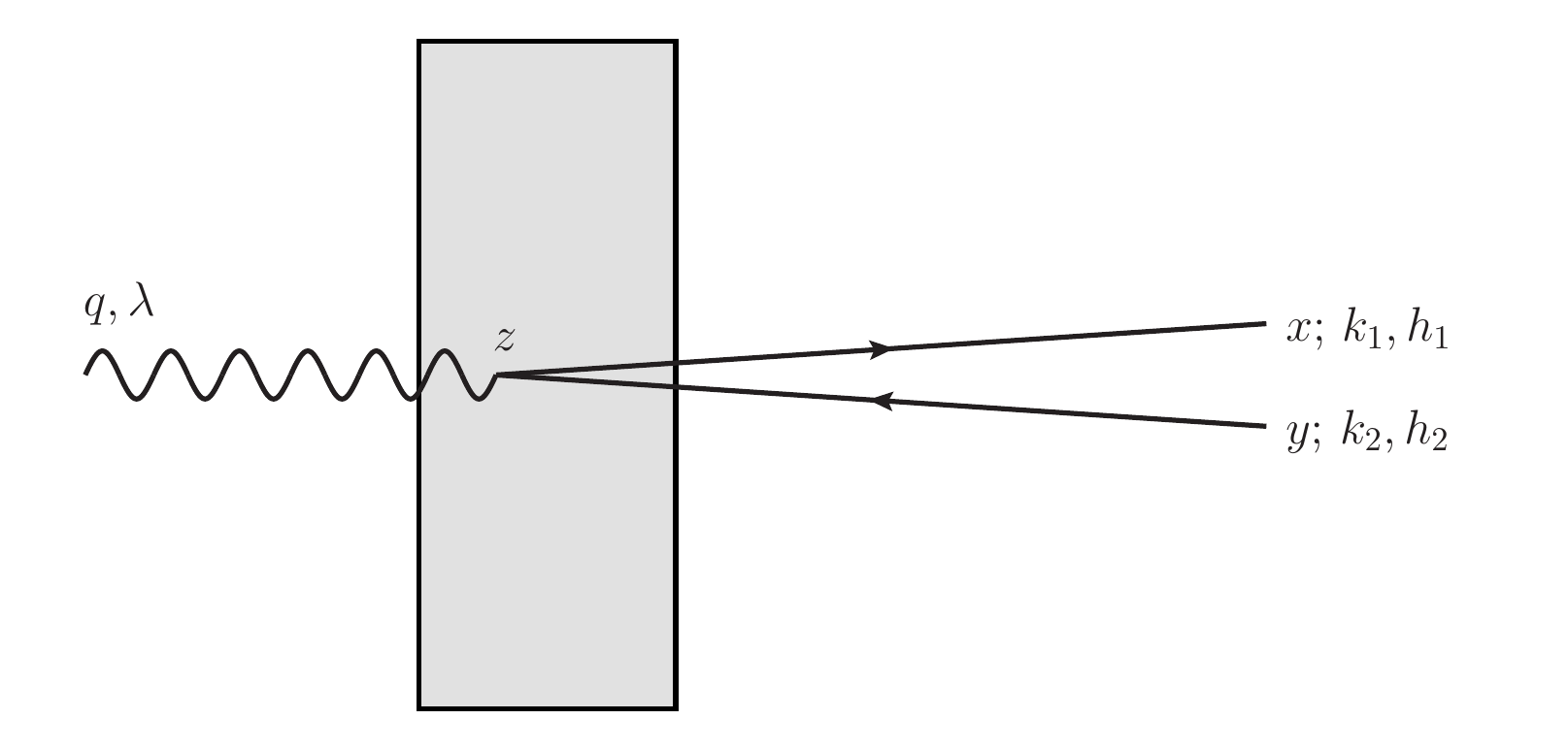}
}
\begin{center}
\resizebox*{10cm}{!}{\hspace{-7cm}\mbox{\box1 \hspace{9cm} \box2}}
\caption{\label{Fig:diagrams} Contributions to dijet production in DIS at next-to-eikonal accuracy: photon splitting into a $q\bar q$ pair before reaching the target (left panel) and photon splitting inside the target (right panel).}
\end{center}
\end{figure}

Let us now consider the process of dijet production in DIS at low $x$, with a gluon background field representing the target. Physically, the gluon field strength of the target should decay faster than any power for $z^+\rightarrow \pm \infty$ due to confinement along the longitudinal direction. For simplicity, we assume this background field strength to have a finite support of length $L^+$ along $z^+$, allowing us define the range from $-\infty$ to $-L^+/2$ as before the target, the range from $-L^+/2$ to $+L^+/2$ as inside the target, and the range from $+L^+/2$ to $+\infty$ as after the target. Moreover we choose a gauge in which not only the field strength but also the gauge field is vanishing outside of the target, for example the light-cone gauge $A^+=0$. 
The S-matrix element from Eq.~\eqref{S_red} can then be split into three contributions as
\ba
 S_{q_1 \bar q_2 \leftarrow \gamma^*} &=& -i e e_f \, \epsilon^\lambda_\mu (q) 
\int d^2\z \int dz^- \int_{-\infty}^{-L^+/2} dz^+ 
e^{-i q \cdot z} \tilde S_F^q(z)_{\alpha\delta}
\gamma^\mu
\tilde S_F^{\bar q}(z)_{\delta\beta} 
\nonumber\\
&&-i e e_f \, \epsilon^\lambda_\mu (q) 
\int d^2\z \int dz^- \int_{-L^+/2}^{L^+/2} dz^+ 
e^{-i q \cdot z} \tilde S_F^q(z)_{\alpha\delta}
\gamma^\mu
\tilde S_F^{\bar q}(z)_{\delta\beta}
\nonumber\\
&&
-i e e_f \, \epsilon^\lambda_\mu (q) 
\int d^2\z \int dz^- \int_{L^+/2}^{+\infty} dz^+ 
e^{-i q \cdot z} \tilde S_{0,F}^q(z)_{\alpha\delta}
\gamma^\mu
\tilde S_{0,F}^{\bar q}(z)_{\delta\beta} 
\, , 
\label{S_red_1}
\ea
according to the region in which the photon splitting happens. In the third term, the photon crosses the medium and splits afterwards. At the LO accuracy in QED, that we are considering here, the photon does not interact with the gluon field, so that this third term is a vacuum contribution. By contrast, the first two terms in Eq.~\eqref{S_red_1} contain both vacuum and medium-induced contributions. 
As we have recalled in the previous subsection, the total vacuum contribution is zero. Hence, we can subtract the total vacuum contribution region by region from the S-matrix element, and obtain 
%
\begin{align}
& S_{q_1 \bar q_2 \leftarrow \gamma^*} = S^{\rm bef}_{q_1 \bar q_2 \leftarrow \gamma^*}  
+ S^{\rm in}_{q_1 \bar q_2 \leftarrow \gamma^*} 
\end{align}
where
\ba
\label{S-sep-bef} 
S^{\rm bef}_{q_1 \bar q_2 \leftarrow \gamma^*} &=& -i e e_f \, \epsilon^\lambda_\mu (q) 
\int d^2\z \int dz^- \int_{-\infty}^{-L^+/2} dz^+ 
e^{-i q \cdot z}
 \bigg[\tilde S_F^q(z)
\gamma^\mu 
 \tilde S_F^{\bar q}(z)
- 
 e^{i(\ck_1+\ck_2) \cdot z}\: 
\bar u(1)\gamma^\mu v(2)
\bigg]\\
S^{\rm in}_{q_1 \bar q_2 \leftarrow \gamma^*} &=&  -i e e_f \, \epsilon^\lambda_\mu (q) 
\int d^2\z \int dz^- \int_{-L^+/2}^{L^+/2} dz^+ 
e^{-i q \cdot z} 
 \bigg[\tilde S_F^q(z)
\gamma^\mu 
\tilde S_F^{\bar q}(z)
- 
e^{i(\ck_1+\ck_2) \cdot z}\: 
\bar u(1)\gamma^\mu v(2)
\bigg]
\label{S-sep-in}
\ea
are the medium-induced contributions corresponding to photon spliting before (see Fig. \ref{Fig:diagrams} left panel) or inside (see Fig. \ref{Fig:diagrams} right panel) the target, respectively.
%
%

\subsection{Power counting beyond the eikonal approximation\label{sec:power_counting}}

Since the aim of this study is the calculation of the first sub-leading corrections beyond the eikonal approximation, the NEik corrections, let us remind how we define the power counting in the high-energy limit in order to have an unambiguous definition of the NEik corrections. This issue has been discussed in-depth in Ref.~\cite{Altinoluk:2021lvu}, and we will provide here only a short reminder.

The high-energy limit of a collision process can be understood for example as the limit of infinite Lorentz boost of the target. In that case, one can classify contributions according to their scaling with the Lorentz boost factor $\gamma_t$ of the target. 
\begin{description}
\item[Scaling of the background field]  Under a large active boost of the target along the $x^-$ direction, the components of the background field strength are transformed following the standard Lorentz transformation rules for tensors. In light-cone coordinates, the tensor components are enhanced by a factor $\gamma_t$ for each upper ``$-$" or lower ``$+$" index under such a boost, and are suppressed by a factor $1/\gamma_t$ for each lower ``$-$" or upper ``$+$" index. Hence, the components of the background field strength scale as 
\ba
{\cal F}^{-j}&\propto&\gamma_t\gg1\\
{\cal F}^{ij}&\propto& (\gamma_t)^0=1\\
{\cal F}^{-+}&\propto&(\gamma_t)^0=1\\
{\cal F}^{+j}&\propto&\frac{1}{\gamma_t}\ll1
\ea
under a large boost of the target along $x^-$ direction. These transformation rules can be extended to the background gauge field of the target, resulting in the following scaling: 
\begin{align}
\mathcal{A}^-(x) \propto &\, \gamma_t\gg1
\label{Aminus_scaling}
\\
\mathcal{A}^j(x)\propto &\,(\gamma_t)^0=1
\label{Aperp_scaling}
\\
\mathcal{A}^+(x) \propto &\, \frac{1}{\gamma_t}\ll1
\label{Aplus_scaling}
\end{align}
%
%
\item[Scaling of derivatives] All the components of the momenta associated with the projectile photon or the produced jets are defined to be invariant under a large boost of the target. By contrast, components of the momenta associated with the target, or equivalently derivatives acting on the background field, follow the same scaling rules based on the counting of ``$+$" and ``$-$" indices. Indeed, the action of a partial derivative on a tensor leads to a higher rank tensor, so that  
\ba
\partial_-{\cal F}^{\mu\nu}&\propto&\frac{1}{\gamma_t} \, {\cal F}^{\mu\nu} \ll {\cal F}^{\mu\nu}\\
\partial_+{\cal F}^{\mu\nu}&\propto&\gamma_t\, {\cal F}^{\mu\nu} \gg{\cal F}^{\mu\nu}\\
\partial_i{\cal F}^{\mu\nu}&\propto&(\gamma_t)^0\, {\cal F}^{\mu\nu} 
\ea
and similar rules apply for partial derivative acting on the background gauge field. Moreover, due to the scaling rules of the components of the background gauge field given in Eqs. \eqref{Aminus_scaling}, \eqref{Aperp_scaling} and \eqref{Aplus_scaling}, background covariant derivatives follow the scaling rules as partial derivatives when acting on the background field. 
\item[Scaling of the width of the target] Since the background field strength ${\cal F}^{\mu\nu}(x)$ represents a hadronic or nuclear target subject to confinement, it should decay faster than a power for $x^+\rightarrow \pm \infty$. Hence, the profile of ${\cal F}^{\mu\nu}(x)$ along $x^+$ has a finite width, that we note $L^+$. Under a large boost of the target along $x^-$ direction, that width scales as  
\begin{align}
L^+ =&\, O\left(\frac{1}{\gamma_t}\right)
\end{align}
due to Lorentz contraction. In particular, in the limit of infinite boost, ${\cal F}^{\mu\nu}(x)$ becomes a shockwave of vanishing width along the $x^+$ direction.
For the purpose of power counting, the finite width $L^+$ of ${\cal F}^{\mu\nu}(x)$ can be assimilated as a finite support for ${\cal F}^{\mu\nu}(x)$ along the $x^+$. In a generic gauge where $A^-(x)\neq0$ (for example the light-cone gauge $A^+(x)=0$), the background gauge field ${\cal A}^{\mu}(x)$ has a finite width of $O(L^+)$ along the  $x^+$ direction, which becomes a shockwave in the infinite boost limit. Therefore, each integration over the position $x$ of a background field strength or gauge field insertion is effectively restricted to the small width $L^+$ along the $x^+$ direction, and can thus be counted as a suppression by a factor of $L^+$ or equivalently of $1/\gamma_t$ at large $\gamma_t$. 

\end{description}
With these power counting rules, eikonal and NEik contributions to propagators can be defined as follows: 

\begin{itemize}
\item
In the eikonal approximation, only the leading component $\mathcal{A}^-(x)$ of the gauge field is kept. Due to its enhancement \eqref{Aminus_scaling} with $\gamma_t$, each power of  $\mathcal{A}^-(x)$ compensates the suppression due to the integration over its $x^+$ coordinate. Hence, multiple insertions of $\mathcal{A}^-(x)$ have to be resummed in the eikonal approximation, leading to the Wilson lines describing the interaction of each parton in the projectile with the target in the CGC formalism in the eikonal approximation. Due to the parametrically small $L^+$ width of the target, each $\mathcal{A}^-(x)$ insertion in a Wilson line has the same transverse position $\x$ in the eikonal approximation. 
Finally, expanding $\mathcal{A}^-(x)$ as a series in  $x^-$, each further term comes with an extra $\partial_-$
acting on the background field, and is thus further suppressed by an extra $1/\gamma_t$ factor under large boosts. Therefore, in the eikonal approximation only the zeroth order term in this gradient expansion is kept, so that the $x^-$ dependence of $\mathcal{A}^-(x)$ is neglected. 
\item
Going from eikonal to NEik accuracy for a propagator, one includes contributions which are suppressed by one power of $1/\gamma_t$ at large $\gamma_t$ compared to the eikonal contribution. Such terms arise as follows:
\begin{itemize}
\item by replacing an enhanced  $\mathcal{A}^-(x)$ insertion with a non-enhanced  $\mathcal{A}^j(x)$ insertion, 

\item or by accounting for the transverse motion of the projectile parton over the duration $L^+$ of the interaction with the target,

\item or by including terms with one $\partial_{x^-}$ derivative acting on the background field in the gradient expansion of $\mathcal{A}^-(x)$ in $x^-$. 

\end{itemize}

\end{itemize}
In practice, when calculating a cross section, one may encounter technical difficulties when squaring the amplitude if the gradient expansion of all $\mathcal{A}^-(x)$ insertions is performed around a fixed value of $x^-$ like $x^-=0$. In order to bypass this issue, one can instead perform the gradient expansion of all $\mathcal{A}^-(x)$ insertions around a common value of  $x^-$ and write the amplitude (or S-matrix) as integral over that variable. This amounts to resumming a subset of non-eikonal corrections together with the eikonal term and one obtains what we refer to as {\it generalized eikonal approximation}. 

In this study, the aim is to calculate the dijet production in DIS, including all types of NEik corrections in the gluon background field of the target. 
In the contribution \eqref{S-sep-bef}  to the S-matrix element, the photon splits at light-cone time $z^+$ before the medium, outside of the background field. Hence, the integration over $z^+$ does not bring a suppression at large $\gamma_t$ in that case.  
In order to calculate both the eikonal and NEik terms in the contribution \eqref{S-sep-bef}  to the S-matrix element, one should thus include both eikonal and NEik terms in the reduced propagators  $\tilde S_F^q(z)$ and $\tilde S_F^{\bar q}(z)$.

By contrast, in the contribution \eqref{S-sep-in} to the S-matrix element, the photon splits at light-cone time $z^+$ inside the medium, so that
the integration over $z^+$ brings a suppression by a factor $L^+$, and thus $1/\gamma_t$. 
Hence, the expression \eqref{S-sep-in} does not contribute at eikonal accuracy, and starts contributing only at NEik accuracy.
In order to calculate the NEik terms in the contribution \eqref{S-sep-in}  to the S-matrix element, we should thus restrict ourselves to the eikonal expression for the reduced propagators  $\tilde S_F^q(z)$ and $\tilde S_F^{\bar q}(z)$.

%
%


\section{Quark propagator in gluon background field and eikonal expansion\label{sec:quark_prop}}


The quark propagator in gluon background field has been studied at NEik accuracy in Refs.~\cite{Altinoluk:2020oyd} and \cite{Altinoluk:2021lvu}, in particular the case of propagator from before to after the target. We will remind the results of these studies in this section. In Sec.~\ref{sec:IA_propag}, we will also calculate explicitly the quark propagator from inside to after the target at Eikonal accuracy, required in order to evaluate the contribution  \eqref{S-sep-in} with photon splitting inside the target.

The quark propagator in a background field with only the $\mathcal{A}^-$ component is
\begin{align}
&{S}_F(x,y)_{\beta\alpha} \bigg|_{\textrm{pure }\mathcal{A}^{-}\textrm{, Eik.}}=
\mathbf{1}_{\beta\alpha}\; \delta^{(3)}(\ux\!-\!\uy)\; \gamma^+ 
\int \frac{dk^+}{2\pi} \frac{i}{2k^+} e^{-ik^+(x^--y^-)}
\nn\\
& 
+\int \frac{d^3\up}{(2\pi)^3} \int \frac{d^3\uk}{(2\pi)^3}\,
e^{-i x\cdot \check{p}+i y\cdot \check{k}}\,
\int dz^- e^{iz^-(p^+ - k^+)} \int d^2\z\, e^{-i \z\cdot (\p-\k)}\, \frac{(\slashed{\check{p}}+m)}{2p^+}\gamma^+
\nn\\
& \;\;\;
\times\, 
\bigg\{
\theta(x^+\!-\!y^+)\, \theta(p^+) \theta(k^+)\,
\mathcal{U}_F(x^+,y^+;\z,z^-)_{\beta\alpha} 
-
\theta(y^+\!-\!x^+)\, \theta(-p^+) \theta(-k^+)\,
\mathcal{U}_F^{\dagger}(y^+,x^+;\z,z^-)_{\beta\alpha}
\bigg\}
\frac{(\slashed{\check{k}}+m)}{2k^+}
\label{full_q_prop_Eik_generic_position_result}
\end{align}
at Eikonal accuracy, but with an overall $z^-$ dependence retained in the Wilson lines for further convenience. This corresponds to what we  call the generalized eikonal approximation for the propagator.  The first term on the right hand side of Eq.~\eqref{full_q_prop_Eik_generic_position_result} corresponds to an instantaneous quark exchange. By contrast, the two terms in the bracket correspond respectively to the propagation of a quark or of an antiquark in the background field. 
Here, the $z^-$ dependent Wilson line is defined as 
\begin{align}
\label{def:zminusWilson}
&{\cal U}_F(x^+,y^+; \z, z^-)\equiv {\cal P}_+\exp\left\{-ig\int_{y^+}^{x^+}dz^+ {\cal A}^-(z)\right\}
\end{align}
where ${\cal P}_+$ denotes ordering of color matrices along $z^+$ direction. In the following, when the $z^-$ dependence of Wilson line is dropped, it corresponds to evaluating Eq. \eqref{def:zminusWilson} at $z^-=0$, such that 
\begin{align}
&{\cal U}_F(x^+,y^+; \z)\equiv{\cal U}_F(x^+,y^+; \z,z^-=0)\, .
\end{align}
The expression \eqref{full_q_prop_Eik_generic_position_result} contains the full eikonal result for any kinematics, provided the background field only has a $\mathcal{A}^-$ component.
By contrast, if the background field has transverse components $\mathcal{A}_j$ as well, new contributions to the quark propagator arise already at Eikonal accuracy if at least one of the endpoints $x$ or $y$ is inside the medium. Such contributions will be derived in Sec.~\ref{sec:IA_propag}.


\subsection{Propagators from before to after the medium at NEik accuracy}

If we restrict ourselves to propagation trough the whole medium, with $y$ before and $x$ after (resp.~$x$ before and $y$ after) the support of the background field, the transverse components  $\mathcal{A}_j$ only matter at NEik accuracy, and the first term in the bracket in Eq.~\eqref{full_q_prop_Eik_generic_position_result} (resp. the second term in the bracket in Eq.~\eqref{full_q_prop_Eik_generic_position_result}) gives the entire result at Eikonal accuracy. 
In such cases, the full NEik corrections were calculated in Refs.~\cite{Altinoluk:2020oyd,Altinoluk:2021lvu}. For the case of quark propagating through the whole medium, meaning $x^+>L^+/2$ and  $y^+<-L^+/2$, one has\footnote{In the case of Wilson lines traversing the whole medium, in order to avoid unnecessary cluttering, we adopt the compact notation 
$\mathcal{U}_F(\z, z^-)\equiv \mathcal{U}_F\Big(\frac{L^+}{2},-\frac{L^+}{2};\z, z^-\Big)$ 
, and similarly we will use the notation $\mathcal{U}_F(\z)\equiv \mathcal{U}_F\Big(\frac{L^+}{2},-\frac{L^+}{2};\z\Big)$.}
\begin{align}
 S_F&(x,y) 
=
\int \frac{d^3\up}{(2\pi)^3} \int \frac{d^3\uk}{(2\pi)^3}\, \theta(p^+) \theta(k^+)
e^{-i x\cdot \check{p}}\, e^{i y\cdot \check{k}}\,
\int d z^- e^{iz^-(p^+-k^+)}
\, \int d^2 \z\, e^{-i\z\cdot(\p\!-\!\k)} \,
\nn\\
&
\times
\frac{(\slashed{\check{p}}+m)}{2p^+} \gamma^+
\Bigg\{\mathcal{U}_F(\z, z^-)
\nn\\
&
-\frac{(\p^j\!+\!\k^j)}{2(p^++k^+)}\int_{-\frac{L^+}{2}}^{\frac{L^+}{2}}dz^+\, 
\left[\mathcal{U}_F\Big(\frac{L^+}{2},z^+;\z, z^-\Big)\,
\overleftrightarrow{\mathcal{D}_{\z^j}}\,
\mathcal{U}_F\Big(z^+,-\frac{L^+}{2};\z, z^-\Big)\right]
\nn\\
&
-\frac{i}{(p^+ + k^+)}\int_{-\frac{L^+}{2}}^{\frac{L^+}{2}}dz^+\,  
\left[\mathcal{U}_F\Big(\frac{L^+}{2},z^+;\z, z^-\Big)\,
\overleftarrow{\mathcal{D}_{\z^j}}\, \overrightarrow{\mathcal{D}_{\z^j}}\, 
\mathcal{U}_F\Big(z^+,-\frac{L^+}{2};\z, z^- \Big)\right] 
\nn \\
&
+ \frac{[\gamma^i,\gamma^j]}{4(p^++k^+)} 
\int_{-\frac{L^+}{2}}^{\frac{L^+}{2}}dz^+\,
\mathcal{U}_F\Big(\frac{L^+}{2},z^+;\z,z^-\Big)\,
g t\!\cdot\!\mathcal{F}_{ij}(z)\,
\mathcal{U}_F\Big(z^+,-\frac{L^+}{2};\z,z^-\Big) 
\Bigg\} \frac{(\slashed{\check{k}}+m)}{2k^+} 
+O(\textrm{NNEik})
\label{q_prop_full}
\, ,
\end{align}
and for the case of antiquark propagating through the whole medium, meaning $y^+>L^+/2$ and  $x^+<-L^+/2$, one has
\begin{align}
 S_F&(x,y) 
=
\int \frac{d^3\up}{(2\pi)^3} \int \frac{d^3\uk}{(2\pi)^3}\, \theta(-p^+)\theta(-k^+)
e^{-i x\cdot \check{p}}\, e^{i y\cdot \check{k}}\,
\, \int dz^- e^{iz^-(p^+-k^+)}
\int d^2 \z\, e^{-i\z\cdot(\p\!-\!\k)} 
\nn\\
&
\times
\frac{(\slashed{\check{p}}+m)}{2p^+}\gamma^+
\Bigg\{-\mathcal{U}_F^{\dag}(\z,z^-)
\nn\\
&
-\frac{(\p^j\!+\!\k^j)}{2(p^++k^+)}\int_{-\frac{L^+}{2}}^{\frac{L^+}{2}}dz^+\,
\left[\mathcal{U}_F^{\dag}\Big(z^+,-\frac{L^+}{2};\z,z^-\Big)\,
\overleftrightarrow{\mathcal{D}_{\z^j}}\,
\mathcal{U}_F^{\dag}\Big(\frac{L^+}{2},z^+;\z,z^-\Big)\right]
\nn\\
&
-\frac{i}{(p^++k^+)}\int_{-\frac{L^+}{2}}^{\frac{L^+}{2}}dz^+\,
\left[\mathcal{U}_F^{\dag}\Big(z^+,-\frac{L^+}{2};\z,z^-\Big)\,
\overleftarrow{\mathcal{D}_{\z^j}}\, \overrightarrow{\mathcal{D}_{\z^j}}\,
\mathcal{U}_F^{\dag}\Big(\frac{L^+}{2},z^+;\z,z^-\Big)\right] 
\nn \\
&
+ \frac{[\gamma^i,\gamma^j]}{4(p^++k^+)}\,
\int_{-\frac{L^+}{2}}^{\frac{L^+}{2}}dz^+\,
\mathcal{U}_F^{\dag}\Big(z^+,-\frac{L^+}{2};\z,z^-\Big)\,
g t\!\cdot\!\mathcal{F}_{ij}(z)\,
\mathcal{U}_F^{\dag}\Big(\frac{L^+}{2},z^+;\z,z^-\Big) 
\Bigg\} \frac{(\slashed{\check{k}}+m)}{2k^+}
+O(\textrm{NNEik})
\, .
\label{qbar_prop_full}
\end{align}
In the above formulas we use the notation
\begin{align}
\overrightarrow{\mathcal{D}_{z^{\mu}}} \equiv &\, \overrightarrow{\d_{z^{\mu}}} +igt\!\cdot\!\mathcal{A}_{\mu}(z) \\
\overleftarrow{\mathcal{D}_{z^{\mu}}} \equiv &\, \overleftarrow{\d_{z^{\mu}}} -igt\!\cdot\!\mathcal{A}_{\mu}(z) \\
\overleftrightarrow{\mathcal{D}_{z^{\mu}}} \equiv &\, \overrightarrow{\mathcal{D}_{z^{\mu}}} - \overleftarrow{\mathcal{D}_{z^{\mu}}} = \overleftrightarrow{\d_{z^{\mu}}} +2igt\!\cdot\!\mathcal{A}_{\mu}(z)\\
\mathcal{F}^a_{\mu\nu}(z)\equiv &\, \d_{\z^{\mu}}\mathcal{A}^a_{\nu}(z)-\d_{\z^{\nu}}\mathcal{A}^a_{\mu}(z) -g f^{abc}\, \mathcal{A}^b_{\mu}(z)\, \mathcal{A}^c_{\nu}(z)
\end{align}
for the background covariant derivatives and field strength, where $t^a$ are the $SU(N_c)$ generators in the fundamental representation. 
Note that in Eqs.~\eqref{q_prop_full} and  \eqref{qbar_prop_full}, the overall $z^-$ dependence of the  Wilson lines is kept, so  that in particular, the first term in the bracket in both equations corresponds to the generalized eikonal approximation of the propagator.


\subsection{Propagators from inside to after the medium at Eik accuracy\label{sec:IA_propag}}

In order to evaluate the inside contribution \eqref{S-sep-in} to the S-matrix element, we need the quark propagator $S_F(x,y)$ from inside to after the medium at Eikonal accuracy, in the outgoing quark kinematics $x^+>L^+/2$ and $-L^+/2<y^+<L^+/2$ as well as in the outgoing antiquark kinematics $y^+>L^+/2$ and $-L^+/2<x^+<L^+/2$. Let us first focus on the former case. From Eq.~\eqref{full_q_prop_Eik_generic_position_result}, one can read off the quark propagator in pure $A^-$ background at Eikonal accuracy in the inside-after quark kinematics $x^+>L^+/2$ and $-L^+/2<y^+<L^+/2$, and find
\ba
{S}_F(x,y)_{\beta\alpha} \bigg|^{\textrm{IA, }q}_{\textrm{pure }\mathcal{A}^{-}\textrm{, Eik.}}
&=&
\int \frac{d^3\up}{(2\pi)^3} \int \frac{d^3\uk}{(2\pi)^3}\, \theta(p^+)\theta(k^+)\, 
e^{-i x\cdot \check{p}}e^{i y^- k^+ -i \y\cdot\k}\,
\frac{(\slashed{\check{p}}+m)}{2p^+}\gamma^+\frac{(\slashed{\check{k}}+m)}{2k^+}\,
\nn\\
&& \: \times \:
\int dz^- e^{iz^-(p^+-k^+)}\int d^2\z\, e^{-i \z\cdot (\p-\k)}\, 
\mathcal{U}_F(x^+,y^+;\z,z^-)_{\beta\alpha} 
\label{f1}
\, .
\ea
In Eq.~\eqref{f1}, we have dropped the phase factor $e^{iy^+\check{k}^-}$ since it would contribute only beyond Eikonal accuracy, because $|y^+| < L^+/2$. 
Expressing the Dirac structure of that contribution as 
\begin{align}
(\slashed{\check{p}}+m)\gamma^+(\slashed{\check{k}}+m)
=&\,
(\slashed{\check{p}}+m)\Big[ \{\gamma^+,\slashed{\check{k}}\} -   (\slashed{\check{k}}\!-\!m)\gamma^+\Big]
=(\slashed{\check{p}}+m)\Big[ 2k^+ -   (\slashed{\check{p}}\!-\!m)\gamma^+-   (\slashed{\check{k}}\!-\!\slashed{\check{p}})\gamma^+\Big]
\nn\\
=&\,
(\slashed{\check{p}}+m)\Big[ 2k^+ 
-     (k^+\!-\!p^+)\gamma^-\gamma^+
+     (\k^i\!-\!\p^i)\gamma^i \gamma^+\Big]
\end{align}
thanks to the identity $(\slashed{\check{p}}+m)(\slashed{\check{p}}\!-\!m)=(\check{p}^2\!-\!m^2)=0$,
one obtains
\ba
{S}_F(x,y)_{\beta\alpha} \bigg|^{\textrm{IA, }q}_{\textrm{pure }\mathcal{A}^{-}\textrm{, Eik.}}
&=&
\int \frac{d^3\up}{(2\pi)^3} \int \frac{d^3\uk}{(2\pi)^3}\, \frac{\theta(p^+)}{2p^+}\, \theta(k^+)\, 
e^{-i x\cdot \check{p}}e^{i y^- k^+ -i \y\cdot\k}\,
\int dz^- e^{iz^-(p^+-k^+)}\int d^2\z\, e^{-i \z\cdot (\p-\k)}\, 
\nn\\
&& \: \times \:
(\slashed{\check{p}}+m)\left[ 1 
+    \frac{ (p^+\!-\!k^+)}{2k^+}\gamma^-\gamma^+
+    \frac{ (\p^i\!-\!\k^i)}{2k^+}\gamma^+ \gamma^i\right]\,
\mathcal{U}_F(x^+,y^+;\z,z^-)_{\beta\alpha} 
\nn\\
&=&
\int \frac{d^3\up}{(2\pi)^3} \int \frac{d^3\uk}{(2\pi)^3}\, \frac{\theta(p^+)}{2p^+}\, \theta(k^+)\, 
e^{-i x\cdot \check{p}}e^{i y^- k^+ -i \y\cdot\k}\,
\int dz^- e^{iz^-(p^+-k^+)}\int d^2\z\, e^{-i \z\cdot (\p-\k)}\, 
\nn\\
&& \: \times \:
(\slashed{\check{p}}+m)\left[ 1 
+    \frac{ \gamma^-\gamma^+}{2k^+}\, i \overrightarrow{\partial_{z^-}}
-    \frac{ \gamma^+ \gamma^i}{2k^+}\, i \overrightarrow{\partial_{\z^i}}\right]\,
\mathcal{U}_F(x^+,y^+;\z,z^-)_{\beta\alpha} 
\label{f2}
\, .
\ea
As explained in Sec.~\ref{sec:gen_struct}, the $z^-$ dependence of the background field and of $\mathcal{U}_F(x^+,y^+;\z,z^-)$ is parametrically slow for a highly boosted target due to Lorentz time dilation. For that reason, the term in $\partial_{z^-}\mathcal{U}_F(x^+,y^+;\z,z^-)$ is a NEik correction. In order to evaluate Eq.~\eqref{S-sep-in} at NEik accuracy, we need the propagator~\eqref{f2} only at strict Eikonal accuracy. In Eq.~\eqref{f2}, it is thus safe to neglect the term in $\partial_{z^-}\mathcal{U}_F(x^+,y^+;\z,z^-)$. Moreover, the whole dependence of $\mathcal{U}_F(x^+,y^+;\z,z^-)$  on $z^-$ can be neglected as well.
It is then possible to perform the integral over $z^-$ in addition to the integral over $\k$. One finds
\ba
{S}_F(x,y)_{\beta\alpha} \bigg|^{\textrm{IA, }q}_{\textrm{pure }\mathcal{A}^{-}\textrm{, Eik.}}
&=&
\int \frac{d^3\up}{(2\pi)^3} \int \frac{dk^+}{2\pi}\, \frac{\theta(p^+)}{2p^+}\, \theta(k^+)\, 
e^{-i x\cdot \check{p}}e^{i y^- k^+}\,
2\pi\delta(p^+\!-\!k^+)\int d^2\z\, e^{-i \z\cdot \p}\, \delta^{(2)}(\z\!-\!\y)\,
\nn\\
&& \: \times \:
(\slashed{\check{p}}+m)\left[ 1 
-    \frac{ \gamma^+ \gamma^i}{2k^+}\, i \overrightarrow{\partial_{\z^i}}\right]\,
\mathcal{U}_F(x^+,y^+;\z)_{\beta\alpha} 
\label{f3}
\, ,
\ea
and finally
\ba
{S}_F(x,y)_{\beta\alpha} \bigg|^{\textrm{IA, }q}_{\textrm{pure }\mathcal{A}^{-}\textrm{, Eik.}}
&=&
\int \frac{d^3\up}{(2\pi)^3}\, \frac{\theta(p^+)}{2p^+}\, 
e^{-i x\cdot \check{p}}\, e^{i y^- p^+}\,
 e^{-i \y\cdot \p}\, 
(\slashed{\check{p}}+m)\left[ 1 
-    \frac{ \gamma^+ \gamma^i}{2p^+}\, i \overrightarrow{\partial_{\y^i}}\right]\,
\mathcal{U}_F(x^+,y^+;\y)_{\beta\alpha} 
\label{f4}
\, .
\ea

In addition to this pure $\mathcal{A}^{-}$ contribution, the quark propagator in that inside-after kinematics also receives an Eikonal contribution from single interaction with the transverse components of the gauge field. In general, the contribution to the propagator due to single insertion of the transverse field is 
\begin{align}
\delta S_F(x,y)
\bigg|_{\textrm{single }\mathcal{A}_{\perp}}
& = \int d^4w\; S_F(x,w) \bigg|_{\textrm{pure }\mathcal{A}^{-}}\;
[-ig\, \gamma^j\, t^a]\; \mathcal{A}^{a}_{j}(w)\;\;\;
S_F(w,y) \bigg|_{\textrm{pure }\mathcal{A}^{-}}
\, .
\label{corr}
\end{align}
Naively, due to the integration over $w^+$ restricted to the inside region, in which $\mathcal{A}^{a}_{j}(w)$ is non-trivial, the contribution \eqref{corr} seems to be a NEik correction. However, due to the instantaneous term present in the propagator in pure $\mathcal{A}^{-}$ field \eqref{full_q_prop_Eik_generic_position_result}, this integration can be removed, and an Eikonal contribution can indeed be obtained\footnote{Because the Dirac structure of the instantaneous contribution to the propagator is simply $\gamma^+$, we would get $\gamma^+\gamma^j\gamma^+=0$ in the case of two instantaneous propagators separated by a transverse field insertion. Because of this observation, there is no Eikonal contribution to the quark propagator in inside-after kinematics with more than one transverse gauge field insertion.} from Eq.~\eqref{corr}, with the transverse gauge field inserted either at $w^+=x^+$ or at $w^+=y^+$. In the inside-after kinematics $x^+>L^+/2$ and $-L^+/2<y^+<L^+/2$ under consideration, the gauge field vanishes at $x^+$ but not at $y^+$, so that Eq.~\eqref{corr} provides an Eikonal contribution
\begin{align}
\delta S_F(x,y)
\bigg|^{\textrm{IA, }q}_{\textrm{single }\mathcal{A}_{\perp}\textrm{, Eik.}}
& = 
\int d^4w\; S_F(x,w)  \bigg|^{\textrm{IA, q}}_{\textrm{pure }\mathcal{A}^{-}\textrm{, Eik.}}\;
[-ig\, \gamma^j\, t^a]\; \mathcal{A}^{a}_{j}(w)\;\;\;
S_F(w,y) \bigg|_{\textrm{pure }\mathcal{A}^{-},\textrm{ instant.} }
\nn\\
& = \int d^4w\; 
\int \frac{d^3\up}{(2\pi)^3}\, \frac{\theta(p^+)}{2p^+}\, 
e^{-i x\cdot \check{p}}\, e^{i w^- p^+}\,
 e^{-i \w\cdot \p}\, 
(\slashed{\check{p}}+m)\bigg\{\left[ 1 
-    \frac{ \gamma^+ \gamma^i}{2p^+}\, i \overrightarrow{\partial_{\w^i}}\right]\,
\mathcal{U}_F(x^+,w^+;\w)\bigg\}
\nn\\
& \;\;\;\;\;\; \times\;
[-ig\, \gamma^j\, t^a]\; \mathcal{A}^{a}_{j}(\uw)\;\;\;
\delta^{(3)}(\uw\!-\!\uy)\; \gamma^+ 
\int \frac{dq^+}{2\pi} \frac{i}{2q^+} e^{-iq^+(w^--y^-)}
\, ,
\label{corr_1}
\end{align}
where we have neglected the $w^-$ dependence of the $ \mathcal{A}^{a}_{j}$, following the Eikonal approximation. In Eq.~\eqref{corr_1}, the term containing the $\partial_{w^i}$ derivative comes with a factor $ \gamma^+ \gamma^i\gamma^j\gamma^+ =\gamma^+\gamma^+ \gamma^i\gamma^j=0$, and thus disappears.
Performing the integration over $w$, and then over $q^+$, one finds
\begin{align}
\delta S_F(x,y)
\bigg|^{\textrm{IA, }q}_{\textrm{single }\mathcal{A}_{\perp}\textrm{, Eik.}}
& = 
\int \frac{d^3\up}{(2\pi)^3}\, \frac{\theta(p^+)}{2p^+}\, 
e^{-i x\cdot \check{p}}\, 
 e^{i y^- p^+}\, e^{-i \y\cdot \p}\, 
(\slashed{\check{p}}+m)(-1) \frac{\gamma^+\gamma^j}{2p^+} 
\mathcal{U}_F(x^+,y^+;\y) \Big[g t\!\cdot\! \mathcal{A}_{j}(\uy) \Big]
\, .
\label{corr_2}
\end{align}

The full quark propagator at NEik accuracy in the inside-after kinematics with outgoing quark, $x^+>L^+/2$ and $-L^+/2<y^+<L^+/2$, is then the sum of the expressions \eqref{f4} and \eqref{corr_2}, which can be written as
\ba
{S}_F(x,y) \bigg|^{\textrm{IA, }q}_{\textrm{Eik.}}
&=&
\int \frac{d^3\up}{(2\pi)^3}\, \frac{\theta(p^+)}{2p^+}\, 
e^{-i x\cdot \check{p}}\, 
(\slashed{\check{p}}+m)\; \mathcal{U}_F(x^+,y^+;\y)\;
\left[ 1 -    \frac{ \gamma^+ \gamma^i}{2p^+}\, i\, \overleftarrow{{\cal D}_{\y^i}}\right]\,
 \, e^{i y^- p^+}\,  e^{-i \y\cdot \p}
\label{IA_q_propag_Eik}
\, .
\ea
We would like to emphasize that the propagator contributions given in Eqs. \eqref{f4} and \eqref{corr_2} are of order $(\gamma_t)^0$ according to the power counting rules introduced in Sec. \ref{sec:power_counting} and therefore corresponding to eikonal order.

Following the same steps, one can also calculate the quark propagator at Eikonal accuracy in the inside-after kinematics with outgoing antiquark, meaning $y^+>L^+/2$ and $-L^+/2<x^+<L^+/2$. One finds
\ba
{S}_F(x,y) \bigg|^{\textrm{IA, }\bar{q}}_{\textrm{Eik.}}
&=&
\int \frac{d^3\uk}{(2\pi)^3}\, (-1)\, \frac{\theta(-k^+)}{2k^+}\, 
e^{i y\cdot \check{k}}\, 
 e^{-i x^- k^+}\,  e^{i \x\cdot \k}\;
\left[ 1 -    \frac{ \gamma^+ \gamma^i}{2k^+}\, i\, \overrightarrow{{\cal D}_{\x^i}}\right]
\; \mathcal{U}^\dagger_F(y^+,x^+;\x)\,
 \,
 (\slashed{\check{k}}+m)
\label{IA_qbar_propag_Eik}
\, .
\ea

\section{Photon splitting inside the medium\label{sec:in}}

In order to calculate the contribution to the S-matrix element at NEik accuracy from the diagram with photon splitting inside the target (see Fig. \ref{Fig:diagrams} right panel), Eq.~\eqref{S-sep-in}, we need the expression for the reduced propagators  $\tilde S_F^q(z)$ and $\tilde S_F^{\bar q}(z)$ at Eikonal accuracy when the vertex location $z^{\mu}$ is inside the target, meaning $-L^+/2<z^+<L^+/2$. In the case of the reduced quark propagator $\tilde S_F^q(z)$, it simply amounts to insert the expression \eqref{IA_q_propag_Eik} into the definition \eqref{integrated_q_prop}, as
\begin{align}
\tilde S^{q}_F(z) \bigg|^{\textrm{In}}_{\textrm{Eik.}}
=&\, 
{\rm lim}_{x^+ \to +\infty} \int d^2\x \int dx^- 
e^{i\ck_1 \cdot x} 
\bar u(1) \gamma^+ S_F(x,z) \bigg|^{\textrm{IA, }q}_{\textrm{Eik.}}
\nonumber\\
=&\, 
\frac{\theta(k_1^+)}{2k_1^+}\, 
\bar u(1) \gamma^+ 
(\slashed{\check{k}_1}+m)\; \mathcal{U}_F(+\infty,z^+;\z)\;
\left[ 1 -    \frac{ \gamma^+ \gamma^i}{2k_1^+}\, i\, \overleftarrow{{\cal D}_{\z^i}}\right]\,
 \, e^{i z^- k_1^+}\,  e^{-i \z\cdot \k_1}
 \nonumber\\
=&\, 
 \mathcal{U}_F\left(\frac{L^+}{2},z^+;\z\right)\; \bar u(1)
\left[ 1 -    \frac{ \gamma^+ \gamma^i}{2k_1^+}\, i\, \overleftarrow{{\cal D}_{\z^i}}\right]\,
 \, e^{i z^- k_1^+}\,  e^{-i \z\cdot \k_1}
\label{reduced_q_prop_in}
\, ,
\end{align}
using the identity $\bar u(1) (\slashed{\check{k}_1}\!-\!m)\equiv \bar u(k_1,h_1) (\slashed{\check{k}_1}\!-\!m)=0$. The Heaviside $\theta(k_1^+)$ can be dropped since the produced quark has by definition $k_1^+>0$. Moreover, since the gauge field is assumed to vanish outside of the target, it does not matter if the end point of the Wilson line is taken at $+\infty$ or at $+L^+/2$.  

Similarly, inserting the expression \eqref{IA_qbar_propag_Eik} into the definition \eqref{integrated_qbar_prop}, one finds the reduced antiquark propagator at Eikonal accuracy from $z^{\mu}$ inside the target as 
\begin{align}
\tilde S^{\bar q}_F(z) \bigg|^{\textrm{In}}_{\textrm{Eik.}}
=&\, 
{\rm lim}_{y^+ \to +\infty} \int d^2\y \int dy^-  
e^{i \ck_2 \cdot y} \; (-1)\; S_F(z,y)\bigg|^{\textrm{IA, }\bar{q}}_{\textrm{Eik.}}
 \gamma^+ v(2)
 \nonumber\\
=&\, 
e^{i z^- k_2^+}\,  e^{-i \z\cdot \k_2} \, 
\left[ 1 +    \frac{ \gamma^+ \gamma^j}{2k_2^+}\, i\, \overrightarrow{{\cal D}_{\z^j}}\right]
 v(2)\, 
 \mathcal{U}_F^{\dag}\left(\frac{L^+}{2},z^+;\z\right)\; 
\label{reduced_qbar_prop_in}
\, .
\end{align}

Now, we can calculate the contribution to the S-matrix from photon splitting inside the target by inserting the expressions \eqref{reduced_q_prop_in} and \eqref{reduced_qbar_prop_in} into Eq.~\eqref{S-sep-in}, and dropping the $z^+$ dependent phase factors, which would contribute only at NNEik accuracy in the S-matrix and cross section.
Hence,
\begin{align}
S^{\rm in}_{q_1 \bar q_2 \leftarrow \gamma^*} 
=&\,
  -i e e_f \, \epsilon^\lambda_\mu (q) 
\int d^2\z \int dz^-
e^{iz^-(k_1^++k_2^+-q^+)} \int_{-L^+/2}^{L^+/2} dz^+ 
\bar u(1)\bigg\{- e^{-i(\k_1+\k_2-\q) \cdot \z}\: 
\gamma^\mu
\nonumber\\
&\, +
 \mathcal{U}_F\left(\frac{L^+}{2},z^+;\z\right)\,
\left[ 1 -    \frac{ \gamma^+ \gamma^i}{2k_1^+}\, i\, \overleftarrow{{\cal D}_{\z^i}}\right]\,
e^{-i(\k_1+\k_2-\q) \cdot \z}\: \gamma^\mu\,
\left[ 1 +    \frac{ \gamma^+ \gamma^j}{2k_2^+}\, i\, \overrightarrow{{\cal D}_{\z^j}}\right]\, 
 \mathcal{U}_F^{\dag}\left(\frac{L^+}{2},z^+;\z\right)
\bigg\}  v(2)
\nonumber\\
=&\,
  -i e e_f \, \epsilon^\lambda_\mu (q)\, 2\pi \delta(k_1^+\!+\!k_2^+\!-\!q^+) 
\int d^2\z  \int_{-L^+/2}^{L^+/2} dz^+ 
\bar u(1)\, \mathcal{U}_F\left(\frac{L^+}{2},z^+;\z\right)
\nonumber\\
&\, \times \, \bigg\{
- \frac{ \gamma^+ \gamma^i  \gamma^\mu}{2k_1^+}\, i\, \overleftarrow{{\cal D}_{\z^i}}\, 
e^{-i(\k_1+\k_2-\q) \cdot \z}\:
+e^{-i(\k_1+\k_2-\q) \cdot \z}\: 
 \frac{\gamma^\mu \gamma^+ \gamma^j}{2k_2^+}\, i\, \overrightarrow{{\cal D}_{\z^j}}
\nonumber\\
&\, \hspace{1cm}
+
 \frac{ \gamma^+ \gamma^i \gamma^\mu \gamma^+ \gamma^j}{(2k_1^+)(2k_2^+)}\,  
 \overleftarrow{{\cal D}_{\z^i}}\, e^{-i(\k_1+\k_2-\q) \cdot \z}\:\, \overrightarrow{{\cal D}_{\z^j}}\, 
\bigg\} \mathcal{U}_F^{\dag}\left(\frac{L^+}{2},z^+;\z\right)\, v(2)
\label{S-mat-in_1}
\, .
\end{align}
The color structure from the second term can be simplified as 
\ba
\bigg[ \mathcal{U}_F\Big(\frac{L^+}{2},z^+;\z\Big) \overrightarrow{ {\cal D}_{\z^j}}  \mathcal{U}_F^{\dagger}\Big(\frac{L^+}{2},z^+;\z\Big)\bigg] 
&=&
\frac{1}{2}\bigg[ \mathcal{U}_F\Big(\frac{L^+}{2},z^+;\z\Big) \Big(\overrightarrow{ {\cal D}_{\z^j}}\!-\!\overleftarrow{ {\cal D}_{\z^j}} \Big)  \mathcal{U}_F^{\dagger}\Big(\frac{L^+}{2},z^+;\z\Big)\bigg] \nonumber\\
&
+&\frac{1}{2} \bigg[ \mathcal{U}_F\Big(\frac{L^+}{2},z^+;\z\Big) \Big(\overrightarrow{ {\cal D}_{\z^j}}\!+\!\overleftarrow{  {\cal D}_{\z^j}} \Big) \mathcal{U}_F^{\dagger}\Big(\frac{L^+}{2},z^+;\z\Big)\bigg] \nonumber\\
&=&\frac{1}{2}\bigg[ \mathcal{U}_F\Big(\frac{L^+}{2},z^+;\z\Big) \overleftrightarrow{ {\cal D}_{\z^j}}\,  \mathcal{U}_F^{\dagger}\Big(\frac{L^+}{2},z^+;\z\Big)\bigg] 
\nonumber\\
&+&
\frac{1}{2}\partial_{\z^j}\bigg[ \mathcal{U}_F\Big(\frac{L^+}{2},z^+;\z\Big)\mathcal{U}_F^{\dagger}\Big(\frac{L^+}{2},z^+;\z\Big)\bigg] \nonumber\\
&=&\frac{1}{2}\bigg[ \mathcal{U}_F\Big(\frac{L^+}{2},z^+;\z\Big) \overleftrightarrow{ {\cal D}_{\z^j}}  \mathcal{U}_F^{\dagger}\Big(\frac{L^+}{2},z^+;\z\Big)\bigg]
\ea
and similarly one gets for the first term
\ba
\bigg[ \mathcal{U}_F\Big(\frac{L^+}{2},z^+;\z\Big) \overleftarrow{ {\cal D}_{\z^j}}  \mathcal{U}_F^{\dagger}\Big(\frac{L^+}{2},z^+;\z\Big)\bigg] 
&=&-\frac{1}{2}\bigg[ \mathcal{U}_F\Big(\frac{L^+}{2},z^+;\z\Big) \overleftrightarrow{ {\cal D}_{\z^j}}  \mathcal{U}_F^{\dagger}\Big(\frac{L^+}{2},z^+;\z\Big)\bigg]
\, .
\ea
Using these relations, one arrives at
\begin{align}
S^{\rm in}_{q_1 \bar q_2 \leftarrow \gamma^*} 
=&\,
  -i e e_f \, \epsilon^\lambda_\mu (q)\, 2\pi \delta(k_1^+\!+\!k_2^+\!-\!q^+) 
\int d^2\z\,  e^{-i(\k_1+\k_2-\q) \cdot \z}\: \int_{-L^+/2}^{L^+/2} dz^+ 
\nonumber\\
&\, \times \, 
\bigg\{
\left[\frac{i}{4k_1^+}\, \bar u(1) \gamma^+ \gamma^j \gamma^\mu v(2)
+\frac{i}{4k_2^+}\, \bar u(1) \gamma^\mu \gamma^+ \gamma^j v(2)
\right]\,
\bigg[ \mathcal{U}_F\Big(\frac{L^+}{2},z^+;\z\Big) \overleftrightarrow{ {\cal D}_{\z^j}}  \mathcal{U}_F^{\dagger}\Big(\frac{L^+}{2},z^+;\z\Big)\bigg]
\nonumber\\
&\, \hspace{1cm}
+\frac{g^{\mu +}}{2k_1^+\, k_2^+}\, \bar u(1) \gamma^+ \gamma^i \gamma^j v(2)\,
\bigg[ \mathcal{U}_F\Big(\frac{L^+}{2},z^+;\z\Big) \overleftarrow{ {\cal D}_{\z^i}}
\overrightarrow{ {\cal D}_{\z^j}}  \mathcal{U}_F^{\dagger}\Big(\frac{L^+}{2},z^+;\z\Big)\bigg]
\bigg\}
\label{S-mat-in_2}
\, ,
\end{align}
where the covariant derivatives now act only on the Wilson lines, within the square brackets, and not on the phase factor.

In light-cone gauge, both the longitudinal and the transverse polarization vectors obey $\epsilon^+_\lambda(q)=0$. Hence, the last term in the bracket in  Eq. \eqref{S-mat-in_2} can not contribute to the scattering processes.

\paragraph{Longitudinal photon case :\\}

 In light cone gauge,  the longitudinal polarization vector  can be chosen as 
\ba
\epsilon^\lambda_\mu(q) \to \epsilon^L_\mu (q) \equiv \frac{Q}{q^+} g^+_\mu \; . 
\label{L_pol_vect}
\ea
Upon inserting this polarization vector in Eq.~\eqref{S-mat-in_2}, the first two Dirac structures vanish as well, due to the identities $\gamma^+\gamma^+=0$ and $\{\gamma^+,\gamma^j\}=0$. Therefore, in the case of longitudinally polarized photon, the inside contribution to the S-matrix element vanishes at NEik order:
\ba
&S^{\rm in}_{q_1 \bar q_2 \leftarrow \gamma^*_L}= 0+O({\rm NNEik})
\label{S-mat-in_L}
\, .
\ea

\paragraph{Transverse photon case :\\}
  
For a transverse photon, the two possible polarization vectors in light cone gauge can be written as 
\ba
\epsilon_{\lambda}^+(q)&=&0\nn\\
\epsilon_{\lambda}^i(q)&=& \varepsilon_{\lambda}^i\nn\\
\epsilon_{\lambda}^-(q)&=& \frac{\q^i\varepsilon^i_{\lambda}}{q^+}
\label{T_pol_vect}
\ea 
in terms of two dimensional polarization vectors $\varepsilon^i_{\lambda}$ that are momentum independent. Upon inserting these polarization vectors in Eq.~\eqref{S-mat-in_2}, one gets 
\begin{align}
S^{\rm in}_{q_1 \bar q_2 \leftarrow \gamma^*_T} 
=&\,
  e e_f \, \varepsilon_{\lambda}^i\, 2\pi \delta(k_1^+\!+\!k_2^+\!-\!q^+) 
\left[-\frac{1}{4k_1^+}\, \bar u(1) \gamma^+ \gamma^j \gamma^i v(2)
-\frac{1}{4k_2^+}\, \bar u(1) \gamma^i \gamma^+ \gamma^j v(2)
\right]\,
\nonumber\\
&\, \times \, 
\int d^2\z\,  e^{-i(\k_1+\k_2-\q) \cdot \z}\: \int_{-L^+/2}^{L^+/2} dz^+\, 
\bigg[ \mathcal{U}_F\Big(\frac{L^+}{2},z^+;\z\Big) \overleftrightarrow{ {\cal D}_{\z^j}}  \mathcal{U}_F^{\dagger}\Big(\frac{L^+}{2},z^+;\z\Big)\bigg]
\label{S-mat-in_T_1}
\, .
\end{align}
It is convenient to extract the parts of the Dirac structures which are either symmetric or antisymmetric under the exchange of $i$ and $j$. Then, one finds
\begin{align}
S^{\rm in}_{q_1 \bar q_2 \leftarrow \gamma^*_T} 
=&\,
  e e_f \, \varepsilon_{\lambda}^i\, 2\pi \delta(k_1^+\!+\!k_2^+\!-\!q^+) \; \frac{q^+}{4k_1^+k_2^+}\;
\bar u(1) \gamma^+\!\left(\frac{(k_2^+\!-\!k_1^+)}{q^+}\, \delta^{i j}
+\frac{1}{2}\, [ \gamma^i,  \gamma^j] 
\right) v(2)
\nonumber\\
&\, \times \, 
\int d^2\z\,  e^{-i(\k_1+\k_2-\q) \cdot \z}\: \int_{-L^+/2}^{L^+/2} dz^+\, 
\bigg[ \mathcal{U}_F\Big(\frac{L^+}{2},z^+;\z\Big) \overleftrightarrow{ {\cal D}_{\z^j}}  \mathcal{U}_F^{\dagger}\Big(\frac{L^+}{2},z^+;\z\Big)\bigg]
\label{S-mat-in_T}
\, ,
\end{align}
where, as a reminder, the covariant derivative acts only on the Wilson lines, within the square brackets, and not on the phase factor.

\section{Photon splitting before the medium\label{sec:bef}}

It remains now to calculate the contribution \eqref{S-sep-bef} to the S-matrix, in which the photon splits before the target, at NEik accuracy  (see Fig. \ref{Fig:diagrams} left panel). For that purpose, we should first calculate the reduced quark and antiquark propagators $\tilde S_F^q(z)$ and $\tilde S_F^{\bar q}(z)$ at NEik accuracy, using the results of Sec.~\ref{sec:quark_prop}.

\subsection{ Calculation of the reduced propagators from before to after the medium}

When the photon splits before the target, we use, for the quark propagator in the formula (\ref{integrated_q_prop}), the expression (\ref{q_prop_full}). Then we get
\begin{align}
\label{integrated_1}
&\tilde S^{q}_F(z)
={\rm lim}_{x^+ \to +\infty} \int d^2\x \int dx^- 
e^{ik_1 \cdot x} 
\bar u(1) \gamma^+ 
\int \frac{d^3\up}{(2\pi)^3} \int \frac{d^3\uk}{(2\pi)^3}\, \int dv^- e^{iv^-(p^+-k^+)} 
\theta(k^+) \theta(p^+)
e^{-i x\cdot \check{p}}\, e^{i z\cdot \check{k}}\,
\nn\\
&
\times
\frac{(\slashed{\check{p}}\!+
\!m)}{2p^+} \gamma^+
\, \int d^2 \v \, e^{-i\v \cdot(\p\!-\!\k)}
\Bigg\{\mathcal{U}_F(\v,v^-)
\nn\\
&
\!\!
+\frac{1}{p^+\!+
\!k^+}\int_{-\frac{L^+}{2}}^{\frac{L^+}{2}}dv^+\!\!\!
\left[\mathcal{U}_F\Big(\frac{L^+}{2},v^+;\v,v^-\Big)
\left(\frac{[\gamma^i,\gamma^j]}{4} gt \cdot \mathcal{F}_{ij}(v) - \frac{(\p^j\!+\!\k^j)}{2}\overleftrightarrow{\mathcal{D}_{\v^j}} 
- i \overleftarrow{\mathcal{D}_{\v^j}} \overrightarrow{\mathcal{D}_{\v^j}} \right)
\mathcal{U}_F\Big(v^+,-\frac{L^+}{2};\v,v^-\Big)\right]
\Bigg\}
 \frac{(\slashed{\check{k}}\!+\!m)}{2k^+}
\end{align}
Next we use $\gamma^+ (\slashed{\check{p}}+m)\gamma^+ = 2p^+\gamma^+$ and integrate over $\x$ and $x^-$ to get $(2\pi)^2 \delta^{(2)}(\k_1-\p)$ and $2\pi \delta (p^+-k_1^+)$. Performing the integration over $\up$ we obtain
\begin{align}
\label{integrated_2}
&\tilde S^{q}_F(z)
=
\bar u(1)  \gamma^+
\int \frac{d^3\uk}{(2\pi)^3}\, 
\theta(k^+)\, e^{i z \cdot  \check{k}}
\int dv^- e^{iv^-(k_1^+-k^+)}
\int d^2 \v \, e^{-i\v \cdot(\k_1\!-\!\k)} \,
\Bigg\{\mathcal{U}_F(\v,v^-)
\nn\\
&
+\frac{1}{k_1^+\!+\!k^+}\int_{-\frac{L^+}{2}}^{\frac{L^+}{2}}dv^+\, 
\left[\mathcal{U}_F\Big(\frac{L^+}{2},v^+;\v,v^-\Big)\,
\left(\frac{[\gamma^i,\gamma^j]}{4} gt \cdot \mathcal{F}_{ij}(v) - \frac{(\k_1^j\!+\!\k^j)}{2}\overleftrightarrow{\mathcal{D}_{\v^j}} 
- i \overleftarrow{\mathcal{D}_{\v^j}}\, \overrightarrow{\mathcal{D}_{\v^j}}\, \right)
\mathcal{U}_F\Big(v^+,-\frac{L^+}{2};\v,v^-\Big)\right]
\Bigg\} \nn \\
&
\times \frac{(\slashed{\check{k}}\!+\!m)}{2k^+}
\, .
\end{align}
where we used $e^{ix^+( \check{k}_1^--\check{p}^-)} \to e^{ix^+ (\check{k}_1^--\check{k}_1^-)}=1$.

For the anti-quark propagator in the formula (\ref{integrated_qbar_prop}) we use the expression (\ref{qbar_prop_full}). We have
\begin{align}
\label{integrated_3}
&\tilde S^{\bar q}_F(z)
=
{\rm lim}_{y^+ \to +\infty} \int d^2\y \int dy^-  e^{i k_2 \cdot y} \; 
\int \frac{d^3\up}{(2\pi)^3} \int \frac{d^3\uk}{(2\pi)^3}\, \int dw^- e^{iw^-(p^+-k^+)} 
\theta(-p^+)\theta(-k^+)
e^{-i z \cdot \check{p}}\, e^{i y\cdot \check{k}}\,
\nn\\
&
\times \, 
\frac{(\slashed{\check{p}}\!+\!m)}{2p^+}\gamma^+
\int d^2 \w\, e^{-i\w\cdot(\p\!-\!\k)} \,
\Bigg\{\mathcal{U}_F^{\dag}(\w,w^-)
\nn\\
&
-\frac{1}{p^+\!+\!k^+}\int_{-\frac{L^+}{2}}^{\frac{L^+}{2}}dw^+\,
\left[\mathcal{U}_F^{\dag}\Big(w^+,-\frac{L^+}{2};\w,w^-\Big)\,
\left(\frac{[\gamma^i,\gamma^j]}{4} gt\cdot \mathcal{F}_{ij}(w) -\frac{(\p^j\!+\!\k^j)}{2} \overleftrightarrow{\mathcal{D}_{\w^j}} 
-i \overleftarrow{\mathcal{D}_{\w^j}}\, \overrightarrow{\mathcal{D}_{\w^j}} \right)
\mathcal{U}_F^{\dag}\Big(\frac{L^+}{2},w^+;\w,w^-\Big)\right]
\Bigg\} \nn \\
&
\times \frac{(\slashed{\check{k}}\!+\!m)}{2k^+} \gamma^+ v(2)
\end{align}
Since $\gamma^+$ commutes with $[\gamma^i,\gamma^j]$ we can use $\gamma^+ (\slashed{\check{k}}+m) \gamma^+ = 2k^+\gamma^+$. We also integrate over $\y$ and $y^-$ and get $(2\pi)^2 \delta^{(2)}(\k_2+\k)$ and $2\pi \delta(k^++k_2^+)$. Then we perform integration over $\uk$ and obtain
\begin{align}
\label{integrated_4}
&\tilde S^{\bar q}_F(z)
=
\int \frac{d^3\up}{(2\pi)^3} \theta(-p^+) \theta(k_2^+)
\int dw^- e^{iw^-(p^++k_2^+)}
e^{- i z \cdot \check{p} } \;
\frac{(\slashed{\check{p}}\!+\!m)}{2p^+} \int d^2 \w\, e^{-i\w\cdot(\p\!+\!\k_2)} \,
\Bigg\{\mathcal{U}_F^{\dag}(\w,w^-)
\nn\\
&
+\frac{1}{k_2^+\!-\!p^+}\int_{-\frac{L^+}{2}}^{\frac{L^+}{2}}dw^+\,
\left[\mathcal{U}_F^{\dag}\Big(w^+,-\frac{L^+}{2};\w,w^-\Big)\,
\left(\frac{[\gamma^i,\gamma^j]}{4} gt\cdot \mathcal{F}_{ij}(w) -\frac{(\p^j\!-\!\k_2^j)}{2} \overleftrightarrow{\mathcal{D}_{\w^j}} 
-i \overleftarrow{\mathcal{D}_{\w^j}}\, \overrightarrow{\mathcal{D}_{\w^j}} \right)
\mathcal{U}_F^{\dag}\Big(\frac{L^+}{2},w^+;\w,w^-\Big)\right]
\Bigg\} \nn \\
& \times \gamma^+ v(2)
\end{align}
where we used $e^{iy^+(\check{k}^- + \check{k}_2^-)} \to e^{iy^+ (-\check{k}_2^- +\check{k}_2^-)}=1$.


\subsection{Before contribution to the S-matrix: general photon polarization\label{sec:bef_gen_pol}}

Let us make the first steps in the calculation of the diagram with photon splitting before the target without specifying the photon polarization.
Inserting the propagators (\ref{integrated_2}) and (\ref{integrated_4}) to Eq.(\ref{S-sep-bef}) we get
\begin{align}
&S^{\rm bef}_{q_1 \bar q_2 \leftarrow \gamma^*} = -i e e_f \, \epsilon^\lambda_\mu (q) 
\int \frac{d^3\uk}{(2\pi)^3}  \, \theta(k^+) \int \frac{d^3\up}{(2\pi)^3}\, \theta(-p^+)
\int d^2 \z \int dz^- \int_{-\infty}^{-L^+/2} dz^+
e^{-iz \cdot (q - \check{k} + \check{p})}
 \nn \\
&
\times
\int dv^-\, e^{iv^-(k_1^+-k^+)}\int dw^-\,   e^{iw^-(k_2^++p^+)}
\int d^2 \v \, e^{-i\v \cdot(\k_1\!-\!\k)} \,
\int d^2 \w\, e^{-i\w\cdot(\p\!+\!\k_2)} \,
\bar u(1) \gamma^+  
\Bigg\{\mathcal{U}_F(\v,v^-)
\nn\\
&
+\frac{1}{k_1^+\!+\!k^+} \int_{-\frac{L^+}{2}}^{\frac{L^+}{2}}dv^+\, 
\mathcal{U}_F\Big(\frac{L^+}{2},v^+;\v,v^-\Big)\,
\left(\frac{[\gamma^i,\gamma^j]}{4} gt \cdot \mathcal{F}_{ij}(v) - \frac{(\k_1^j\!+\!\k^j)}{2}\overleftrightarrow{\mathcal{D}_{\v^j}} 
- i \overleftarrow{\mathcal{D}_{\v^j}}\, \overrightarrow{\mathcal{D}_{\v^j}}\, \right)
\mathcal{U}_F\Big(v^+,-\frac{L^+}{2};\v,v^-\Big)
\Bigg\}
 \nn \\
& \times
\frac{(\slashed{\check{k}}\!+\!m)}{2k^+} \gamma^\mu 
\frac{(\slashed{\check{p}}\!+\!m)}{2p^+}
\Bigg\{\mathcal{U}_F^{\dag}(\w,w^-)
\nn\\
&
+\frac{1}{k_2^+\!-\!p^+}\int_{-\frac{L^+}{2}}^{\frac{L^+}{2}}dw^+\,
\mathcal{U}_F^{\dag}\Big(w^+,-\frac{L^+}{2};\w,w^-\Big)\,
\left(\frac{[\gamma^{i'},\gamma^{j'}]}{4} gt\cdot \mathcal{F}_{i'j'}(w) -\frac{(\p^{j'}\!-\!\k_2^{j'})}{2} \overleftrightarrow{\mathcal{D}_{\w^{j'}}} 
-i \overleftarrow{\mathcal{D}_{\w^{j'}}}\, \overrightarrow{\mathcal{D}_{\w^{j'}}} \right)
\mathcal{U}_F^{\dag}\Big(\frac{L^+}{2},w^+;\w,w^-\Big)
\Bigg\}
\nn \\
& \times
\gamma^+ v(2) 
\hspace{1cm}
+\, i e e_f \, \epsilon^\lambda_\mu (q) \, \bar u(1)\gamma^{\mu}  v(2) \,
\int d^2 \z \int dz^- \int_{-\infty}^{-L^+/2} dz^+\, e^{-iz\cdot(q-\ck_1-\ck_2)}
\label{S-bef-1} 
\end{align}
The integrals over the location $z^{\mu}$ of the photon splitting vertex before the target can be performed explicitly: the unrestricted integrations over $\z$ and $z^-$ enforce the conservation of the transverse and $+$ components of the momentum at the vertex, whereas the integrations over $z^+$ are of the form
\begin{align}
\int_{-\infty}^{-L^+/2} dz^+\, e^{-iz^+ E^-} 
=&\,
\frac{i}{(E^-\!+\!i\epsilon)}\, 
 e^{i\frac{L^+}{2} E^-} 
 \, ,
\end{align}
with $E^-=(q^- \!-\! \check{k}^- \!+\! \check{p}^-)$, except in the vacuum subtraction term, in which 
$E^-=(q^- \!-\! \ck_1^-\!-\!\ck_2^-)$.
We thus obtain
\begin{align}
&S^{\rm bef}_{q_1 \bar q_2 \leftarrow \gamma^*} = -i e e_f \, \epsilon^\lambda_\mu (q) 
\int \frac{d^3\uk}{(2\pi)^3}  \, \theta(k^+) \int \frac{d^3\up}{(2\pi)^3}\, \theta(-p^+)\,
 (2\pi)^3 \delta^{(3)}(\uq - \uk+\up)\:
 \frac{i}{(q^- \!-\! \check{k}^- \!+\! \check{p}^-\!+\!i\epsilon)}\, 
 e^{i\frac{L^+}{2} (q^- - \check{k}^- + \check{p}^-)}  
 \nn \\
&
\times
\int dv^-\, e^{iv^-(k_1^+-k^+)}\int dw^-\,   e^{iw^-(k_2^++p^+)}
\int d^2 \v \, e^{-i\v \cdot(\k_1\!-\!\k)} \,
\int d^2 \w\, e^{-i\w\cdot(\p\!+\!\k_2)} \,
\bar u(1) \gamma^+  
\Bigg\{\mathcal{U}_F(\v,v^-)
\nn\\
&
+\frac{1}{k_1^+\!+\!k^+} \int_{-\frac{L^+}{2}}^{\frac{L^+}{2}}dv^+\, 
\mathcal{U}_F\Big(\frac{L^+}{2},v^+;\v,v^-\Big)\,
\left(\frac{[\gamma^i,\gamma^j]}{4} gt \cdot \mathcal{F}_{ij}(v) - \frac{(\k_1^j\!+\!\k^j)}{2}\overleftrightarrow{\mathcal{D}_{\v^j}} 
- i \overleftarrow{\mathcal{D}_{\v^j}}\, \overrightarrow{\mathcal{D}_{\v^j}}\, \right)
\mathcal{U}_F\Big(v^+,-\frac{L^+}{2};\v,v^-\Big)
\Bigg\}
 \nn \\
& \times
\frac{(\slashed{\check{k}}\!+\!m)}{2k^+} \gamma^\mu 
\frac{(\slashed{\check{p}}\!+\!m)}{2p^+}
\Bigg\{\mathcal{U}_F^{\dag}(\w,w^-)
\nn\\
&
+\frac{1}{k_2^+\!-\!p^+}\int_{-\frac{L^+}{2}}^{\frac{L^+}{2}}dw^+\,
\mathcal{U}_F^{\dag}\Big(w^+,-\frac{L^+}{2};\w,w^-\Big)\,
\left(\frac{[\gamma^i,\gamma^j]}{4} gt\cdot \mathcal{F}_{ij}(w) -\frac{(\p^j\!-\!\k_2^j)}{2} \overleftrightarrow{\mathcal{D}_{\w^j}} 
-i \overleftarrow{\mathcal{D}_{\w^j}}\, \overrightarrow{\mathcal{D}_{\w^j}} \right)
\mathcal{U}_F^{\dag}\Big(\frac{L^+}{2},w^+;\w,w^-\Big)
\Bigg\}
\nn \\
& \times
\gamma^+ v(2) 
\hspace{1cm}
+\, i e e_f \, \epsilon^\lambda_\mu (q) \, \bar u(1)\gamma^{\mu}  v(2) \;
 (2\pi)^3 \delta^{(3)}(\uq - \underline{k_1}-\underline{k_2})\:
 \frac{i}{(q^- \!-\! \ck_1^-\!-\!\ck_2^-\!+\!i\epsilon)}\, 
 e^{i\frac{L^+}{2} (q^-- \ck_1^--\ck_2^-)}  
\label{S-bef-2} 
\, .
\end{align}
The expression \eqref{S-bef-2} contains effects beyond the Eikonal approximation of various types and origins : decorations on the Wilson line associated with the quark or with the antiquark, phase factor dependent on the target width $L^+$, and dependence on $v^-$ or $w^-$ of the background field and Wilson lines. Since our aim is to calculate the S-matrix element and cross section at NEik accuracy, we only need to extract the NEik correction associated with each non-Eikonal effect separately.  We would need to take into account two non-Eikonal effects simultaneously only in order to calculate the observable at NNEik accuracy and beyond.

First, let us consider the contribution to the expression \eqref{S-bef-2} with decorations inserted on the quark Wilson line, but not on the antiquark Wilson line, which is
\begin{align}
&S^{\rm bef}_{q_1 \bar q_2 \leftarrow \gamma^*}\bigg|_{\textrm{dec. on }q} = -i e e_f \, \epsilon^\lambda_\mu (q) 
\int \frac{d^3\uk}{(2\pi)^3}  \, \theta(k^+) \int \frac{d^3\up}{(2\pi)^3}\, \theta(-p^+)\,
 (2\pi)^3 \delta^{(3)}(\uq - \uk+\up)\:
 \frac{i}{(q^- \!-\! \check{k}^- \!+\! \check{p}^-\!+\!i\epsilon)}\, 
 e^{i\frac{L^+}{2} (q^- - \check{k}^- + \check{p}^-)}  
 \nn \\
&
\times
\int dv^-\, e^{iv^-(k_1^+-k^+)}\int dw^-\,   e^{iw^-(k_2^++p^+)}
\int d^2 \v \, e^{-i\v \cdot(\k_1\!-\!\k)} \,
\int d^2 \w\, e^{-i\w\cdot(\p\!+\!\k_2)} \,
\bar u(1) \gamma^+  \: \frac{1}{k_1^+\!+\!k^+}
\nn\\
&  \times
 \int_{-\frac{L^+}{2}}^{\frac{L^+}{2}}dv^+\, 
\bigg[\mathcal{U}_F\Big(\frac{L^+}{2},v^+;\v,v^-\Big)\,
\left(\frac{[\gamma^i,\gamma^j]}{4} gt \cdot \mathcal{F}_{ij}(v) - \frac{(\k_1^j\!+\!\k^j)}{2}\overleftrightarrow{\mathcal{D}_{\v^j}} 
- i \overleftarrow{\mathcal{D}_{\v^j}}\, \overrightarrow{\mathcal{D}_{\v^j}}\, \right)
\mathcal{U}_F\Big(v^+,-\frac{L^+}{2};\v,v^-\Big)
\bigg] 
 \nn \\
& \times
\frac{(\slashed{\check{k}}\!+\!m)}{2k^+} \gamma^\mu 
\frac{(\slashed{\check{p}}\!+\!m)}{2p^+}
\mathcal{U}_F^{\dag}(\w,w^-)
\gamma^+ v(2) 
\label{S-bef-q_dec_1} 
\, .
\end{align}
 Due to the integration over the $v^+$ at which the decorations are inserted, the contribution \eqref{S-bef-q_dec_1} only starts at NEik accuracy. We can thus neglect in that expression any further non-Eikonal effect: 
the phase factor dependent on $L^+$, and the dependence on $v^-$ and $w^-$ of the background field and Wilson lines. The latter allows us to perform the integrations over $v^-$ and $w^-$ analytically, and to obtain\footnote{Note that in this case $\theta(k^+)=\theta(k^+_1)=1$ and $\theta(-p^+)=\theta(k^+_2)=1$, since $k^+_1>0$ and $k^+_2>0$ by definition for the produced particles.}
\begin{align}
&S^{\rm bef}_{q_1 \bar q_2 \leftarrow \gamma^*}\bigg|_{\textrm{dec. on }q} 
= 
2\pi\delta(k_1^+\!+\!k_2^+\!-\!q^+)\:
(-i) e e_f \, \epsilon^\lambda_\mu (q) 
\int \frac{d^3\uk}{(2\pi)^3}  \, 
2\pi\delta(k^+\!-\!k_1^+)
\int \frac{d^3\up}{(2\pi)^3}\, 
 (2\pi)^3 \delta^{(3)}(\uq - \uk+\up)\:
 \nn \\
&
\times
 \frac{1}{2k_1^+}\:  \frac{i}{(q^- \!-\! \check{k}^- \!+\! \check{p}^-\!+\!i\epsilon)}\, 
\int d^2 \v \, e^{-i\v \cdot(\k_1\!-\!\k)} \,
\int d^2 \w\, e^{-i\w\cdot(\p\!+\!\k_2)} \,\;
\bar u(1)   \,
\nn\\
& \times
 \int_{-\frac{L^+}{2}}^{\frac{L^+}{2}}\!\!\!\!dv^+\, 
\bigg[\mathcal{U}_F\Big(\frac{L^+}{2},v^+;\v\Big)\,
\left(\frac{[\gamma^i,\gamma^j]}{4} gt \cdot \mathcal{F}_{ij}(\uv) - \frac{(\k_1^j\!+\!\k^j)}{2}\overleftrightarrow{\mathcal{D}_{\v^j}} 
- i \overleftarrow{\mathcal{D}_{\v^j}}\, \overrightarrow{\mathcal{D}_{\v^j}}\, \right)
\mathcal{U}_F\Big(v^+,-\frac{L^+}{2};\v\Big)
\mathcal{U}_F^{\dag}(\w)
\bigg] 
 \nn \\
& \times
\gamma^+\frac{(\slashed{\check{k}}\!+\!m)}{2k^+} \gamma^\mu 
\frac{(\slashed{\check{p}}\!+\!m)}{2p^+}
\gamma^+ v(2) 
\label{S-bef-q_dec} 
\, .
\end{align}
Similarly, one can extract from Eq.~\eqref{S-bef-2} the NEik contribution associated with decorations inserted on the antiquark Wilson line, and simplify it into
\begin{align}
&S^{\rm bef}_{q_1 \bar q_2 \leftarrow \gamma^*} \bigg|_{\textrm{dec. on }\bar{q}}
=
2\pi\delta(k_1^+\!+\!k_2^+\!-\!q^+)\:
(-i) e e_f \, \epsilon^\lambda_\mu (q) 
\int \frac{d^3\uk}{(2\pi)^3}  \, 2\pi\delta(k^+\!-\!k_1^+)
 \int \frac{d^3\up}{(2\pi)^3}\, 
 (2\pi)^3 \delta^{(3)}(\uq - \uk+\up)\:
 \nn \\
&
\times
\frac{1}{2k_2^+}\; \frac{i}{(q^- \!-\! \check{k}^- \!+\! \check{p}^-\!+\!i\epsilon)}\, 
\bar u(1) \gamma^+ \frac{(\slashed{\check{k}}\!+\!m)}{2k^+} \gamma^\mu 
\frac{(\slashed{\check{p}}\!+\!m)}{2p^+}\gamma^+
\int d^2 \v \, e^{-i\v \cdot(\k_1\!-\!\k)} \,
\int d^2 \w\, e^{-i\w\cdot(\p\!+\!\k_2)} \, 
\int_{-\frac{L^+}{2}}^{\frac{L^+}{2}}\!\!\!\!dw^+
\nn\\
& \times
\bigg[\mathcal{U}_F(\v)
\mathcal{U}_F^{\dag}\Big(w^+,-\frac{L^+}{2};\w\Big)
\left(\frac{[\gamma^i,\gamma^j]}{4} gt\cdot \mathcal{F}_{ij}(\uw) \!-\!\frac{(\p^j\!-\!\k_2^j)}{2} \overleftrightarrow{\mathcal{D}_{\w^j}} 
\!-\!i \overleftarrow{\mathcal{D}_{\w^j}}\, \overrightarrow{\mathcal{D}_{\w^j}} \right)
\mathcal{U}_F^{\dag}\Big(\frac{L^+}{2},w^+;\w\Big)
\bigg]
 v(2) 
\label{S-bef-qbar_dec}
\, .
\end{align}

After extracting from Eq.~\eqref{S-bef-2}  the contributions \eqref{S-bef-q_dec}  and \eqref{S-bef-qbar_dec} of decorations on the quark or antiquark Wilson lines, the leftover is
\begin{align}
&S^{\rm bef}_{q_1 \bar q_2 \leftarrow \gamma^*} 
-S^{\rm bef}_{q_1 \bar q_2 \leftarrow \gamma^*}\bigg|_{\textrm{dec. on }q}
-S^{\rm bef}_{q_1 \bar q_2 \leftarrow \gamma^*} \bigg|_{\textrm{dec. on }\bar{q}}
= -i e e_f \, \epsilon^\lambda_\mu (q) 
\int \frac{d^3\uk}{(2\pi)^3}  \, \theta(k^+) \int \frac{d^3\up}{(2\pi)^3}\, \theta(-p^+)\,
 (2\pi)^3 \delta^{(3)}(\uq - \uk+\up)\:
 \nn \\
&
\times
 \frac{i}{(q^- \!-\! \check{k}^- \!+\! \check{p}^-\!+\!i\epsilon)}\, 
 e^{i\frac{L^+}{2} (q^- - \check{k}^- + \check{p}^-)}\;
   \bar u(1) \gamma^+  \frac{(\slashed{\check{k}}\!+\!m)}{2k^+} \gamma^\mu 
\frac{(\slashed{\check{p}}\!+\!m)}{2p^+}
\gamma^+ v(2) 
 \nn \\
& \times
\int d^2 \v \, e^{-i\v \cdot(\k_1\!-\!\k)}
\int d^2 \w\, e^{-i\w\cdot(\p\!+\!\k_2)}
\int dv^-\, e^{iv^-(k_1^+-k^+)}\int dw^-\,   e^{iw^-(k_2^++p^+)}\;
\mathcal{U}_F(\v,v^-)
\mathcal{U}_F^{\dag}(\w,w^-)
\nn \\
& 
+\, i e e_f \, \epsilon^\lambda_\mu (q) \, \bar u(1)\gamma^{\mu}  v(2) \;
 (2\pi)^3 \delta^{(3)}(\uq - \underline{k_1}-\underline{k_2})\:
 \frac{i}{(q^- \!-\! \ck_1^-\!-\!\ck_2^-\!+\!i\epsilon)}\, 
 e^{i\frac{L^+}{2} (q^- - \ck_1^--\ck_2^-)}  
\label{S-bef-leftover_1} 
\, .
\end{align}
Noting that
\begin{align}
&
\int \frac{d^3\uk}{(2\pi)^3}  \, \theta(k^+) \int \frac{d^3\up}{(2\pi)^3}\, \theta(-p^+)\,
 (2\pi)^3 \delta^{(3)}(\uq - \uk+\up)\:
  \frac{i}{(q^- \!-\! \check{k}^- \!+\! \check{p}^-\!+\!i\epsilon)}\, 
 e^{i\frac{L^+}{2} (q^- - \check{k}^- + \check{p}^-)}\;
 \nn \\
&
\times
   \bar u(1) \gamma^+  \frac{(\slashed{\check{k}}\!+\!m)}{2k^+} \gamma^\mu 
\frac{(\slashed{\check{p}}\!+\!m)}{2p^+}
\gamma^+ v(2)
\int d^2 \v \, e^{-i\v \cdot(\k_1\!-\!\k)}
\int d^2 \w\, e^{-i\w\cdot(\p\!+\!\k_2)}
\int dv^-\, e^{iv^-(k_1^+-k^+)}\int dw^-\,   e^{iw^-(k_2^++p^+)}\; 
 \nn \\
= &\,
 \int \frac{d^3\uk}{(2\pi)^3}  \, \theta(k^+) \int \frac{d^3\up}{(2\pi)^3}\, \theta(-p^+)\,
 (2\pi)^3 \delta^{(3)}(\uq - \uk+\up)\:
  \frac{i}{(q^- \!-\! \check{k}^- \!+\! \check{p}^-\!+\!i\epsilon)}\, 
 e^{i\frac{L^+}{2} (q^- - \check{k}^- + \check{p}^-)}\;
 \nn \\
&
\times
   \bar u(1) \gamma^+  \frac{(\slashed{\check{k}}\!+\!m)}{2k^+} \gamma^\mu 
\frac{(\slashed{\check{p}}\!+\!m)}{2p^+}
\gamma^+ v(2)\;
(2\pi)^3 \delta^{(3)}(\uk\!-\!\underline{k_1})\;
(2\pi)^3 \delta^{(3)}(\up\!+\!\underline{k_2})\;
 \nn \\
= &\,
(2\pi)^3 \delta^{(3)}(\uq - \underline{k_1}-\underline{k_2})\:
 \frac{i}{(q^- \!-\! \ck_1^-\!-\!\ck_2^-\!+\!i\epsilon)}\, 
 e^{i\frac{L^+}{2} (q^- - \ck_1^--\ck_2^-)}\;
\bar u(1) \gamma^+  \frac{(\slashed{\check{k}_1}\!+\!m)}{2k^+_1} \gamma^\mu 
\frac{(-\slashed{\check{k}_2}\!+\!m)}{(-2k^+_2)}
\gamma^+ v(2)
 \nn \\
= &\,
(2\pi)^3 \delta^{(3)}(\uq - \underline{k_1}-\underline{k_2})\:
 \frac{i}{(q^- \!-\! \ck_1^-\!-\!\ck_2^-\!+\!i\epsilon)}\, 
 e^{i\frac{L^+}{2} (q^- - \ck_1^--\ck_2^-)}\;
\bar u(1) \gamma^\mu  v(2)\;
\label{S-bef-leftover_vac_int} 
\, ,
\end{align}
it is possible to rewrite Eq.~\eqref{S-bef-leftover_1}  as
\begin{align}
&S^{\rm bef}_{q_1 \bar q_2 \leftarrow \gamma^*} 
-S^{\rm bef}_{q_1 \bar q_2 \leftarrow \gamma^*}\bigg|_{\textrm{dec. on }q}
-S^{\rm bef}_{q_1 \bar q_2 \leftarrow \gamma^*} \bigg|_{\textrm{dec. on }\bar{q}}
= -i e e_f \, \epsilon^\lambda_\mu (q) 
\int \frac{d^3\uk}{(2\pi)^3}  \, \theta(k^+) \int \frac{d^3\up}{(2\pi)^3}\, \theta(-p^+)\,
 (2\pi)^3 \delta^{(3)}(\uq - \uk+\up)\:
 \nn \\
&
\times
 \frac{i}{(q^- \!-\! \check{k}^- \!+\! \check{p}^-\!+\!i\epsilon)}\, 
 e^{i\frac{L^+}{2} (q^- - \check{k}^- + \check{p}^-)}\;
   \bar u(1) \gamma^+  \frac{(\slashed{\check{k}}\!+\!m)}{2k^+} \gamma^\mu 
\frac{(\slashed{\check{p}}\!+\!m)}{2p^+}
\gamma^+ v(2) 
 \nn \\
& \times
\int d^2 \v \, e^{-i\v \cdot(\k_1\!-\!\k)}
\int d^2 \w\, e^{-i\w\cdot(\p\!+\!\k_2)}
\int dv^-\, e^{iv^-(k_1^+-k^+)}\int dw^-\,   e^{iw^-(k_2^++p^+)}\;
\Big[\mathcal{U}_F(\v,v^-)
\mathcal{U}_F^{\dag}(\w,w^-)
-1\Big]
\label{S-bef-leftover_2} 
\, .
\end{align}
Expanding the $L^+$ dependent phase factor at small target width $L^+$, one finds the contribution linear  in $L^+$ to be
\begin{align}
&S^{\rm bef}_{q_1 \bar q_2 \leftarrow \gamma^*} \bigg|_{L^+\textrm{ phase}}
= -i e e_f \, \epsilon^\lambda_\mu (q) 
\int \frac{d^3\uk}{(2\pi)^3}  \, \theta(k^+) \int \frac{d^3\up}{(2\pi)^3}\, \theta(-p^+)\,
 (2\pi)^3 \delta^{(3)}(\uq - \uk+\up)\:
  \bar u(1) \gamma^+  \frac{(\slashed{\check{k}}\!+\!m)}{2k^+} \gamma^\mu 
\frac{(\slashed{\check{p}}\!+\!m)}{2p^+}
\gamma^+ v(2) 
 \nn \\
&
\times
 (-1)\,\frac{L^+}{2} 
\int d^2 \v \, e^{-i\v \cdot(\k_1\!-\!\k)}
\int d^2 \w\, e^{-i\w\cdot(\p\!+\!\k_2)}
\int dv^-\, e^{iv^-(k_1^+-k^+)}\int dw^-\,   e^{iw^-(k_2^++p^+)}\;
\Big[\mathcal{U}_F(\v,v^-)
\mathcal{U}_F^{\dag}(\w,w^-)
-1\Big]
\label{S-bef-Lplus_1} 
\, .
\end{align}
The contribution \eqref{S-bef-leftover_2} starts at NEik accuracy due to the overall $L^+$ factor. We can thus neglect the dependence of the Wilson lines on $v^-$ or $w^-$ in that case, and find
\begin{align}
&S^{\rm bef}_{q_1 \bar q_2 \leftarrow \gamma^*} \bigg|_{L^+\textrm{ phase}}
= 2\pi\delta(k_1^+\!+\!k_2^+\!-\!q^+)\: (-i) e e_f \, \epsilon^\lambda_\mu (q) 
\int \frac{d^3\uk}{(2\pi)^3}  \, 2\pi\delta(k^+\!-\!k_1^+) \int \frac{d^3\up}{(2\pi)^3}\, 
 (2\pi)^3 \delta^{(3)}(\uq - \uk+\up)\;
 \nn \\
&
\times
 (-1)\, \frac{L^+}{2}\;
   \bar u(1) \gamma^+  \frac{(\slashed{\check{k}}\!+\!m)}{2k^+} \gamma^\mu 
\frac{(\slashed{\check{p}}\!+\!m)}{2p^+}
\gamma^+ v(2) 
\int d^2 \v \, e^{-i\v \cdot(\k_1\!-\!\k)}
\int d^2 \w\, e^{-i\w\cdot(\p\!+\!\k_2)}\;
\Big[\mathcal{U}_F(\v)
\mathcal{U}_F^{\dag}(\w)
-1\Big]
\label{S-bef-Lplus} 
\, .
\end{align}
Subtracting the contribution \eqref{S-bef-Lplus} as well from Eq.~\eqref{S-bef-leftover_2}, one has, at NEik accuracy
\begin{align}
&S^{\rm bef}_{q_1 \bar q_2 \leftarrow \gamma^*} 
-S^{\rm bef}_{q_1 \bar q_2 \leftarrow \gamma^*}\bigg|_{\textrm{dec. on }q}
-S^{\rm bef}_{q_1 \bar q_2 \leftarrow \gamma^*} \bigg|_{\textrm{dec. on }\bar{q}}
-S^{\rm bef}_{q_1 \bar q_2 \leftarrow \gamma^*} \bigg|_{L^+\textrm{ phase}}
= -i e e_f \, \epsilon^\lambda_\mu (q) 
\int \frac{d^3\uk}{(2\pi)^3}  \, \theta(k^+) \int \frac{d^3\up}{(2\pi)^3}\, \theta(-p^+)\,
 \nn \\
&
\times
(2\pi)^3 \delta^{(3)}(\uq - \uk+\up)\:
 \frac{i}{(q^- \!-\! \check{k}^- \!+\! \check{p}^-\!+\!i\epsilon)}\;
   \bar u(1) \gamma^+  \frac{(\slashed{\check{k}}\!+\!m)}{2k^+} \gamma^\mu 
\frac{(\slashed{\check{p}}\!+\!m)}{2p^+}
\gamma^+ v(2) 
 \nn \\
& \times
\int d^2 \v \, e^{-i\v \cdot(\k_1\!-\!\k)}
\int d^2 \w\, e^{-i\w\cdot(\p\!+\!\k_2)}
\int dv^-\, e^{iv^-(k_1^+-k^+)}\int dw^-\,   e^{iw^-(k_2^++p^+)}\;
\Big[\mathcal{U}_F(\v,v^-)
\mathcal{U}_F^{\dag}(\w,w^-)
-1\Big]
\label{S-bef-leftover_3} 
\, .
\end{align}
In Eq.~\eqref{S-bef-leftover_3}, the only leftover effect beyond the Eikonal approximation is the dependence on $v^-$ or $w^-$ of the Wilson lines. Using the change of variables $(v^-,w^-)\mapsto(b^-,r^-)$ defined as 
\begin{align}
&\, b^- = \frac{(v^-+w^-)}{2}  \, , \hspace{1cm} r^- =(w^--v^-)
\end{align}
so that 
\begin{align}
&\, w^- = b^- +\frac{r^-}{2}  \, , \hspace{1cm} v^- =b^- -\frac{r^-}{2}\, ,
\end{align}
and Taylor-expanding the Wilson lines around $b^-$, one finds
\begin{align}
&
\int dv^-\, e^{iv^-(k_1^+-k^+)}\int dw^-\,   e^{iw^-(k_2^++k^+-q^+)}\;
\Big[\mathcal{U}_F(\v,v^-)
\mathcal{U}_F^{\dag}(\w,w^-)
-1\Big]
\nn\\
=&\, 
 \int db^-\, e^{ib^-(k_1^++k_2^+-q^+)}\int dr^-\,   e^{ir^-\left[k^+-\frac{1}{2}(k_1^+-k_2^++q^+)\right]}\;
\bigg[\mathcal{U}_F\Big(\v,b^- \!-\!\frac{r^-}{2}\Big)
\mathcal{U}_F^{\dag}\Big(\w,b^- \!+\!\frac{r^-}{2}\Big)
-1\bigg]
\nn\\
=&\, 
 \int db^-\, e^{ib^-(k_1^++k_2^+-q^+)}\int dr^-\,   e^{ir^-\left[k^+-\frac{1}{2}(k_1^+-k_2^++q^+)\right]}\;
\bigg\{\bigg[\mathcal{U}_F\Big(\v,b^- \Big)
\mathcal{U}_F^{\dag}\Big(\w,b^- \Big)
-1\bigg]
\nn\\
&\,  \;
-\frac{r^-}{2}\, \left(\partial_{b^-}\mathcal{U}_F\Big(\v,b^- \Big)\right)
\mathcal{U}_F^{\dag}\Big(\w,b^- \Big)
+\frac{r^-}{2}\, \mathcal{U}_F\Big(\v,b^- \Big)
\left(\partial_{b^-}\mathcal{U}_F^{\dag}\Big(\w,b^- \Big)\right)
+O(NNEik)
\bigg\}
\nn\\
=&\, 
2\pi\delta\left(k^+\!-\!\frac{1}{2}(k_1^+\!-\!k_2^+\!+\!q^+)\right)
 \int db^-\, e^{ib^-(k_1^++k_2^+-q^+)}\;
\bigg[\mathcal{U}_F\Big(\v,b^- \Big)
\mathcal{U}_F^{\dag}\Big(\w,b^- \Big)
-1\bigg]
\nn\\
&\,  \;
-\frac{i}{2}\: 2\pi\delta'\left(k^+\!-\!\frac{1}{2}(k_1^+\!-\!k_2^+\!+\!q^+)\right)
\int db^-\, e^{ib^-(k_1^++k_2^+-q^+)}\;
\bigg[\mathcal{U}_F\Big(\v,b^- \Big)\overleftrightarrow{\partial_{b^-}}
\mathcal{U}_F^{\dag}\Big(\w,b^- \Big)
\bigg]
+O(NNEik)
\nn\\
=&\, 
2\pi\delta\left(k^+\!-\!\frac{1}{2}(k_1^+\!-\!k_2^+\!+\!q^+)\right)
 \int db^-\, e^{ib^-(k_1^++k_2^+-q^+)}\;
\bigg[\mathcal{U}_F\Big(\v,b^- \Big)
\mathcal{U}_F^{\dag}\Big(\w,b^- \Big)
-1\bigg]
\nn\\
&\,  \;
-\frac{i}{2}\: 2\pi\delta'\left(k^+\!-\!\frac{1}{2}(k_1^+\!-\!k_2^+\!+\!q^+)\right)
2\pi\delta(k_1^+\!+\!k_2^+\!-\!q^+)\;
\bigg[\mathcal{U}_F\Big(\v,b^- \Big)\overleftrightarrow{\partial_{b^-}}
\mathcal{U}_F^{\dag}\Big(\w,b^- \Big)
\bigg]\bigg|_{b^-=0}
+O(NNEik)
\label{relative_z_minus_dep} 
\, .
\end{align}
Indeed, each derivative along the $-$ direction acting on a Wilson line provides an extra power suppression in the high-energy limit. In the last step of Eq.~\eqref{relative_z_minus_dep}, in the second term which starts only at NEik order due to the derivative in $b^-$, we have neglected the overall $b^-$ of the color object after taking the derivative, since it would matter only at NNEik accuracy. 
By contrast, in the first term in \eqref{relative_z_minus_dep} we have kept the dependence of the Wilson lines on the average $b^-$ instead of expanding it, in order to facilitate the calculation on the cross section. This corresponds  to the Generalized Eikonal  approximation  at the  S-matrix  level.   
Hence, at NEik accuracy, one finds two contributions in the expression \eqref{S-bef-leftover_3}, corresponding to the two terms in Eq.~\eqref{relative_z_minus_dep}.
First, we have the Generalized Eikonal contribution
\begin{align}
S^{\rm bef}_{q_1 \bar q_2 \leftarrow \gamma^*} \bigg|_{\textrm{Gen. Eik}}
=&\, -i e e_f \, \epsilon^\lambda_\mu (q) 
\int d^2 \v \int d^2 \w
 \int db^-\, e^{ib^-(k_1^++k_2^+-q^+)}\;
\bigg[\mathcal{U}_F\Big(\v,b^- \Big)
\mathcal{U}_F^{\dag}\Big(\w,b^- \Big)
-1\bigg]
 \nn \\
&
\times
\int \frac{d^3\uk}{(2\pi)^3}  \, \theta(k^+) 
2\pi\delta\left(k^+\!-\!\frac{1}{2}(k_1^+\!-\!k_2^+\!+\!q^+)\right)
\int \frac{d^3\up}{(2\pi)^3}\, \theta(-p^+)\,
(2\pi)^3 \delta^{(3)}(\uq - \uk+\up)\:
  \,   
 \nn \\
& \times
e^{-i\v \cdot(\k_1\!-\!\k)}
\, e^{-i\w\cdot(\p\!+\!\k_2)}\:
 \frac{i}{(q^- \!-\! \check{k}^- \!+\! \check{p}^-\!+\!i\epsilon)}\; 
  \bar u(1) \gamma^+  \frac{(\slashed{\check{k}}\!+\!m)}{2k^+} \gamma^\mu 
\frac{(\slashed{\check{p}}\!+\!m)}{2p^+}
\gamma^+ v(2) 
\label{S-bef-GenEik} 
\, ,
\end{align}
which contains the dependence of the Wilson lines on a common $b^-$ as the only leftover effect beyond the strict Eikonal approximation. Second, we have an explicit NEik correction\footnote{Note that due to the choice of the light cone gauge $A^+=0$, the ordinary derivative $\overleftrightarrow{\partial_{b^-}}$ can equivalently be written as a covariant derivative $\overleftrightarrow{D_{b^-}}$.} 
\begin{align}
S^{\rm bef}_{q_1 \bar q_2 \leftarrow \gamma^*} \bigg|_{\textrm{dyn. target}}
= &\,
2\pi\delta(k_1^+\!+\!k_2^+\!-\!q^+)\;
(-i) e e_f \, \epsilon^\lambda_\mu (q) 
\int d^2 \v \int d^2 \w
\bigg[\mathcal{U}_F\Big(\v,b^- \Big)\overleftrightarrow{\partial_{b^-}}
\mathcal{U}_F^{\dag}\Big(\w,b^- \Big)
\bigg]\bigg|_{b^-=0}
 \nn \\
&
\times
\int \frac{d^3\uk}{(2\pi)^3}  \, \theta(k^+)\, 
(-1)\, \frac{i}{2}\: 2\pi\delta'\!(k^+\!-\!k_1^+)
 \int \frac{d^3\up}{(2\pi)^3}\, \theta(-p^+)\,
(2\pi)^3 \delta^{(3)}(\uq - \uk+\up)\:
 \nn \\
&
\times
 e^{-i\v \cdot(\k_1\!-\!\k)}
\, e^{-i\w\cdot(\p\!+\!\k_2)}\,
\frac{i}{(q^- \!-\! \check{k}^- \!+\! \check{p}^-\!+\!i\epsilon)}\;
   \bar u(1) \gamma^+  \frac{(\slashed{\check{k}}\!+\!m)}{2k^+} \gamma^\mu 
\frac{(\slashed{\check{p}}\!+\!m)}{2p^+}
\gamma^+ v(2) 
 \, 
\label{S-bef-dyn} 
\, .
\end{align}
As a reminder, the $-$ axis in light-cone coordinates can be interpreted as the longitudinal direction for the right-moving projectile and as the time direction for the left-moving target. If the target background field is dynamical, meaning $z^-$ dependent, the quark and antiquark from the projectile will probe a different value of the field not only because of their transverse separation $\w\!-\!\v$ but also because of their longitudinal separation $r^-=w^-\!-\!v^-$. The contribution \eqref{S-bef-dyn} is then the NEik correction induced by the tidal-like force exerted by the dynamical background field on the quark and antiquark due to their longitudinal separation $r^-$.

All in all, we have written the S-matrix element for the before diagram at NEik accuracy as
\begin{align}
S^{\rm bef}_{q_1 \bar q_2 \leftarrow \gamma^*}
=&\, 
S^{\rm bef}_{q_1 \bar q_2 \leftarrow \gamma^*} \bigg|_{\textrm{Gen. Eik}}
+S^{\rm bef}_{q_1 \bar q_2 \leftarrow \gamma^*}\bigg|_{\textrm{dec. on }q}
+S^{\rm bef}_{q_1 \bar q_2 \leftarrow \gamma^*} \bigg|_{\textrm{dec. on }\bar{q}}
+S^{\rm bef}_{q_1 \bar q_2 \leftarrow \gamma^*} \bigg|_{L^+\textrm{ phase}}
+S^{\rm bef}_{q_1 \bar q_2 \leftarrow \gamma^*} \bigg|_{\textrm{dyn. target}}
\nn\\
&\;
 +O(NNEik)
\label{S-bef-total} 
\, ,
\end{align}
with the Generalized Eikonal contribution given in Eq.~\eqref{S-bef-GenEik}, and the four types of further NEik corrections given respectively by Eqs.~\eqref{S-bef-q_dec}, \eqref{S-bef-qbar_dec}, \eqref{S-bef-Lplus} and \eqref{S-bef-dyn}.  

Most of these contributions contain the same energy denominator, which is
\begin{align}
(q^- \!-\! \check{k}^- \!+\! \check{p}^-\!+\!i\epsilon)
=&\,
\frac{(\q^2\!-\!Q^2)}{2q^+}-\frac{(\k^2\!+\!m^2)}{2k^+}+\frac{(\p^2\!+\!m^2)}{2p^+}+i\epsilon
\label{ED_1} 
\, ,
\end{align}
introducing as usual the photon virtuality $Q^2\equiv -q^{\mu}q_{\mu}$. Using the momentum conservation constraint $\up=\uk-\uq$ at the photon splitting vertex, valid for all these contributions, one finds
\begin{align}
(q^- \!-\! \check{k}^- \!+\! \check{p}^-\!+\!i\epsilon)
=&\,
\frac{(\q^2-Q^2)}{2q^+}-\frac{(\k^2+m^2)}{2k^+}-\frac{\left((\q\!-\!\k)^2+m^2\right)}{2(q^+\!-\!k^+)}+i\epsilon
\nn\\
=&\,
-\frac{Q^2}{2q^+}-\frac{q^+}{2k^+(q^+\!-\!k^+)}\bigg[\left(\k\!-\!\frac{k^+}{q^+}\q\right)^2 +m^2 \bigg]+i\epsilon
\label{ED_2} 
\, .
\end{align}
Since the real part of this energy denominator cannot change sign, the $+i\epsilon$ has no effect and can be always dropped. 
Using the expression \eqref{ED_2} in the Generalized Eikonal contribution \eqref{S-bef-GenEik} , one has
\begin{align}
S^{\rm bef}_{q_1 \bar q_2 \leftarrow \gamma^*} \bigg|_{\textrm{Gen. Eik}}
=&\,  e e_f \, \epsilon^\lambda_\mu (q) 
\int d^2 \v \int d^2 \w
 \int db^-\, e^{ib^-(k_1^++k_2^+-q^+)}\;
\bigg[\mathcal{U}_F\Big(\v,b^- \Big)
\mathcal{U}_F^{\dag}\Big(\w,b^- \Big)
-1\bigg]
 \nn \\
&
\times
\frac{1}{2q^+}\int \frac{d^3\uk}{(2\pi)^3}  \, \theta(k^+) \, \theta(q^+\!-\!k^+) 
2\pi\delta\left(k^+\!-\!\frac{1}{2}(k_1^+\!-\!k_2^+\!+\!q^+)\right)
e^{-i\v \cdot(\k_1-\k)}\, e^{-i\w\cdot(\k_2+\k-\q)}\:
  \,   
 \nn \\
& \times
 \frac{
 \bar u(1) \gamma^+  (\slashed{\check{k}}\!+\!m) \gamma^\mu 
(\slashed{\check{k}}\!-\!\slashed{q}\!+\!m)
\gamma^+ v(2) 
 }{\left[\left(\k\!-\!\frac{k^+}{q^+}\q\right)^2 +m^2
 +\frac{k^+(q^+-k^+)}{(q^+)^2}\, Q^2\right]}\; 
\label{S-bef-GenEik_fin} 
\, .
\end{align}
Similarly, one obtains 
\begin{align}
S^{\rm bef}_{q_1 \bar q_2 \leftarrow \gamma^*} \bigg|_{\textrm{dyn. target}}
= &\,
2\pi\delta(k_1^+\!+\!k_2^+\!-\!q^+)\;
 e e_f \, \epsilon^\lambda_\mu (q)\,
\int d^2 \v \int d^2 \w
\bigg[\mathcal{U}_F\Big(\v,b^- \Big)\overleftrightarrow{\partial_{b^-}}
\mathcal{U}_F^{\dag}\Big(\w,b^- \Big)
\bigg]\bigg|_{b^-=0}
 \nn \\
&
\times
\frac{(-i)}{4q^+}\int \frac{d^3\uk}{(2\pi)^3}  \, \theta(k^+) \, \theta(q^+\!-\!k^+) 
2\pi\delta'\!(k^+\!-\!k_1^+)
e^{-i\v \cdot(\k_1-\k)}\, e^{-i\w\cdot(\k_2+\k-\q)}\:
 \,   
 \nn \\
& \times
 \frac{
 \bar u(1) \gamma^+  (\slashed{\check{k}}\!+\!m) \gamma^\mu 
(\slashed{\check{k}}\!-\!\slashed{q}\!+\!m)
\gamma^+ v(2) 
 }{\left[\left(\k\!-\!\frac{k^+}{q^+}\q\right)^2 +m^2
 +\frac{k^+(q^+-k^+)}{(q^+)^2}\, Q^2\right]}\; 
 \, 
\label{S-bef-dyn_fin} 
\end{align}
  from Eq.~\eqref{S-bef-dyn} and 
\begin{align}
&S^{\rm bef}_{q_1 \bar q_2 \leftarrow \gamma^*} \bigg|_{L^+\textrm{ phase}}
= 2\pi\delta(k_1^+\!+\!k_2^+\!-\!q^+)\:  e e_f \, \epsilon^\lambda_\mu (q) 
\int d^2 \v \int d^2 \w\, \Big[\mathcal{U}_F(\v)
\mathcal{U}_F^{\dag}(\w)
-1\Big]
\,   
 \nn \\
& \times\,
 \frac{(-i)L^+}{8k_1^+k_2^+}
\int \frac{d^3\uk}{(2\pi)^3}  \, 2\pi\delta(k^+\!-\!k_1^+) 
e^{-i\v \cdot(\k_1-\k)}\, e^{-i\w\cdot(\k_2+\k-\q)}\:
 \bar u(1) \gamma^+  (\slashed{\check{k}}\!+\!m) \gamma^\mu 
(\slashed{\check{k}}\!-\!\slashed{q}\!+\!m)
\gamma^+ v(2) 
\label{S-bef-Lplus_fin} 
\end{align}
 from Eq.~\eqref{S-bef-Lplus}. Then, from Eq.~\eqref{S-bef-q_dec} , one finds
\begin{align}
&S^{\rm bef}_{q_1 \bar q_2 \leftarrow \gamma^*}\bigg|_{\textrm{dec. on }q} 
= 
2\pi\delta(k_1^+\!+\!k_2^+\!-\!q^+)\:
 e e_f \, \epsilon^\lambda_\mu (q) 
\int d^2 \v\int d^2 \w\,  \frac{1}{4q^+ k_1^+}\, \int_{-\frac{L^+}{2}}^{\frac{L^+}{2}}\!\!\!\!dv^+\, 
\int \frac{d^3\uk}{(2\pi)^3}  \, 
2\pi\delta(k^+\!-\!k_1^+)
 \nn \\
& \times
\bar u(1)   \, 
\bigg[\mathcal{U}_F\Big(\frac{L^+}{2},v^+;\v\Big)\,
\left(\frac{[\gamma^i,\gamma^j]}{4} gt \cdot \mathcal{F}_{ij}(\uv) - \frac{(\k_1^j\!+\!\k^j)}{2}\overleftrightarrow{\mathcal{D}_{\v^j}} 
- i \overleftarrow{\mathcal{D}_{\v^j}}\, \overrightarrow{\mathcal{D}_{\v^j}}\, \right)
\mathcal{U}_F\Big(v^+,-\frac{L^+}{2};\v\Big)
\mathcal{U}_F^{\dag}(\w)
\bigg] 
 \nn \\
& \times
\frac{
 \gamma^+  (\slashed{\check{k}}\!+\!m) \gamma^\mu 
(\slashed{\check{k}}\!-\!\slashed{q}\!+\!m)
\gamma^+ v(2) 
 }{\left[\left(\k\!-\!\frac{k^+}{q^+}\q\right)^2 +m^2
 +\frac{k^+(q^+-k^+)}{(q^+)^2}\, Q^2\right]}\; e^{-i\v \cdot(\k_1-\k)}\, e^{-i\w\cdot(\k_2+\k-\q)}\,\;
\label{S-bef-q_dec_2} 
\, .
\end{align}
 At this stage, let us note that the third line in Eq.~\eqref{S-bef-q_dec_2} is identical to the $\k$ dependent factor in the integrand of the Generalized Eikonal contribution \eqref{S-bef-GenEik_fin} or in the integrand of Eq.~\eqref{S-bef-dyn_fin}. By contrast, in Eq.~\eqref{S-bef-q_dec_2} there is an extra dependence on $\k$ in the bracket in the second line. For later convenience, let us eliminate this extra dependence on $\k$. Since $\k$ appears both in the $\v$ and in the $\w$ dependent phase factors, this extra $\k$ can be traded for a derivative in $\v$ or for a derivative in $\w$ or for a combination of both. All of these choices would lead to equivalent results, related to each other via integration by parts. Choosing to trade $\k$ for a derivative in $\w$, one arrives at
\begin{align}
&S^{\rm bef}_{q_1 \bar q_2 \leftarrow \gamma^*}\bigg|_{\textrm{dec. on }q} 
= 
2\pi\delta(k_1^+\!+\!k_2^+\!-\!q^+)\:
 e e_f \, \epsilon^\lambda_\mu (q) 
\int d^2 \v\int d^2 \w\,  \frac{1}{4q^+ k_1^+}\, \int_{-\frac{L^+}{2}}^{\frac{L^+}{2}}\!\!\!\!dv^+\, 
\int \frac{d^3\uk}{(2\pi)^3}  \, 
2\pi\delta(k^+\!-\!k_1^+)
 \nn \\
& \times
\bar u(1)   \, 
\bigg\{\bigg[\mathcal{U}_F\Big(\frac{L^+}{2},v^+;\v\Big)\,
\left(\frac{[\gamma^i,\gamma^j]}{4} gt \cdot \mathcal{F}_{ij}(\uv) - \frac{(\k_1^j\!-\!\k_2^j\!+\!\q^j)}{2}\overleftrightarrow{\mathcal{D}_{\v^j}} 
- i \overleftarrow{\mathcal{D}_{\v^j}}\, \overrightarrow{\mathcal{D}_{\v^j}}\, \right)
\mathcal{U}_F\Big(v^+,-\frac{L^+}{2};\v\Big)
\mathcal{U}_F^{\dag}(\w)
\bigg] 
 \nn \\
& 
\hspace{2cm}
+\frac{i}{2}
\bigg[\mathcal{U}_F\Big(\frac{L^+}{2},v^+;\v\Big)\,
\overleftrightarrow{\mathcal{D}_{\v^j}}\,
\mathcal{U}_F\Big(v^+,-\frac{L^+}{2};\v\Big)
\big(\partial_{\w^j}\mathcal{U}_F^{\dag}(\w)\big)
\bigg] 
\bigg\}
\nn \\
& \times
\frac{
 \gamma^+  (\slashed{\check{k}}\!+\!m) \gamma^\mu 
(\slashed{\check{k}}\!-\!\slashed{q}\!+\!m)
\gamma^+ v(2) 
 }{\left[\left(\k\!-\!\frac{k^+}{q^+}\q\right)^2 +m^2
 +\frac{k^+(q^+-k^+)}{(q^+)^2}\, Q^2\right]}\; e^{-i\v \cdot(\k_1-\k)}\, e^{-i\w\cdot(\k_2+\k-\q)}\,\;
\label{S-bef-q_dec_fin} 
\, .
\end{align}
 
In order to obtain more compact expressions, let us introduce the notations
\begin{align}
\label{Wilson_dec_1}
\mathcal{U}^{(1)}_{F;j} ( \v) =&\,  \int_{-\frac{L^+}{2}}^{\frac{L^+}{2}}dv^+\, 
\mathcal{U}_F\Big(\frac{L^+}{2},v^+;\v\Big) 
\overleftrightarrow{\mathcal{D}_{\v^j}}
\mathcal{U}_F\Big(v^+,-\frac{L^+}{2};\v\Big) 
\\
\label{Wilson_dec_2}
\mathcal{U}^{(2)}_F ( \v) =&\,
\int_{-\frac{L^+}{2}}^{\frac{L^+}{2}}dv^+\,  
\mathcal{U}_F\Big(\frac{L^+}{2},v^+;\v\Big) 
\overleftarrow{\mathcal{D}_{\v^j}}\, \overrightarrow{\mathcal{D}_{\v^j}} 
\mathcal{U}_F\Big(v^+,-\frac{L^+}{2};\v\Big) 
\\
\mathcal{U}^{(3)}_{F; ij} ( \v)
 =&\,  
\int_{-\frac{L^+}{2}}^{\frac{L^+}{2}}dv^+\,  
\mathcal{U}_F\Big(\frac{L^+}{2},v^+;\v\Big) 
gt \!\cdot\! \mathcal{F}_{ij}(\uv) 
\mathcal{U}_F\Big(v^+,-\frac{L^+}{2};\v\Big) 
\label{Wilson_dec_3}
\end{align}
 for the decorated Wilson lines appearing at NEik accuracy.
Their Hermitian conjugate is
\ba
\label{Wilson_dec_1_hc}
\mathcal{U}^{(1)\dag}_{F;j} ( \v) &=&  
-\int_{-\frac{L^+}{2}}^{\frac{L^+}{2}}dv^+\, 
\mathcal{U}_F^{\dag}\Big(v^+,-\frac{L^+}{2};\v\Big) 
\overleftrightarrow{\mathcal{D}_{\v^j}}
\mathcal{U}_F^{\dag}\Big(\frac{L^+}{2},v^+;\v\Big)  
\\
\label{Wilson_dec_2_hc}
\mathcal{U}^{(2)\dag}_F ( \v) &=& 
\int_{-\frac{L^+}{2}}^{\frac{L^+}{2}}dv^+\,  
\mathcal{U}_F^{\dag}\Big(v^+,-\frac{L^+}{2};\v\Big) 
\overleftarrow{\mathcal{D}_{\v^j}}\, \overrightarrow{\mathcal{D}_{\v^j}} 
\mathcal{U}_F^{\dag}\Big(\frac{L^+}{2},v^+;\v\Big)
\\
\mathcal{U}^{(3)\dag}_{F; ij} ( \v) &=&  
\int_{-\frac{L^+}{2}}^{\frac{L^+}{2}}dv^+\, 
\mathcal{U}_F^{\dag}\Big(v^+,-\frac{L^+}{2};\v\Big) 
gt \!\cdot\! \mathcal{F}_{ij}(\uv)
\mathcal{U}_F^{\dag}\Big(\frac{L^+}{2},v^+;\v\Big) 
\label{Wilson_dec_3_hc}
\, .
\ea
With these notations, the expression \eqref{S-bef-q_dec_fin} becomes
\begin{align}
 S^{\rm bef}_{q_1 \bar q_2 \leftarrow \gamma^*}\bigg|_{\textrm{dec. on }q} 
= &\,
2\pi\delta(k_1^+\!+\!k_2^+\!-\!q^+)\:
 e e_f \, \epsilon^\lambda_\mu (q) 
 \frac{1}{4q^+ k_1^+}
\int d^2 \v\int d^2 \w
\int \frac{d^3\uk}{(2\pi)^3}  \,  2\pi\delta(k^+\!-\!k_1^+)\,  
e^{-i\v \cdot(\k_1-\k)}\, e^{-i\w\cdot(\k_2+\k-\q)}\,
 \nn \\
& \times
\bar u(1)   \! 
\left[
\frac{[\gamma^i,\gamma^j]}{4}\, \mathcal{U}^{(3)}_{F; ij} ( \v)
- i\,  \mathcal{U}^{(2)}_F ( \v) 
+ \, \mathcal{U}^{(1)}_{F;j} ( \v)  
\bigg(- \frac{(\k_1^j\!-\!\k_2^j\!+\!\q^j)}{2}+\frac{i}{2}\, \partial_{\w^j}\bigg)
\right]
\mathcal{U}_F^{\dag}(\w)\,
\nn \\
& \times
\frac{
 \gamma^+  (\slashed{\check{k}}\!+\!m) \gamma^\mu 
(\slashed{\check{k}}\!-\!\slashed{q}\!+\!m)
\gamma^+ v(2) 
 }{\left[\left(\k\!-\!\frac{k^+}{q^+}\q\right)^2 +m^2
 +\frac{k^+(q^+-k^+)}{(q^+)^2}\, Q^2\right]}\; \;
\label{S-bef-q_dec_fin_2} 
\, .
\end{align}

Following the same steps, one can rewrite the contribution with decoration on the antiquark Wilson line \eqref{S-bef-qbar_dec} as  
\begin{align}
&S^{\rm bef}_{q_1 \bar q_2 \leftarrow \gamma^*} \bigg|_{\textrm{dec. on }\bar{q}}
=
2\pi\delta(k_1^+\!+\!k_2^+\!-\!q^+)\:
 e e_f \, \epsilon^\lambda_\mu (q) 
 \int d^2 \v \int d^2 \w\,  \frac{1}{4q^+ k_2^+}\, \int_{-\frac{L^+}{2}}^{\frac{L^+}{2}}\!\!\!\!dw^+
\int \frac{d^3\uk}{(2\pi)^3}  \, 2\pi\delta(k^+\!-\!k_1^+)
 \nn \\
&
\times
e^{-i\v \cdot(\k_1-\k)}\, e^{-i\w\cdot(\k_2+\k-\q)}\;
\frac{\bar u(1)
 \gamma^+  (\slashed{\check{k}}\!+\!m) \gamma^\mu 
(\slashed{\check{k}}\!-\!\slashed{q}\!+\!m)
\gamma^+
 }{\left[\left(\k\!-\!\frac{k^+}{q^+}\q\right)^2 +m^2
 +\frac{k^+(q^+-k^+)}{(q^+)^2}\, Q^2\right]}\;
 \nn \\
&
\times 
\bigg\{
\bigg[\mathcal{U}_F(\v)
\mathcal{U}_F^{\dag}\Big(w^+,-\frac{L^+}{2};\w\Big)
\left(\frac{[\gamma^i,\gamma^j]}{4} gt\cdot \mathcal{F}_{ij}(\uw) \!+\!\frac{(\k_2^j\!-\!\k_1^j\!+\!\q^j)}{2} \overleftrightarrow{\mathcal{D}_{\w^j}} 
\!-\!i \overleftarrow{\mathcal{D}_{\w^j}}\, \overrightarrow{\mathcal{D}_{\w^j}} \right)
\mathcal{U}_F^{\dag}\Big(\frac{L^+}{2},w^+;\w\Big)
\bigg]
\nn \\
& 
\hspace{2cm}
-\frac{i}{2}
\bigg[\big(\partial_{\v^j}\mathcal{U}_F(\v)\big)
\mathcal{U}_F^{\dag}\Big(w^+,-\frac{L^+}{2};\w\Big)
\overleftrightarrow{\mathcal{D}_{\w^j}}
\mathcal{U}_F^{\dag}\Big(\frac{L^+}{2},w^+;\w\Big)
\bigg] 
\bigg\} v(2) 
\label{S-bef-qbar_dec_fin}
\, ,
\end{align}
 or, using the notations \eqref{Wilson_dec_1_hc}, \eqref{Wilson_dec_2_hc} and \eqref{Wilson_dec_3_hc},
\begin{align}
S^{\rm bef}_{q_1 \bar q_2 \leftarrow \gamma^*} \bigg|_{\textrm{dec. on }\bar{q}}
= &\,
2\pi\delta(k_1^+\!+\!k_2^+\!-\!q^+)\:
 e e_f \, \epsilon^\lambda_\mu (q) 
 \frac{1}{4q^+ k_2^+}\, 
 \int d^2 \v \int d^2 \w\,  
 \nn \\
&
\times
\int \frac{d^3\uk}{(2\pi)^3}  \, 2\pi\delta(k^+\!-\!k_1^+)\,
\frac{\bar u(1)
 \gamma^+  (\slashed{\check{k}}\!+\!m) \gamma^\mu 
(\slashed{\check{k}}\!-\!\slashed{q}\!+\!m)
\gamma^+
 }{\left[\left(\k\!-\!\frac{k^+}{q^+}\q\right)^2 +m^2
 +\frac{k^+(q^+-k^+)}{(q^+)^2}\, Q^2\right]}\;
  e^{-i\w\cdot(\k_2+\k-\q)}\,
 \nn \\
&
\hspace{-2cm}
\times 
\Bigg[\mathcal{U}_F(\v)
\left(\frac{[\gamma^i,\gamma^j]}{4}\,  \mathcal{U}^{(3)\dag}_{F; ij} ( \w)
\!-\!i\,  \mathcal{U}^{(2)\dag}_F ( \w) 
\!+\!\bigg(\frac{i}{2}\, \overleftarrow{\partial_{\v^j}} \!-\!\frac{(\k_2^j\!-\!\k_1^j\!+\!\q^j)}{2}\bigg)\mathcal{U}^{(1)\dag}_{F;j} ( \w) 
 \right)
\Bigg]
v(2) \, 
e^{-i\v \cdot(\k_1-\k)}\;
\label{S-bef-qbar_dec_fin_2}
\, ,
\end{align}


\subsection{Before contribution to the S-matrix: longitudinal photon polarization\label{sec:bef_L_pol}}

Let us now evalute our general expressions \eqref{S-bef-GenEik_fin},  \eqref{S-bef-dyn_fin}, \eqref{S-bef-Lplus_fin}, \eqref{S-bef-q_dec_fin_2} and \eqref{S-bef-qbar_dec_fin_2}, in the case of an incoming longitudinal photon. Introducing the longitudinal polarization vector in LC gauge given in \eqref{L_pol_vect}, one finds
\begin{align}
\epsilon^L_\mu (q) \, \gamma^+  (\slashed{\check{k}}\!+\!m) \gamma^\mu 
(\slashed{\check{k}}\!-\!\slashed{q}\!+\!m)
\gamma^+ 
=&\, 
\frac{Q}{q^+}\; \gamma^+  (\slashed{\check{k}}\!+\!m) \gamma^+
(\slashed{\check{k}}\!-\!\slashed{q}\!+\!m)\gamma^+
= 
\frac{Q}{q^+}\; \{\gamma^+,\,  \slashed{\check{k}}\} \gamma^+ \{
\slashed{\check{k}}\!-\!\slashed{q},\, \gamma^+\}
\nn\\
=&\, 
-\frac{4k^+(q^+\!-\!k^+)}{q^+}\, Q\,  \gamma^+
\label{intern_Dirac_struc_L}
\, .
\end{align}
Inserting this result into Eq.~\eqref{S-bef-Lplus_fin} 
\begin{align}
S^{\rm bef}_{q_1 \bar q_2 \leftarrow \gamma^*_L} \bigg|_{L^+\textrm{ phase}}
=&\, 2\pi\delta(k_1^+\!+\!k_2^+\!-\!q^+)\:  e e_f 
\int d^2 \v \int d^2 \w\, \Big[\mathcal{U}_F(\v)
\mathcal{U}_F^{\dag}(\w)
-1\Big]\,  \frac{(-i)L^+}{8k_1^+k_2^+}
\,   
 \nn \\
& \times\,
\int \frac{d^3\uk}{(2\pi)^3}  \, 2\pi\delta(k^+\!-\!k_1^+) 
e^{-i\v \cdot(\k_1-\k)}\, e^{-i\w\cdot(\k_2+\k-\q)}\:
(-1)\, \frac{4k^+(q^+\!-\!k^+)}{q^+}\, Q\,  
 \bar u(1) \gamma^+  v(2) 
 \nn\\
 =&\,
 2\pi\delta(k_1^+\!+\!k_2^+\!-\!q^+)  e e_f
 Q  \bar u(1) \gamma^+  v(2)  \frac{i L^+}{2q^+}
\int \!d^2 \v\! \int \!d^2 \w \Big[\mathcal{U}_F(\v)
\mathcal{U}_F^{\dag}(\v)
\!-\!1\Big] \delta^{(2)}(\v\!-\!\w)
 e^{-i\v\cdot(\k_1+\k_2-\q)}\:
  \nn\\
 =&\, 0
 \, .
\label{S-bef-Lplus_L} 
\end{align}

In order to calculate the cross section, it is sufficient to know the S-matrix element for $\q=0$. Hence, we will assume $\q=0$ from now on, for simplicity. Then, inserting the expression~\eqref{intern_Dirac_struc_L} into \eqref{S-bef-GenEik_fin} and taking $\q=0$, one obtains
\begin{align}
S^{\rm bef}_{q_1 \bar q_2 \leftarrow \gamma^*_L} \bigg|_{\textrm{Gen. Eik}}
=&\, -2Q\,  e e_f \, \bar u(1) \gamma^+ v(2) 
\int d^2 \v\, e^{-i\v \cdot\k_1}\int d^2 \w\, e^{-i\w\cdot\k_2}
 \int db^-\, e^{ib^-(k_1^++k_2^+-q^+)}\;
\bigg[\mathcal{U}_F\Big(\v,b^- \Big)
\mathcal{U}_F^{\dag}\Big(\w,b^- \Big)
-1\bigg]
 \nn \\
&
\times
\int \frac{d^3\uk}{(2\pi)^3}  \, \theta(k^+) \, \theta(q^+\!-\!k^+) 
2\pi\delta\left(k^+\!-\!\frac{1}{2}(k_1^+\!-\!k_2^+\!+\!q^+)\right)\;
\frac{k^+(q^+\!-\!k^+)}{(q^+)^2}\,
 \frac{
e^{i\k \cdot(\v-\w)}
 }{\left[\k^2 +m^2
 +\frac{k^+(q^+-k^+)}{(q^+)^2}\, Q^2\right]}\; 
 \nn\\
 =&\, 
 -2Q\,  \frac{e e_f}{2\pi} \, \bar u(1) \gamma^+ v(2)\, 
 \frac{(q^+\!+\!k_1^+\!-\!k_2^+)(q^+\!+\!k_2^+\!-\!k_1^+)}{4(q^+)^2}\,
 \theta(q^+\!+\!k_1^+\!-\!k_2^+)\,
 \theta(q^+\!+\!k_2^+\!-\!k_1^+)\,
 \int d^2 \v\, e^{-i\v \cdot\k_1}
  \nn \\
&
\times
\int d^2 \w\, e^{-i\w\cdot\k_2}\,
\textrm{K}_0\left(\hat{Q}\, |\w\!-\!\v|\right)
\int db^-\, e^{ib^-(k_1^++k_2^+-q^+)}\;
\bigg[\mathcal{U}_F\Big(\v,b^- \Big)
\mathcal{U}_F^{\dag}\Big(\w,b^- \Big)
-1\bigg]
\label{S-bef-GenEik_L} 
\, ,
\end{align}
using the relation 
\ba
\int \frac{d^2\k}{(2\pi)^2}  \frac{e^{-i \k \cdot \r}}{(\k^2+\Delta)} 
&=& \frac{1}{2\pi} \textrm{K}_0(\sqrt{\Delta}|\r|)
\label{int-1-scal}
\, , 
\ea
where $\textrm{K}_{\alpha}(z)$ is the modified Bessel function of the second kind. In Eq.~\eqref{S-bef-GenEik_L}, we have introduced the notation 
\ba
\hat{Q} = \sqrt{m^2 +  \frac{(q^+\!+\!k_1^+\!-\!k_2^+)(q^+\!-\!k_1^+\!+\!k_2^+)}{4(q^+)^2}\, Q^2}
\label{def_Qhat}
\, .
\ea
Similarly, in the longitudinal photon case, Eq.~\eqref{S-bef-q_dec_fin_2} becomes 
\begin{align}
S^{\rm bef}_{q_1 \bar q_2 \leftarrow \gamma^*_L}\bigg|_{\textrm{dec. on }q} 
=&\, 
2\pi\delta(k_1^+\!+\!k_2^+\!-\!q^+)\:
 \frac{e e_f}{2\pi} \, (-1)Q\, \frac{k_2^+}{(q^+)^2}
\int d^2 \v\, e^{-i\v \cdot\k_1}\int d^2 \w\, e^{-i\w\cdot\k_2}\,
\textrm{K}_0\left(\bar{Q}\, |\w\!-\!\v|\right)
 \nn \\
& \times\,
\bar u(1)  \gamma^+ 
\left[\frac{[\gamma^i,\gamma^j]}{4}\, \mathcal{U}^{(3)}_{F; ij} ( \v) 
- i\, \mathcal{U}^{(2)}_F ( \v)\,
+ \mathcal{U}^{(1)}_{F;j} ( \v) \,  \bigg(\frac{(\k_2^j\!-\!\k_1^j)}{2}+\frac{i}{2}\, \partial_{\w^j}\bigg)
 \right]
\mathcal{U}_F^{\dag}(\w)\,
v(2) 
\label{S-bef-q_dec_L} 
\end{align}
and Eq.~\eqref{S-bef-qbar_dec_fin_2}
\begin{align}
S^{\rm bef}_{q_1 \bar q_2 \leftarrow \gamma^*_L} \bigg|_{\textrm{dec. on }\bar{q}}
=&\,
2\pi\delta(k_1^+\!+\!k_2^+\!-\!q^+)\:
 \frac{e e_f}{2\pi} \, (-1)Q\, \frac{k_1^+}{(q^+)^2}
\int d^2 \v\, e^{-i\v \cdot\k_1}\int d^2 \w\, e^{-i\w\cdot\k_2}\,
\textrm{K}_0\left(\bar{Q}\, |\w\!-\!\v|\right)
 \nn \\
& \times\,
\bar u(1)  \gamma^+ 
\Bigg[
\mathcal{U}_F(\v)
\left(\frac{[\gamma^i,\gamma^j]}{4}\,  \mathcal{U}^{(3)\dag}_{F; ij} ( \w) 
\!-\!i\, \mathcal{U}^{(2)\dag}_F ( \w) 
\!+\!\bigg(\frac{i}{2}\, \overleftarrow{\partial_{\v^j}} -\frac{(\k_2^j\!-\!\k_1^j)}{2} \bigg)\mathcal{U}^{(1)\dag}_{F;j} ( \w)
\right)
\Bigg]
v(2) 
\label{S-bef-qbar_dec_L}
\, ,
\end{align}
where $\bar Q$ is defined as\footnote{Note that $\hat{Q}$, as defined in Eq.~\eqref{def_Qhat}, collapses to $\bar{Q}$ if $k_1^+\!+\!k_2^+=q^+$, which is the case in most terms, apart from the generalized eikonal contribution \eqref{S-bef-GenEik_L}. Still, we keep a separate notation for $\bar{Q}$, since that quantity is the one commonly used in the literature about dipole factorization for DIS processes in the eikonal limit. }
\ba
\label{Q-bar}
\bar{Q} \equiv \sqrt{m^2 + Q^2 \frac{k_1^+k_2^+}{(q^+)^2}}
\, .
\ea
Finally, integrating by part in $k^+$, Eq.~\eqref{S-bef-dyn_fin} leads to
\begin{align}
S^{\rm bef}_{q_1 \bar q_2 \leftarrow \gamma^*_L} \bigg|_{\textrm{dyn. target}}
= &\,
2\pi\delta(k_1^+\!+\!k_2^+\!-\!q^+)\;
 i Q\, e e_f \,  \bar u(1) \gamma^+ v(2) 
 \int d^2 \v\, e^{-i\v \cdot\k_1}\int d^2 \w\, e^{-i\w\cdot\k_2}\,
\bigg[\mathcal{U}_F\Big(\v,b^- \Big)\overleftrightarrow{\partial_{b^-}}
\mathcal{U}_F^{\dag}\Big(\w,b^- \Big)
\bigg]\bigg|_{b^-=0}
 \nn \\
&
\times
\int \frac{d^3\uk}{(2\pi)^3}  \, 
2\pi\delta\!(k^+\!-\!k_1^+)
e^{i\k \cdot(\v-\w)}\:
(-\partial_{k^+})\left\{
 \theta(k^+) \, \theta(q^+\!-\!k^+) \frac{\left(\frac{k^+(q^+-k^+)}{(q^+)^2}\right)
 }{\left[\k^2 +m^2
 +\frac{k^+(q^+-k^+)}{(q^+)^2}\, Q^2\right]}
 \right\}\; 
 \nn\\
 = &\,
2\pi\delta(k_1^+\!+\!k_2^+\!-\!q^+)\;
 i Q\, e e_f \,  \bar u(1) \gamma^+ v(2) 
 \int d^2 \v\, e^{-i\v \cdot\k_1}\int d^2 \w\, e^{-i\w\cdot\k_2}\,
\bigg[\mathcal{U}_F\Big(\v,b^- \Big)\overleftrightarrow{\partial_{b^-}}
\mathcal{U}_F^{\dag}\Big(\w,b^- \Big)
\bigg]\bigg|_{b^-=0}
 \nn \\
&
\times\,
\frac{(k_1^+\!-\!k_2^+)}{(q^+)^2}\,
\int \frac{d^2\k}{(2\pi)^2}  \, e^{i\k \cdot(\v-\w)}\:
\frac{\left[\k^2 + m^2\right]}{\left[\k^2 + \bar{Q}^2\right]^2}
\nn\\
 = &\,
2\pi\delta(k_1^+\!+\!k_2^+\!-\!q^+)\;
 i Q\, \frac{e e_f}{2\pi} \,  \bar u(1) \gamma^+ v(2) \,
 \frac{(k_1^+\!-\!k_2^+)}{(q^+)^2}\,
 \int d^2 \v\, e^{-i\v \cdot\k_1}\int d^2 \w\, e^{-i\w\cdot\k_2}\,
 \nn \\
&
\times\,
\left[
\textrm{K}_0\left(\bar{Q}\, |\w\!-\!\v|\right)
-\frac{\left(\bar{Q}^2\!-\!m^2\right)}{2\bar{Q}}\, |\w\!-\!\v|\, 
\textrm{K}_1\left(\bar{Q}\, |\w\!-\!\v|\right)
\right]
\bigg[\mathcal{U}_F\Big(\v,b^- \Big)\overleftrightarrow{\partial_{b^-}}
\mathcal{U}_F^{\dag}\Big(\w,b^- \Big)
\bigg]\bigg|_{b^-=0}
 \, ,
\label{S-bef-dyn_L} 
\end{align}
discarding zero mode contributions at $k_1^+=0$ or $k_2^+=0$, which would not contribute to the cross section of dijet production in the experimentally meaningful range. In the last step of Eq.~\eqref{S-bef-dyn_L}, we have used both the relation~\eqref{int-1-scal} and    
\begin{align}
 \int \frac{d^2\k}{(2\pi)^2}  \frac{e^{-i \k \cdot \r}}{(\k^2+\Delta)^2} 
 =&\, 
 \frac{1}{4\pi} \frac{|\r|}{\sqrt{\Delta}} K_{1} (\sqrt{\Delta}|\r|) 
 \label{int-2-scal}
 \, .
\end{align}

In summary, in the longitudinal photon polarization case, the S-matrix element at NEik accuracy is given by 
\begin{align}
 S_{q_1 \bar q_2 \leftarrow \gamma^*_L}= &\,
S^{\rm bef}_{q_1 \bar q_2 \leftarrow \gamma^*_L} \bigg|_{\textrm{Gen. Eik}}+
S^{\rm bef}_{q_1 \bar q_2 \leftarrow \gamma^*_L}\bigg|_{\textrm{dec. on }q} +
S^{\rm bef}_{q_1 \bar q_2 \leftarrow \gamma^*_L} \bigg|_{\textrm{dec. on }\bar{q}}
+
S^{\rm bef}_{q_1 \bar q_2 \leftarrow \gamma^*_L} \bigg|_{\textrm{dyn. target}}
\end{align}
where explicit expressions for each contribution are given in Eqs. ~\eqref{S-bef-GenEik_L},~\eqref{S-bef-q_dec_L},~\eqref{S-bef-qbar_dec_L} and~\eqref{S-bef-dyn_L}. We would like to remind that the contribution ~\eqref{S-bef-Lplus_L} vanishes, as well as the the contribution from photon splitting inside the target \eqref{S-mat-in_L}. 

By contrast, the strict Eikonal approximation for the S-matrix element can be obtained from the Generalized Eikonal contribution~\eqref{S-bef-GenEik_L} by neglecting the $b^-$ dependence of the Wilson lines. In such a way, one recovers the standard result
\begin{align}
S^{\rm bef}_{q_1 \bar q_2 \leftarrow \gamma^*_L} \bigg|_{\textrm{Strict Eik}}
 =&\, 2\pi\delta(k_1^+\!+\!k_2^+\!-\!q^+)\;
 (-2)Q\,  \frac{e e_f}{2\pi} \, \bar u(1) \gamma^+ v(2)\, 
 \frac{k_1^+k_2^+}{(q^+)^2}\,
 \int d^2 \v\, e^{-i\v \cdot\k_1}
  \nn \\
&
\times
\int d^2 \w\, e^{-i\w\cdot\k_2}\,
\textrm{K}_0\left(\bar{Q}\, |\w\!-\!\v|\right)
\Big[\mathcal{U}_F(\v )
\mathcal{U}_F^{\dag}(\w )
-1\Big]
\label{S-bef-StrictEik_L} 
\, .
\end{align}


\subsection{Before contribution to the S-matrix: transverse photon polarization\label{sec:bef_T_pol}}

Let us now consider the case of transverse photon polarization, with polarization vectors as given in \eqref{T_pol_vect}. The part of the Dirac structure associated with the photon splitting before the target, is then
\begin{align}
 \gamma^+  (\slashed{\check{k}}\!+\!m)\slashed{\epsilon}_{\lambda}(q)
(\slashed{\check{k}}\!-\!\slashed{q}\!+\!m)
\gamma^+ 
=&\, 
 \gamma^+  \Big[k^+\gamma^- \!-\!\k^j\gamma^j \!+\!m\Big] 
 \left[ -\varepsilon_{\lambda}^i \gamma^i +\varepsilon_{\lambda}^i \frac{\q^i}{q^+} \gamma^+\right] 
 \Big[(k^+\!-\!q^+)\gamma^- \!-\!(\k^l\!-\!\q^l)\gamma^l \!+\!m\Big]\gamma^+
 \nn\\
 =&\, 
 \gamma^+  \Big[-\k^j\gamma^j \!+\!m\Big] 
 \left[ -\varepsilon_{\lambda}^i \gamma^i \right] 
(k^+\!-\!q^+)\gamma^- \gamma^+
+
 \gamma^+  k^+\gamma^-  
 \left[ -\varepsilon_{\lambda}^i \gamma^i \right] 
 \Big[-(\k^l\!-\!\q^l)\gamma^l \!+\!m\Big]\gamma^+
 \nn\\
 &\, 
 +
  \gamma^+  k^+\gamma^-  
 \left[ \varepsilon_{\lambda}^i \frac{\q^i}{q^+} \gamma^+\right] 
(k^+\!-\!q^+)\gamma^- \gamma^+
 \nn\\
 =&\, 
\varepsilon_{\lambda}^i  \gamma^+\bigg\{
  2(q^+\!-\!k^+)\Big[-\k^j\gamma^j \!+\!m\Big]\gamma^i
  +2k^+\gamma^i\Big[(\k^j\!-\!\q^j)\gamma^j \!+\!m\Big]
  -\frac{4k^+(q^+\!-\!k^+)}{q^+}\, \q^i
\bigg\}
\nn\\
 =&\, 
 2q^+
 \varepsilon_{\lambda}^i  \gamma^+\bigg\{
 \left[\frac{(q^+\!-\!2k^+)}{q^+}\, \delta^{ij}+\frac{[\gamma^i,\gamma^j]}{2}\right]
 \left[\k^j\!-\!\frac{k^+}{q^+}\,\q^j\right]
 +m\, \gamma^i
 \bigg\}
\label{intern_Dirac_struc_T}
\, .
\end{align}
Inserting the result~\eqref{intern_Dirac_struc_T} into the expression Eq.~\eqref{S-bef-GenEik_fin} for the generalized eikonal contribution, and taking $\q=0$, one finds
\begin{align}
S^{\rm bef}_{q_1 \bar q_2 \leftarrow \gamma^*_T} \bigg|_{\textrm{Gen. Eik}}
=&\,  e e_f \, 
\int d^2 \v\, e^{-i\v \cdot\k_1}\int d^2 \w\, e^{-i\w\cdot\k_2}
 \int db^-\, e^{ib^-(k_1^++k_2^+-q^+)}\;
\bigg[\mathcal{U}_F\Big(\v,b^- \Big)
\mathcal{U}_F^{\dag}\Big(\w,b^- \Big)
-1\bigg]
 \nn \\
&
\times
\frac{1}{2q^+}\int \frac{d^3\uk}{(2\pi)^3}  \, \theta(k^+) \, \theta(q^+\!-\!k^+) 
2\pi\delta\left(k^+\!-\!\frac{1}{2}(k_1^+\!-\!k_2^+\!+\!q^+)\right)
 \frac{
e^{-i\k \cdot(\w-\v)}
 }{\left[\k^2 +m^2
 +\frac{k^+(q^+-k^+)}{(q^+)^2}\, Q^2\right]}\:
  \,   
 \nn \\
& \times
2q^+
 \varepsilon_{\lambda}^i \,
   \bar u(1)\gamma^+\bigg\{
 \left[\frac{(q^+\!-\!2k^+)}{q^+}\, \delta^{ij}+\frac{[\gamma^i,\gamma^j]}{2}\right]
 \k^j
 +m\, \gamma^i
 \bigg\}
 v(2) 
 \;
 \nn\\ 
 =&\, 
 \frac{e e_f}{2\pi} \, 
  \varepsilon_{\lambda}^i \,
 \theta(q^+\!+\!k_1^+\!-\!k_2^+)\,
 \theta(q^+\!+\!k_2^+\!-\!k_1^+)\,
 \int d^2 \v\, e^{-i\v \cdot\k_1}\int d^2 \w\, e^{-i\w\cdot\k_2}\,
  \nn \\
&
\times
\bigg\{
-i\, \frac{(\w^j\!-\!\v^j)}{|\w\!-\!\v|}\, \hat{Q}\, \textrm{K}_1\left(\hat{Q}\, |\w\!-\!\v|\right)\, 
 \bar u(1) \gamma^+ \! \left[\frac{(k_2^+\!-\!k_1^+)}{q^+}\, \delta^{ij}
 +\frac{[\gamma^i,\gamma^j]}{2}\right] v(2)\, 
 \nn \\
&\,
+\textrm{K}_0\left(\hat{Q}\, |\w\!-\!\v|\right)\, 
m\, \bar u(1) \gamma^+  \gamma^i v(2)\, 
\bigg\}
\int db^-\, e^{ib^-(k_1^++k_2^+-q^+)}\;
\bigg[\mathcal{U}_F\Big(\v,b^- \Big)
\mathcal{U}_F^{\dag}\Big(\w,b^- \Big)
-1\bigg]
\label{S-bef-GenEik_T} 
\, ,
\end{align}
using the identities~\eqref{int-1-scal} and
\begin{align}
\int \frac{d^2\k}{(2\pi)^2}  \frac{e^{-i \k \cdot \r}}{(\k^2+\Delta)} \, \k^j
=&\,
 \frac{(-i)}{2\pi}\, \frac{\r^j}{|\r|}\, \sqrt{\Delta}\; \textrm{K}_1(\sqrt{\Delta}|\r|)
\label{int-1-vect}
\, .
\end{align}
In the same way,  in the transverse photon case and for $\q=0$, the contribution~\eqref{S-bef-q_dec_fin_2} 
becomes 
\begin{align}
&
S^{\rm bef}_{q_1 \bar q_2 \leftarrow \gamma^*_T}\bigg|_{\textrm{dec. on }q} 
= 
2\pi\delta(k_1^+\!+\!k_2^+\!-\!q^+)\: 
 e e_f \,  \varepsilon_{\lambda}^i \, \frac{1}{2 k_1^+}
\int d^2 \v\, e^{-i\v \cdot\k_1}\int d^2 \w\, e^{-i\w\cdot\k_2}
\int \frac{d^2\k}{(2\pi)^2}  \, 
\frac{e^{-i\k \cdot(\w-\v)} 
 }{\left[\k^2 +\bar{Q}^2\right]}\,  
 \bar u(1) \gamma^+
 \nn \\
& \times
\bigg[
\frac{[\gamma^l,\gamma^m]}{4}\,  \mathcal{U}^{(3)}_{F; lm} ( \v) 
- i\, \mathcal{U}^{(2)}_F ( \v)\, 
+ \mathcal{U}^{(1)}_{F;l} ( \v)\,\bigg(\frac{(\k_2^l\!-\!\k_1^l)}{2}+\frac{i}{2}\, \partial_{\w^l}\bigg) 
\bigg]  
\mathcal{U}_F^{\dag}(\w)
\bigg\{
 \left[\frac{(k_2^+\!-\!k_1^+)}{q^+}\, \delta^{ij}+\frac{[\gamma^i,\gamma^j]}{2}\right]
 \k^j
 +m\, \gamma^i
 \bigg\}  v(2)\;
 \;
 \nn\\
 =& \,
 2\pi\delta(k_1^+\!+\!k_2^+\!-\!q^+)\: 
 \frac{e e_f}{2\pi} \,  \varepsilon_{\lambda}^i \,
 \frac{1}{2 k_1^+}
\int d^2 \v\, e^{-i\v \cdot\k_1}\int d^2 \w\, e^{-i\w\cdot\k_2}\,
\bar u(1) \gamma^+   \, 
 \nn \\
& \times
\bigg[
\frac{[\gamma^l,\gamma^m]}{4}\,  \mathcal{U}^{(3)}_{F; lm} ( \v) 
- i\, \mathcal{U}^{(2)}_F ( \v)\, 
+ \mathcal{U}^{(1)}_{F;l} ( \v)\,\bigg(\frac{(\k_2^l\!-\!\k_1^l)}{2}+\frac{i}{2}\, \partial_{\w^l}\bigg) 
\bigg]  
\mathcal{U}_F^{\dag}(\w)
\nn \\
& \times
\bigg\{
-i\, \frac{(\w^j\!-\!\v^j)}{|\w\!-\!\v|}\, \bar{Q}\, \textrm{K}_1\left(\bar{Q}\, |\w\!-\!\v|\right)\, 
 \left[\frac{(k_2^+\!-\!k_1^+)}{q^+}\, \delta^{ij}+\frac{[\gamma^i,\gamma^j]}{2}\right]
 +\textrm{K}_0\left(\bar{Q}\, |\w\!-\!\v|\right)\; m\, \gamma^i
 \bigg\}  v(2)\;
\label{S-bef-q_dec_T} 
\end{align}
and the contribution~\eqref{S-bef-qbar_dec_fin_2} 
becomes 
\begin{align}
&S^{\rm bef}_{q_1 \bar q_2 \leftarrow \gamma^*_T} \bigg|_{\textrm{dec. on }\bar{q}}
=
2\pi\delta(k_1^+\!+\!k_2^+\!-\!q^+)\:
 e e_f \, \varepsilon_{\lambda}^i \, \frac{1}{2 k_2^+}
\int d^2 \v\, e^{-i\v \cdot\k_1}\int d^2 \w\, e^{-i\w\cdot\k_2}\,
\int \frac{d^2\k}{(2\pi)^2}  \, 
\frac{e^{-i\k \cdot(\w-\v)} 
 }{\left[\k^2 +\bar{Q}^2\right]}\;
\bar u(1)
 \gamma^+
 \nn \\
&
\times
\bigg\{
 \left[\frac{(k_2^+\!-\!k_1^+)}{q^+}\, \delta^{ij}+\frac{[\gamma^i,\gamma^j]}{2}\right]
 \k^j
 +m\, \gamma^i
 \bigg\}
\bigg[\mathcal{U}_F(\v)
\left(\frac{[\gamma^l,\gamma^m]}{4}\, \mathcal{U}^{(3)\dag}_{F; lm} ( \w)
\!-\!i\, \mathcal{U}^{(2)\dag}_F ( \w) 
\!+\!\bigg( \frac{i}{2} \overleftarrow{\partial_{\v^l}}-\frac{(\k_2^l\!-\!\k_1^l)}{2} \bigg)\mathcal{U}^{(1)\dag}_{F;l} ( \w)
\right)
\bigg]
 v(2) 
 \nn\\
 &=\, 
 2\pi\delta(k_1^+\!+\!k_2^+\!-\!q^+)\:
 \frac{e e_f}{2\pi} \, \varepsilon_{\lambda}^i \, 
  \frac{1}{2 k_2^+}
\int d^2 \v\, e^{-i\v \cdot\k_1}\int d^2 \w\, e^{-i\w\cdot\k_2}\,
 \nn \\
&
\times\;
\bar u(1)
 \gamma^+
\bigg\{
-i\, \frac{(\w^j\!-\!\v^j)}{|\w\!-\!\v|}\, \bar{Q}\, \textrm{K}_1\left(\bar{Q}\, |\w\!-\!\v|\right)\, 
 \left[\frac{(k_2^+\!-\!k_1^+)}{q^+}\, \delta^{ij}+\frac{[\gamma^i,\gamma^j]}{2}\right]
 +\textrm{K}_0\left(\bar{Q}\, |\w\!-\!\v|\right)\; m\, \gamma^i
 \bigg\}
 \nn \\
&
\times 
\bigg[\mathcal{U}_F(\v)
\left(\frac{[\gamma^l,\gamma^m]}{4}\, \mathcal{U}^{(3)\dag}_{F; lm} ( \w)
\!-\!i\, \mathcal{U}^{(2)\dag}_F ( \w) 
\!+\!\bigg( \frac{i}{2} \overleftarrow{\partial_{\v^l}}-\frac{(\k_2^l\!-\!\k_1^l)}{2} \bigg)\mathcal{U}^{(1)\dag}_{F;l} ( \w)
\right)
\bigg]
 v(2)
\label{S-bef-qbar_dec_T}
\, . 
\end{align}
As a reminder, in Eq.~\eqref{S-bef-q_dec_T} and \eqref{S-bef-qbar_dec_T}, the covariant and the ordinary derivatives only act within the square bracket, on the Wilson lines.

Using the expression~\eqref{intern_Dirac_struc_T}, the contribution~\eqref{S-bef-Lplus_fin} can be simplified for transverse photon and $\q=0$, as  
\begin{align}
S^{\rm bef}_{q_1 \bar q_2 \leftarrow \gamma^*_T} \bigg|_{L^+\textrm{ phase}}
= &\,
2\pi\delta(k_1^+\!+\!k_2^+\!-\!q^+)\:  e e_f \, \varepsilon_{\lambda}^i \,
\int d^2 \v\, e^{-i\v \cdot\k_1}\int d^2 \w\, e^{-i\w\cdot\k_2}\,
\Big[\mathcal{U}_F(\v)
\mathcal{U}_F^{\dag}(\w)
-1\Big]
\,   
 \nn \\
& \times\,
 \frac{(-i)L^+q^+}{4k_1^+k_2^+}
\int \frac{d^2\k}{(2\pi)^2}  \, 
e^{-i\k \cdot(\w-\v)}\:
 \bar u(1) \gamma^+  
\bigg\{
 \left[\frac{(k_2^+\!-\!k_1^+)}{q^+}\, \delta^{ij}+\frac{[\gamma^i,\gamma^j]}{2}\right]
 \k^j
 +m\, \gamma^i
 \bigg\} 
v(2) 
 \nn \\
 =&\, 
 2\pi\delta(k_1^+\!+\!k_2^+\!-\!q^+)\:  e e_f \, \varepsilon_{\lambda}^i \,
\int d^2 \v\, e^{-i\v \cdot\k_1}\int d^2 \w\, e^{-i\w\cdot\k_2}\,
\Big[\mathcal{U}_F(\v)
\mathcal{U}_F^{\dag}(\w)
-1\Big]
\,   
 \nn \\
& \times\,
 \frac{(-i)L^+q^+}{4k_1^+k_2^+}
\int \frac{d^2\k}{(2\pi)^2}  \:
 \bar u(1) \gamma^+  
\bigg\{
 \left[\frac{(k_2^+\!-\!k_1^+)}{q^+}\, \delta^{ij}+\frac{[\gamma^i,\gamma^j]}{2}\right]
 \frac{i}{2}\big(\overrightarrow{\partial_{\w^j}}-\overrightarrow{\partial_{\v^j}}\big)
 +m\, \gamma^i
 \bigg\} 
v(2) \: 
e^{-i\k \cdot(\w-\v)}
\nn \\
 = &\, 
 2\pi\delta(k_1^+\!+\!k_2^+\!-\!q^+)\:  e e_f \, \varepsilon_{\lambda}^i \,
\int d^2 \v\, e^{-i\v \cdot\k_1}\int d^2 \w\, e^{-i\w\cdot\k_2}\,
 \frac{(-i)L^+q^+}{4k_1^+k_2^+}\;
\delta^{(2)}(\w\!-\!\v)\:
\,   
 \nn \\
& \hspace{-1.5cm}
\times\,
 \bar u(1) \gamma^+  
\bigg\{
 \left[\frac{(k_2^+\!-\!k_1^+)}{q^+}\, \delta^{ij}+\frac{[\gamma^i,\gamma^j]}{2}\right]
 \bigg(-\frac{i}{2}\big(\overrightarrow{\partial_{\w^j}}-\overrightarrow{\partial_{\v^j}}\big)
 -\frac{(\k_2^j\!-\!\k_1^j)}{2}\bigg)
 +m\, \gamma^i
 \bigg\} 
v(2) \: 
\Big[\mathcal{U}_F(\v)
\mathcal{U}_F^{\dag}(\w)
-1\Big]
\nn \\
 =  &\, 
 2\pi\delta(k_1^+\!+\!k_2^+\!-\!q^+)\:  e e_f \, \varepsilon_{\lambda}^i \,
  \frac{(-1)L^+q^+}{8k_1^+k_2^+}\;
  \bar u(1) \gamma^+\!
  \left[\frac{(k_2^+\!-\!k_1^+)}{q^+}\, \delta^{ij}+\frac{[\gamma^i,\gamma^j]}{2}\right]  
  v(2) \: 
 \nn \\
& \times\,
 \int d^2 \v\, e^{-i\v \cdot(\k_1+\k_2)}\:
\Big[\mathcal{U}_F(\v)\overleftrightarrow{\partial_{\v^j}}
\mathcal{U}_F^{\dag}(\v)
\Big]
\, ,
\label{S-bef-Lplus_T} 
\end{align}
where, in the final expression, the partial derivative acts only within the square bracket, on the Wilson lines.

Using Eq.~\eqref{intern_Dirac_struc_T} and taking $\q=0$, the contribution \eqref{S-bef-dyn_fin} associated with the dynamics of the target can be evaluated as
\begin{align}
S^{\rm bef}_{q_1 \bar q_2 \leftarrow \gamma^*_T} \bigg|_{\textrm{dyn. target}}
= &\,
2\pi\delta(k_1^+\!+\!k_2^+\!-\!q^+)\;
 e e_f \, \varepsilon_{\lambda}^i \,
\int d^2 \v\, e^{-i\v \cdot\k_1}\int d^2 \w\, e^{-i\w\cdot\k_2}\,
\bigg[\mathcal{U}_F\Big(\v,b^- \Big)\overleftrightarrow{\partial_{b^-}}
\mathcal{U}_F^{\dag}\Big(\w,b^- \Big)
\bigg]\bigg|_{b^-=0}
 \nn \\
&
\times
\frac{(-i)}{2}\int \frac{d^3\uk}{(2\pi)^3}  \, \theta(k^+) \, \theta(q^+\!-\!k^+) 
2\pi\delta'\!(k^+\!-\!k_1^+)\;
\frac{e^{-i\k \cdot(\w-\v)} }{\left[\k^2 +m^2
 +\frac{k^+(q^+-k^+)}{(q^+)^2}\, Q^2\right]} \:
 \,   
 \nn \\
& \times
 \bar u(1) \gamma^+  \bigg\{
 \left[\frac{(q^+\!-\!2k^+)}{q^+}\, \delta^{ij}+\frac{[\gamma^i,\gamma^j]}{2}\right]
 \k^j
 +m\, \gamma^i
 \bigg\}
 v(2) 
 \, 
\nn \\
= &\,
2\pi\delta(k_1^+\!+\!k_2^+\!-\!q^+)\;
 e e_f \, \varepsilon_{\lambda}^i \,
\int d^2 \v\, e^{-i\v \cdot\k_1}\int d^2 \w\, e^{-i\w\cdot\k_2}\,
\bigg[\mathcal{U}_F\Big(\v,b^- \Big)\overleftrightarrow{\partial_{b^-}}
\mathcal{U}_F^{\dag}\Big(\w,b^- \Big)
\bigg]\bigg|_{b^-=0}
 \nn \\
&
\times
\frac{(-i)}{2}\int \frac{d^2\k}{(2\pi)^2}\, e^{-i\k \cdot(\w-\v)}\,
(-1)\partial_{k^+} 
\Bigg\{
\frac{\theta(k^+) \, \theta(q^+\!-\!k^+)}{\left[\k^2 +m^2
 +\frac{k^+(q^+-k^+)}{(q^+)^2}\, Q^2\right]} \:
 \nn\\
&
\times
\bar u(1) \gamma^+  \bigg(
 \left[\frac{(q^+\!-\!2k^+)}{q^+}\, \delta^{ij}+\frac{[\gamma^i,\gamma^j]}{2}\right]
 \k^j
 +m\, \gamma^i
 \bigg)
v(2)
\Bigg\}\Bigg|_{k^+=k_1^+}
\nn \\
= &\,
2\pi\delta(k_1^+\!+\!k_2^+\!-\!q^+)\;
 \frac{e e_f}{2\pi}\, \, \varepsilon_{\lambda}^i \,
\int d^2 \v\, e^{-i\v \cdot\k_1}\int d^2 \w\, e^{-i\w\cdot\k_2}\,
\bigg[\mathcal{U}_F\Big(\v,b^- \Big)\overleftrightarrow{\partial_{b^-}}
\mathcal{U}_F^{\dag}\Big(\w,b^- \Big)
\bigg]\bigg|_{b^-=0}\;  
 \nn \\
&
\times
\bar u(1) \gamma^+\Bigg\{ 
- \frac{(\w^i\!-\!\v^i)\, \bar{Q}}{|\w\!-\!\v|\, q^+}\, \textrm{K}_1\left(\bar{Q}\, |\w\!-\!\v|\right)\,
-\frac{i(k_2^+\!-\!k_1^+)\, Q^2\, |\w\!-\!\v|}{4(q^+)^2\, \bar{Q}}\: 
 \textrm{K}_1\left(\bar{Q}\, |\w\!-\!\v|\right)\, m\, \gamma^i
 \nn\\
&\;\;\;\;
-\frac{(k_2^+\!-\!k_1^+)\, Q^2}{4(q^+)^2}\, (\w^j\!-\!\v^j) \textrm{K}_0\left(\bar{Q}\, |\w\!-\!\v|\right)  \left[\frac{(k_2^+\!-\!k_1^+)}{q^+}\, \delta^{ij}+\frac{[\gamma^i,\gamma^j]}{2}\right]
\Bigg\}v(2)
\label{S-bef-dyn_T} 
\,  ,
\end{align}
thanks to the identities~\eqref{int-2-scal}, \eqref{int-1-vect}
and 
\begin{align}
\int \frac{d^2\k}{(2\pi)^2}  \frac{e^{-i \k \cdot \r}}{(\k^2+\Delta)^2} \, \k^j
=&\,
 \frac{(-i)}{4\pi}\, \r^j \; \textrm{K}_0(\sqrt{\Delta}|\r|)
\label{int-2-vect}
\, .
\end{align}
In Eq.~\eqref{S-bef-dyn_T} we have once again discarded the terms obtained by acting with $\partial_{k^+}$ on $\theta(k^+)$ or $\theta(q^+\!-\!k^+)$, which are zero modes $k_1^+=0$ or $k_2^+=0$, irrelevant for  DIS dijet production. 

All in all, at NEik accuracy, the S-matrix element for $q\bar{q}$ production from a transverse photon is given by 
\begin{align}
 S_{q_1 \bar q_2 \leftarrow \gamma^*_T}= &\,
S^{\rm bef}_{q_1 \bar q_2 \leftarrow \gamma^*_T} \bigg|_{\textrm{Gen. Eik}}+
S^{\rm bef}_{q_1 \bar q_2 \leftarrow \gamma^*_T}\bigg|_{\textrm{dec. on }q} +
S^{\rm bef}_{q_1 \bar q_2 \leftarrow \gamma^*_T} \bigg|_{\textrm{dec. on }\bar{q}}\nonumber\\
&+
S^{\rm bef}_{q_1 \bar q_2 \leftarrow \gamma^*_T} \bigg|_{L^+\textrm{ phase}}
+
S^{\rm bef}_{q_1 \bar q_2 \leftarrow \gamma^*_T} \bigg|_{\textrm{dyn. target}}+
S^{\rm in}_{q_1 \bar q_2 \leftarrow \gamma^*_T} 
\end{align}
where explicit expressions for each contribution coming from the photon splitting before reaching the target are given in Eqs.  \eqref{S-bef-GenEik_T}, \eqref{S-bef-q_dec_T}, \eqref{S-bef-qbar_dec_T}, \eqref{S-bef-Lplus_T}, and \eqref{S-bef-dyn_T}. The expression for the contribution coming from the photon splitting inside the target is given in Eq. \eqref{S-mat-in_T}.


Again, the strict Eikonal approximation for the S-matrix element is obtained from the Generalized Eikonal contribution~\eqref{S-bef-GenEik_T} by neglecting the $b^-$ dependence of the Wilson lines. In such a way, one recovers the standard result
\begin{align}
& S^{\rm bef}_{q_1 \bar q_2 \leftarrow \gamma^*_T} \bigg|_{\textrm{Strict Eik}}
 = 
 2\pi\delta(k_1^+\!+\!k_2^+\!-\!q^+)\;
 \frac{e e_f}{2\pi} \, 
  \varepsilon_{\lambda}^i \,
 \int d^2 \v\, e^{-i\v \cdot\k_1}\int d^2 \w\, e^{-i\w\cdot\k_2}\,
 \Big[\mathcal{U}_F(\v )
\mathcal{U}_F^{\dag}(\w )
-1\Big]
  \nn \\
&
\times
\bigg\{
-i\, \frac{(\w^j\!-\!\v^j)}{|\w\!-\!\v|}\, \bar{Q}\, \textrm{K}_1\left(\bar{Q}\, |\w\!-\!\v|\right)\, 
 \bar u(1) \gamma^+ \! \left[\frac{(k_2^+\!-\!k_1^+)}{q^+}\, \delta^{ij}
 +\frac{[\gamma^i,\gamma^j]}{2}\right] v(2)\, 
+\textrm{K}_0\left(\bar{Q}\, |\w\!-\!\v|\right)\, 
m\, \bar u(1) \gamma^+  \gamma^i v(2)\, 
\bigg\}
\label{S-bef-StrictEik_T} 
\, .
\end{align}


\section{NEik DIS dijet production cross section via longitudinal photon \label{sec:gammaL_cross_sec}}


\subsection{Relations between S-matrix, amplitude and cross section beyond Eikonal accuracy\label{sec:rel_Smat_ampl_cross_sec}}

In standard CGC framework, where the observables are computed in the eikonal approximation, the background field does not depend on $x^-$. Because of that, no light cone "$+$" momentum can be exchanged with the target.
 In this case, the scattering amplitude can be defined as 
\begin{align}
S_{q_1 \bar q_2 \leftarrow \gamma^*_L} \bigg|_{x^-\textrm{ indep.}}
=& \,   (2q^+)\,2\pi\, \delta(k_1^+\!+\!k_2^+\!-\!q^+)\, i{\cal M}_{q_1 \bar q_2 \leftarrow \gamma^*_L}
\, ,\label{ampl_L_def}
\end{align}
 by factorizing out the obtained Dirac delta function
. 
Then, as discussed in appendix \ref{sec:xmin_indep_cross_sec}, the cross section is obtained by squaring this amplitude and including the correct prefactor, as    
\begin{align}
\frac{d\sigma_{\gamma^*_L \rightarrow q_1 \bar q_2}}{d \textrm{P.S.}}  \bigg|_{x^-\textrm{ indep.}}
=&
\, (2q^+)\, 2\pi\, \delta(k_1^+\!+\!k_2^+\!-\!q^+) \,
\sum_{\rm hel. \,,\,   col. } |{\cal M}_{q_1 \bar q_2 \leftarrow \gamma^*_L}|^2 
\, ,
\label{xsec_L_def}
\end{align}
where the two-particle phase space is defined as
\begin{align}
d {\rm P.S.}= \frac{d^2 \k_1}{(2\pi)^2} \frac{d k^+_1}{2k_1^+(2\pi)}\,  \frac{d^2 \k_2}{(2\pi)^2} \frac{d k^+_2}{2k_2^+(2\pi)} 
\label{2part_phase_space}
\, ,
\end{align} 
and the summation in Eq.~\eqref{xsec_L_def} is over the colors and light-front helicities of the produced quark and anti-quark.

When performing a  strict expansion  of the S-matrix element into Eikonal contribution, NEik contribution and so on, the  gradient expansion of the background field with respect to $x^-$ would be performed entirely. In that case, not all the terms would be of the form~\eqref{ampl_L_def}: some NEik corrections would include $\delta'(k_1^+\!+\!k_2^+\!-\!q^+)$ instead. It is not possible to calculate the contribution of such terms to the cross section without introducing wave-packets, which is a major inconvenience. This is the motivation which has led us to introduce the Generalized Eikonal approximation, in which the dependence of the Wilson lines on a common $x^-$ is kept at the S-matrix level.

Then, one should use the result derived in Appendix \ref{App:X_section_x_minus} (for the scattering amplitude see Eq. \eqref{ampl_2} and for the cross section see Eq. \eqref{xsec_9}) in the case of a  $x^-$ dependent background field. We would like to mention that, at the accuracy considered in the present study, this modified procedure is necessary only to compute the contribution of the  squared Generalized Eikonal amplitude to the cross section. For the rest of the contributions that are explicitly NEik order, the $x^-$ dependence of the background field can be neglected, and one can go back to the standard procedure from eikonal CGC to obtain the cross section. 

Following this argument, the cross section at NEik accuracy for DIS dijet for longitudinal photon can be written as 
%
\begin{align}
 \frac{d\sigma_{\gamma^{*}_L\rightarrow q_1\bar q_2}}{d {\rm P.S.}} =  \frac{d\sigma_{\gamma^{*}_L\rightarrow q_1\bar q_2}}{d {\rm P.S.}}\Bigg|_{\rm Gen. \, Eik}+  \frac{d\sigma_{\gamma^{*}_L\rightarrow q_1\bar q_2}}{d {\rm P.S.}}\Bigg|_{\rm NEik \, corr.} + O({\rm NNEik})
 \label{def_L_cross_sec}
 \, .
\end{align}
%

The Generalized Eikonal contribution to the cross section is then given by 
%
\begin{align}
\label{Cross_Section_GEik}
\frac{d\sigma_{\gamma^{*}_L\rightarrow q_1\bar q_2}}{d {\rm P.S.}}\Bigg|_{\rm Gen. \, Eik}
= 2q^+ \int d (\Delta b^-) e^{i\Delta b^-(k_1^++k_2^+ - q^+)} 
\sum_{\rm hel. \,,\,   col. }
\Big\langle  \Big(\mathbf{M}_{q_1 \bar q_2 \leftarrow \gamma^*_L}^{\rm Gen.\,  Eik} \Big(- \frac{\Delta b^-}{2}\Big)\Big)^\dag \mathbf{M}_{q_1 \bar q_2 \leftarrow \gamma^*_L}^{\rm Gen.\,  Eik} \Big(\frac{\Delta b^-}{2} \Big) \Big\rangle
\end{align}
%
where $\langle \cdots\rangle$ stands for averaging over the background field of the target. The S-matrix and the $b^-$-dependent scattering amplitude $\mathbf{M}_{q_1 \bar q_2 \leftarrow \gamma^*_L}^{\rm Gen.\,  Eik}$ are related via 
\begin{align}
\label{M_long_GEik_amp}
&
S_{q_1 \bar q_2 \leftarrow \gamma^*_L} \bigg|_{\rm Gen.\, Eik }= 
2q^+\int db^- e^{ib^-(k_1^++k_2^+-q^+)} i \mathbf{M}_{q_1 \bar q_2 \leftarrow \gamma^*_L}^{\rm Gen.\,  Eik} (b^-)
\, .
\end{align}

The explicit NEik correction to the cross section in Eq.~\eqref{def_L_cross_sec} corresponds to the interference between the Generalized Eikonal contribution \eqref{S-bef-GenEik_L}  and the NEik corrections (\eqref{S-bef-q_dec_L}, \eqref{S-bef-qbar_dec_L} and \eqref{S-bef-dyn_L} in the longitudinal photon case). In that case, and at the NEik accuracy, the Generalized Eikonal contribution \eqref{S-bef-GenEik_L} can be replaced by the strict Eikonal contribution \eqref{S-bef-StrictEik_L}. Then, the $x^-$-dependence of the background field can be dropped, and
  relations of the type   \eqref{ampl_L_def} and \eqref{xsec_L_def} can be used, leading to
\begin{align}
 \frac{d\sigma_{\gamma^{*}_L\rightarrow q_1\bar q_2}}{d {\rm P.S.}}\Bigg|_{\rm NEik \, corr.}
&
= (2q^+)\, 2\pi \delta(k_1^+\!+\!k_2^+\!-\!q^+) \sum_{\rm hel. \,,\,  col. }
\Big[ \Big\langle  \Big({\cal M}_{q_1 \bar q_2 \leftarrow \gamma^*_L}^{\rm strict~Eik} \Big)^\dag  
{\cal M }_{q_1 \bar q_2 \leftarrow \gamma^*_L}^{\rm NEik \, corr.} \Big\rangle
+ \Big \langle \Big({\cal M}_{q_1 \bar q_2 \leftarrow \gamma^*_L}^{\rm NEik \, corr.} \Big)^\dag {\cal M}_{q_1 \bar q_2 \leftarrow \gamma^*_L}^{\rm strict~Eik} \Big\rangle \Big] \nn \\
&
=  (2q^+)\, 2\pi \delta(k_1^+\!+\!k_2^+\!-\!q^+) \sum_{\rm hel. \,,\,  col. }
2 {\rm Re}  \Big\langle  \Big({\cal M}_{q_1 \bar q_2 \leftarrow \gamma^*_L}^{\rm strict~Eik}\Big)^\dag  {\cal M}_{q_1 \bar q_2 \leftarrow \gamma^*_L}^{\rm NEik \, corr.} \Big\rangle
\, .
\label{Cross_Section_long_NEik}
\end{align}


\subsection{Generalized Eikonal contribution to the cross section for longitudinal photon}

By comparing Eqs.~\eqref{S-bef-GenEik_L}  and \eqref{M_long_GEik_amp}, one can read off the $b^-$-dependent amplitude
\begin{align}
i \mathbf{M}_{q_1 \bar q_2 \leftarrow \gamma^*_L}^{\rm Gen.\,  Eik} (b^-)
=&\, 
 -Q\,  \frac{e e_f}{2\pi} \, \bar u(1) \gamma^+ v(2)\, 
 \frac{(q^+\!+\!k_1^+\!-\!k_2^+)(q^+\!+\!k_2^+\!-\!k_1^+)}{4(q^+)^3}\,
 \theta(q^+\!+\!k_1^+\!-\!k_2^+)\,
 \theta(q^+\!+\!k_2^+\!-\!k_1^+)\,
  \nn \\
&
\times
\int d^2 \v\, e^{-i\v \cdot\k_1}
\int d^2 \w\, e^{-i\w\cdot\k_2}\,
\textrm{K}_0\left(\hat{Q}\, |\w\!-\!\v|\right)
\Big[\mathcal{U}_F(\v,b^- )
\mathcal{U}_F^{\dag}(\w,b^- )
\!-\!1\Big]
\label{bdep_Ampl-GenEik_L} 
\, .
\end{align}
Inserting Eq.~\eqref{bdep_Ampl-GenEik_L} into Eq.~\eqref{Cross_Section_GEik}, one arrives at\footnote{Here we introduce a compact notation for the transverse coordinate integrals $\int_{\x}\cdots \equiv \int d^2\x\cdots$ \, .}
\begin{align}
\frac{d\sigma_{\gamma^{*}_L\rightarrow q_1\bar q_2}}{d {\rm P.S.}}\Bigg|_{\rm Gen. \, Eik}
&= 2q^+ \int d (\Delta b^-) e^{i\Delta b^-(k_1^++k_2^+ - q^+)} 
 \Bigg(\frac{e e_f \, Q}{2\pi} \Bigg)^2 
\theta\Big(q^+\!+\!k_1^+\!-\!k_2^+\Big) \theta\Big(q^+\!-\!k_1^+\!+\!k_2^+\Big) 
\frac{k_1^+k_2^+}{2(q^+)^6} 
\nn \\
&
\times
(q^+\!+\!k_1^+\!-\!k_2^+)^2 (q^+\!-\!k_1^+\!+\!k_2^+)^2 
\int_{\v, \v', \w, \w'}
e^{i \k_1 \cdot (\v'-\v) } \, e^{i\k_2 \cdot (\w'-\w) } 
K_0(\hat{Q}|\w'-\v'|) K_0(\hat{Q}|\w-\v|)
 \nn \\
&
\times
\Bigg\langle
{\rm Tr} \Bigg[
\Big( \mathcal{U}_F \Big(\w',-\frac{\Delta b^-}{2}\Big) 
\mathcal{U}_F^\dag \Big(\v', -\frac{\Delta b^-}{2}\Big) - \mathbf{1} \Big) 
\Big( \mathcal{U}_F \Big(\v, \frac{\Delta b^-}{2}\Big) 
\mathcal{U}_F^\dag \Big(\w,  \frac{\Delta b^-}{2}\Big) - \mathbf{1} \Big) \Bigg] \Bigg\rangle
\label{Cross_Section_GEik_L_2}
\, .
\end{align} 
The required Dirac algebra has been performed in appendix, see Eq.~\eqref{Dirac_trace_1L_res}.  
The color operator in Eq.~\eqref{Cross_Section_GEik_L_2} can be split as
\begin{align}
&
\Bigg\langle
{\rm Tr} \Bigg[
\Big( \mathcal{U}_F \Big(\w',-\frac{\Delta b^-}{2}\Big) 
\mathcal{U}_F^\dag \Big(\v', -\frac{\Delta b^-}{2}\Big) - \mathbf{1} \Big) 
\Big( \mathcal{U}_F \Big(\v, \frac{\Delta b^-}{2}\Big) 
\mathcal{U}_F^\dag \Big(\w,  \frac{\Delta b^-}{2}\Big) - \mathbf{1} \Big) \Bigg] \Bigg\rangle
\nn\\
=&\,
\Bigg\langle
{\rm Tr} \Bigg[
 \mathcal{U}_F \Big(\w',-\frac{\Delta b^-}{2}\Big) 
\mathcal{U}_F^\dag \Big(\v', -\frac{\Delta b^-}{2}\Big) 
\mathcal{U}_F \Big(\v, \frac{\Delta b^-}{2}\Big) 
\mathcal{U}_F^\dag \Big(\w,  \frac{\Delta b^-}{2}\Big) \Bigg] \Bigg\rangle
\nn\\
&\,
-\Bigg\langle
{\rm Tr} \Bigg[
 \mathcal{U}_F \Big(\w',-\frac{\Delta b^-}{2}\Big) 
\mathcal{U}_F^\dag \Big(\v', -\frac{\Delta b^-}{2}\Big) 
 \Bigg] \Bigg\rangle
 -
\Bigg\langle
{\rm Tr} \Bigg[
 \mathcal{U}_F \Big(\v, \frac{\Delta b^-}{2}\Big) 
\mathcal{U}_F^\dag \Big(\w,  \frac{\Delta b^-}{2}\Big)  \Bigg] \Bigg\rangle
+1
\, .
\label{GEik_color_op_expand}
\end{align} 
In the two dipole terms in the expression \eqref{GEik_color_op_expand}, the dependence on $\Delta b^-$ can be removed by using the invariance under translations along $x^-$ which is restored by the target average. Then, using the definitions 
\ba
\label{dipole}
d\big(\v,\w\big) &=& \Big\langle \frac{1}{N_c} {\rm Tr} 
\big[  \mathcal{U}_F(\v) \mathcal{U}_F^\dag ( \w ) \big] \Big\rangle \\
\label{quadrupole}
Q\Big(\w',\v',\v,\w,\frac{\Delta b^-}{2}\Big) &=& \Big\langle \frac{1}{N_c} {\rm Tr} 
\Big[  \mathcal{U}_F \Big( \w',\frac{\Delta b^-}{2}\Big) \mathcal{U}_F^\dag \Big( \v',\frac{\Delta b^-}{2}\Big)  
\mathcal{U}_F \Big( \v,-\frac{\Delta b^-}{2}\Big) \mathcal{U}_F^\dag \Big( \w,-\frac{\Delta b^-}{2}\Big) \Big] \Big\rangle
\, ,
\ea
one can rewrite the expression \eqref{Cross_Section_GEik_L_2} as
\begin{align}
\frac{d\sigma_{\gamma^{*}_L\rightarrow q_1\bar q_2}}{d {\rm P.S.}}\Bigg|_{\rm Gen. \, Eik}
&= N_c\, \frac{\alpha_{\textrm{em}} }{\pi}\, e_f^2\, Q^2\, 
\theta\Big(q^+\!+\!k_1^+\!-\!k_2^+\Big) \theta\Big(q^+\!-\!k_1^+\!+\!k_2^+\Big) 
\frac{k_1^+k_2^+}{(q^+)^5}\;
(q^+\!+\!k_1^+\!-\!k_2^+)^2 (q^+\!-\!k_1^+\!+\!k_2^+)^2  
\nn \\
&
\times
\int_{\v, \v', \w, \w'}
e^{i \k_1 \cdot (\v'-\v) } \, e^{i\k_2 \cdot (\w'-\w) } 
K_0(\hat{Q}|\w'-\v'|) K_0(\hat{Q}|\w-\v|)
 \nn \\
&
\times
\int d (\Delta b^-) e^{i\Delta b^-(k_1^++k_2^+ - q^+)}
\bigg\{
Q\Big(\w',\v',\v,\w,\frac{\Delta b^-}{2}\Big)
-d\big(\w',\v'\big)-d\big(\v,\w\big)+1
\bigg\}
\, 
\label{Cross_Section_GEik_L_3}
\end{align} 
with $\alpha_{\rm em}=e^2/(4\pi)$. 


\subsection{Explicit NEik correction to the cross section for longitudinal photon}

In order to obtain the NEik correction to the cross section from Eq.~\eqref{Cross_Section_long_NEik}, we first need the strict Eikonal amplitude, and the full NEik correction to the the amplitude beyond the generalized Eikonal contribution. They are related to the corresponding contribution to the S-matrix elements as in Eq.~\eqref{ampl_L_def}. From Eq.~\eqref{S-bef-StrictEik_L}, we then find the strict Eikonal amplitude as  
\begin{align}
i {\cal M}_{q_1 \bar q_2 \leftarrow \gamma^*_L}^{\rm strict~Eik}
 =&\, 
 -Q\,  \frac{e e_f}{2\pi} \, \bar u(1) \gamma^+ v(2)\, 
 \frac{k_1^+k_2^+}{(q^+)^3}\,
 \int d^2 \v\, e^{-i\v \cdot\k_1}
\int d^2 \w\, e^{-i\w\cdot\k_2}\,
\textrm{K}_0\left(\bar{Q}\, |\w\!-\!\v|\right)
\Big[\mathcal{U}_F(\v )
\mathcal{U}_F^{\dag}(\w )
-1\Big]
\label{Ampl-StrictEik_L} 
\, .
\end{align}
The NEik correction to the amplitude can be written as
\begin{align}
i {\cal M }_{q_1 \bar q_2 \leftarrow \gamma^*_L}^{\rm NEik \, corr.}= 
i{\cal M }_{q_1 \bar q_2 \leftarrow \gamma^*_L}^{\textrm{dec. on }q}
 +
i{\cal M }_{q_1 \bar q_2 \leftarrow \gamma^*_L}^{\textrm{dec. on }\bar{q}}
+
i{\cal M }_{q_1 \bar q_2 \leftarrow \gamma^*_L}^{\textrm{dyn. target}}
\, ,
\label{NEik_corr_Ampl_L}
\end{align}   
with the three terms obtained from Eqs.~\eqref{S-bef-q_dec_L}, \eqref{S-bef-qbar_dec_L} and \eqref{S-bef-dyn_L} as
\begin{align}
i{\cal M }_{q_1 \bar q_2 \leftarrow \gamma^*_L}^{\textrm{dec. on }q}
=&\, 
 -Q\,\frac{e e_f}{2\pi} \,  \frac{k_2^+}{2(q^+)^3}
\int d^2 \v\, e^{-i\v \cdot\k_1}\int d^2 \w\, e^{-i\w\cdot\k_2}\,
\textrm{K}_0\left(\bar{Q}\, |\w\!-\!\v|\right)
 \nn \\
& \times
\bar u(1)  \gamma^+ 
\left[\frac{[\gamma^i,\gamma^j]}{4}\, \mathcal{U}^{(3)}_{F; ij} ( \v) 
- i\, \mathcal{U}^{(2)}_F ( \v)\,
+ \mathcal{U}^{(1)}_{F;j} ( \v) \,  \bigg(\frac{(\k_2^j\!-\!\k_1^j)}{2}+\frac{i}{2}\, \partial_{\w^j}\bigg)
 \right]
\mathcal{U}_F^{\dag}(\w)\,
v(2) 
\label{ampl-q_dec_L} 
\, ,
\end{align}

\begin{align}
i{\cal M }_{q_1 \bar q_2 \leftarrow \gamma^*_L}^{\textrm{dec. on }\bar{q}}
=&\,
 -Q\,\frac{e e_f}{2\pi} \,  \frac{k_1^+}{2(q^+)^3}
\int d^2 \v\, e^{-i\v \cdot\k_1}\int d^2 \w\, e^{-i\w\cdot\k_2}\,
\textrm{K}_0\left(\bar{Q}\, |\w\!-\!\v|\right)
 \nn \\
& \times
\bar u(1)  \gamma^+ 
\Bigg[
\mathcal{U}_F(\v)
\left(\frac{[\gamma^i,\gamma^j]}{4}\,  \mathcal{U}^{(3)\dag}_{F; ij} ( \w) 
\!-\!i\, \mathcal{U}^{(2)\dag}_F ( \w) 
\!+\!\bigg(\frac{i}{2}\, \overleftarrow{\partial_{\v^j}} -\frac{(\k_2^j\!-\!\k_1^j)}{2} \bigg)\mathcal{U}^{(1)\dag}_{F;j} ( \w)
\right)
\Bigg]
v(2) 
\label{ampl-qbar_dec_L}
\, ,
\end{align}
and

\begin{align}
i{\cal M }_{q_1 \bar q_2 \leftarrow \gamma^*_L}^{\textrm{dyn. target}}
 = &\,
 i Q\, \frac{e e_f}{2\pi} \,  \bar u(1) \gamma^+ v(2) \,
 \frac{(k_1^+\!-\!k_2^+)}{2(q^+)^3}\,
 \int d^2 \v\, e^{-i\v \cdot\k_1}\int d^2 \w\, e^{-i\w\cdot\k_2}\,
 \nn \\
&
\times\,
\left[
\textrm{K}_0\left(\bar{Q}\, |\w\!-\!\v|\right)
-\frac{\left(\bar{Q}^2\!-\!m^2\right)}{2\bar{Q}}\, |\w\!-\!\v|\, 
\textrm{K}_1\left(\bar{Q}\, |\w\!-\!\v|\right)
\right]
\bigg[\mathcal{U}_F\Big(\v,b^- \Big)\overleftrightarrow{\partial_{b^-}}
\mathcal{U}_F^{\dag}\Big(\w,b^- \Big)
\bigg]\bigg|_{b^-=0}
 \, .
\label{ampl-dyn_L} 
\end{align}

In the NEik corrections to the amplitude \eqref{ampl-q_dec_L}, \eqref{ampl-qbar_dec_L} and \eqref{ampl-dyn_L}, all the terms apart from the ones with a $\mathcal{F}_{ij}$ decoration have the same Dirac structure $\bar u(1) \gamma^+ v(2)$ as the Eikonal contribution \eqref{Ampl-StrictEik_L}. 
Hence, when calculating the overlap of these NEik corrections with the strict Eikonal amplitude in Eq.~\eqref{Cross_Section_long_NEik}, the same structure as in Eqs.~\eqref{Dirac_trace_1L_def}-\eqref{Dirac_trace_1L_res}  will be encountered. 
On the other hand, the terms with a $\mathcal{F}_{ij}$ decoration (within $\mathcal{U}^{(3)\dag}_{F; ij}$) will lead at the cross section level \eqref{Cross_Section_long_NEik} to a different Dirac structure, \eqref{Dirac_trace_2L_def}, which is shown to vanish in appendix, see Eq.~\eqref{Dirac_trace_2L_res}.
Hence, the terms with a $\mathcal{F}_{ij}$ decoration do not contribute to the cross section at NEik accuracy in the longitudinal photon case.

Then, the contribution of eq.~\eqref{ampl-q_dec_L}  at cross section level is
\begin{align}
 \frac{d\sigma_{\gamma^{*}_L\rightarrow q_1\bar q_2}}{d {\rm P.S.}}\Bigg|_{\rm NEik \, corr.}^{\textrm{dec. on }q}
&
=  (2q^+)\, 2\pi \delta(k_1^+\!+\!k_2^+\!-\!q^+) 
\;   8 k_1^+ k_2^+
 Q^2\,  \left(\frac{e e_f}{2\pi}\right)^2 \,  
 \frac{k_1^+k_2^+}{(q^+)^3}\,
  \frac{k_2^+}{2(q^+)^3}
  \nn \\
&
\times
  2 {\rm Re} \int_{\v, \v', \w, \w'}
e^{i \k_1 \cdot (\v'-\v) } \, e^{i\k_2 \cdot (\w'-\w) } 
\textrm{K}_0\left(\bar{Q}\, |\w'\!-\!\v'|\right)
\textrm{K}_0\left(\bar{Q}\, |\w\!-\!\v|\right)
 \nn \\
&
\times
 {\rm Tr} \bigg\langle
\Big[\mathcal{U}_F(\w' )\mathcal{U}_F^{\dag}(\v' )-1\Big]
\bigg[
- i\, \mathcal{U}^{(2)}_F ( \v)\,
+ \mathcal{U}^{(1)}_{F;j} ( \v) \,  \bigg(\frac{(\k_2^j\!-\!\k_1^j)}{2}+\frac{i}{2}\, \partial_{\w^j}\bigg)
 \bigg]
\mathcal{U}_F^{\dag}(\w)
\bigg\rangle
\, .
\label{Cross_Section_long_NEik_dec_on_q}
\end{align}
Similarly, the contribution from \eqref{ampl-qbar_dec_L} at cross section level is 
\begin{align}
 \frac{d\sigma_{\gamma^{*}_L\rightarrow q_1\bar q_2}}{d {\rm P.S.}}\Bigg|_{\rm NEik \, corr.}^{\textrm{dec. on }\bar{q}}
&
=   (2q^+)\, 2\pi \delta(k_1^+\!+\!k_2^+\!-\!q^+) 
\;   8 k_1^+ k_2^+
 Q^2\,  \left(\frac{e e_f}{2\pi}\right)^2 \,  
 \frac{k_1^+k_2^+}{(q^+)^3}\,
  \frac{k_1^+}{2(q^+)^3}
 \nn \\
 &
 \times
2 {\rm Re} \int_{\v, \v', \w, \w'}
e^{i \k_1 \cdot (\v'-\v) } \, e^{i\k_2 \cdot (\w'-\w) } 
\textrm{K}_0\left(\bar{Q}\, |\w'\!-\!\v'|\right)
\textrm{K}_0\left(\bar{Q}\, |\w\!-\!\v|\right)
\nn \\
 &
 \times
 {\rm Tr} \bigg\langle
\Big[\mathcal{U}_F(\w' )\mathcal{U}_F^{\dag}(\v' )-1\Big]
\bigg[
\mathcal{U}_F(\v)
\bigg(
\!-\!i\, \mathcal{U}^{(2)\dag}_F ( \w) 
\!+\!\bigg(\frac{i}{2}\, \overleftarrow{\partial_{\v^j}} -\frac{(\k_2^j\!-\!\k_1^j)}{2} \bigg)\mathcal{U}^{(1)\dag}_{F;j} ( \w)
\bigg)
\bigg]
\bigg\rangle
\, .
\label{Cross_Section_long_NEik_dec_on_qbar}
\end{align}
Finally, the contribution from \eqref{ampl-dyn_L} at cross section level is
\begin{align}
 \frac{d\sigma_{\gamma^{*}_L\rightarrow q_1\bar q_2}}{d {\rm P.S.}}\Bigg|_{\rm NEik \, corr.}^{\textrm{dyn. target}}
&
=  (2q^+)\, 2\pi \delta(k_1^+\!+\!k_2^+\!-\!q^+) 
\;   8 k_1^+ k_2^+
 Q^2\,  \left(\frac{e e_f}{2\pi}\right)^2 \,  
 \frac{k_1^+k_2^+}{(q^+)^3}\,
 \frac{(k_1^+\!-\!k_2^+)}{2(q^+)^3}\;
 2 {\rm Re} \; ( -i)\,
 \nn \\
 &
\times  
 \int_{\v, \v', \w, \w'}
e^{i \k_1 \cdot (\v'-\v) } \, e^{i\k_2 \cdot (\w'-\w) }\,
 \left[
\textrm{K}_0\left(\bar{Q}\, |\w\!-\!\v|\right)
-\frac{\left(\bar{Q}^2\!-\!m^2\right)}{2\bar{Q}}\, |\w\!-\!\v|\, 
\textrm{K}_1\left(\bar{Q}\, |\w\!-\!\v|\right)
\right]
 \nn \\
 &
\times 
\textrm{K}_0\left(\bar{Q}\, |\w'\!-\!\v'|\right) \;
 {\rm Tr} \Bigg\langle
\Big[\mathcal{U}_F(\w' )
\mathcal{U}_F^{\dag}(\v' )
-1\Big]
\,
\bigg[\mathcal{U}_F\Big(\v,b^- \Big)\overleftrightarrow{\partial_{b^-}}
\mathcal{U}_F^{\dag}\Big(\w,b^- \Big)
\bigg]\bigg|_{b^-=0}
 \Bigg\rangle
\, .
\label{Cross_Section_long_NEik_dyn}
\end{align}

In order to obtain more compact expressions, let us introduce the following notations for
 decorated dipoles and quadrupoles operators
\ba
\label{dipole_1}
d^{(1)}_{j}(\v_*,\w) &=& \Bigg\langle \frac{1}{N_c} {\rm Tr} 
\Big[  \mathcal{U}^{(1)}_{F;j} ( \v)  \mathcal{U}_F^\dag ( \w) \Big] \Bigg\rangle \\
\label{dipole_2}
d^{(2)}(\v_*,\w) &=& \Bigg\langle \frac{1}{N_c} {\rm Tr} 
\Big[  \mathcal{U}_F^{(2)} ( \v) \mathcal{U}_F^\dag ( \w) \Big] \Bigg\rangle \\
\label{quadrupole_1}
Q^{(1)}_{j}(\w',\v',\v_*,\w) &=& \Bigg\langle \frac{1}{N_c} {\rm Tr} 
\Big[  \mathcal{U}_F ( \w') \mathcal{U}_F^\dag ( \v')  \mathcal{U}^{(1)}_{F;j} ( \v)  \mathcal{U}_F^\dag ( \w) \Big] \Bigg\rangle \\
\label{quadrupole_2}
Q^{(2)}(\w',\v',\v_*,\w) &=& \Bigg\langle \frac{1}{N_c} {\rm Tr} 
\Big[  \mathcal{U}_F( \w') \mathcal{U}_F^\dag ( \v')  \mathcal{U}_F^{(2)} ( \v) \mathcal{U}_F^\dag ( \w) \Big] \Bigg\rangle 
\, ,
\ea
with the star indicating the position of the decoration.
Moreover, for the case of $\partial_-$ acting on Wilson lines, we define the following decorated dipoles and quadrupoles operators
\ba
\label{dipole_dec}
\tilde d(\v_*,\w_*) &=& \Bigg\langle \frac{1}{N_c} {\rm Tr}
\Bigg[ \Big( \mathcal{U}_F ( \v,b^-) \overleftrightarrow{\partial_-} 
\mathcal{U}_F^\dag ( \w,b^-) \Big) \Big|_{b^-=0}\Bigg] \Bigg\rangle, \\
\label{quadrupole_dec}
\tilde Q(\w',\v',\v_*,\w_*) &=& \Bigg\langle  \frac{1}{N_c} {\rm Tr} 
\Bigg[  \mathcal{U}_F ( \w') \mathcal{U}_F^\dag ( \v') 
\Big(\mathcal{U}_F ( \v,b^-) \overleftrightarrow{\partial_-}  \mathcal{U}_F^\dag ( \w,b^-) \Big) \Big|_{b^-=0}  \Bigg] \Bigg\rangle
\, . 
\ea 

With these definitions, the expression \eqref{Cross_Section_long_NEik_dec_on_q} can be simplified into
\begin{align}
 \frac{d\sigma_{\gamma^{*}_L\rightarrow q_1\bar q_2}}{d {\rm P.S.}}\Bigg|_{\rm NEik \, corr.}^{\textrm{dec. on }q}
&
=   2\pi \delta(k_1^+\!+\!k_2^+\!-\!q^+) 
\;   8  N_c\,  \frac{\alpha_{\textrm{em}}}{\pi} \,e_f^2\,
 Q^2\,   
 \frac{(k_1^+)^2 (k_2^+)^3}{(q^+)^5}\,
  \nn \\
&
\times
  2 {\rm Re} \int_{\v, \v', \w, \w'}
e^{i \k_1 \cdot (\v'-\v) } \, e^{i\k_2 \cdot (\w'-\w) } 
\textrm{K}_0\left(\bar{Q}\, |\w'\!-\!\v'|\right)
\textrm{K}_0\left(\bar{Q}\, |\w\!-\!\v|\right)
 \nn \\
&
\hspace{-3cm}\times
 \Bigg\{
 \Big[\frac{ (\k_2^j\!-\!\k_1^j)}{2}\,
+\frac{i}{2}\,  \partial_{\w^j} \Big]
 \Big[Q^{(1)}_j(\w',\v',\v_*,\w)-d^{(1)}_j(\v_*,\w)\Big]
 -i \Big[Q^{(2)}(\w',\v',\v_*,\w)-d^{(2)}(\v_*,\w)\Big]
  \Bigg\}
\label{Cross_Section_long_NEik_dec_on_q_2}
\end{align}
and the expression \eqref{Cross_Section_long_NEik_dec_on_qbar} into
\begin{align}
 \frac{d\sigma_{\gamma^{*}_L\rightarrow q_1\bar q_2}}{d {\rm P.S.}}\Bigg|_{\rm NEik \, corr.}^{\textrm{dec. on }\bar{q}}
&
=   2\pi \delta(k_1^+\!+\!k_2^+\!-\!q^+) 
\;   8  N_c\,  \frac{\alpha_{\textrm{em}}}{\pi} \,e_f^2\,
 Q^2\,   
 \frac{(k_1^+)^3 (k_2^+)^2}{(q^+)^5}\,
 \nn \\
 &
 \times
2 {\rm Re} \int_{\v, \v', \w, \w'}
e^{i \k_1 \cdot (\v'-\v) } \, e^{i\k_2 \cdot (\w'-\w) } 
\textrm{K}_0\left(\bar{Q}\, |\w'\!-\!\v'|\right)
\textrm{K}_0\left(\bar{Q}\, |\w\!-\!\v|\right)
\nn \\
&
\hspace{-3cm}\times
 \Bigg\{
 \Big[-\frac{(\k_2^j\!-\!\k_1^j)}{2} 
 +\frac{i}{2}\,\partial_{\v^j}
 \Big]
 \Big[{Q^{(1)}_j(\v',\w',\w_*,\v)}^{\dag}-{d^{(1)}_j(\w_*,\v)}^{\dag} \Big]
 -i  \Big[{Q^{(2)}(\v',\w',\w_*,\v)}^{\dag}-{d^{(2)}(\w_*,\v)}^{\dag}\Big]
   \Bigg\}
\, .
\label{Cross_Section_long_NEik_dec_on_qbar_2}
\end{align}
Finally, the expression \eqref{Cross_Section_long_NEik_dyn} can be simplified into
\begin{align}
 \frac{d\sigma_{\gamma^{*}_L\rightarrow q_1\bar q_2}}{d {\rm P.S.}}\Bigg|_{\rm NEik \, corr.}^{\textrm{dyn. target}}
&
=  2\pi \delta(k_1^+\!+\!k_2^+\!-\!q^+) 
\;   8  N_c\,  \frac{\alpha_{\textrm{em}}}{\pi} \,e_f^2\,
 Q^2\,   
 \frac{(k_1^+)^2 (k_2^+)^2(k_1^+\!-\!k_2^+)}{(q^+)^5}\,
  2 {\rm Re} \, ( -i)\!\!
\int_{\v, \v', \w, \w'}\!\!\!\!\!\!
e^{i \k_1 \cdot (\v'-\v) } \, e^{i\k_2 \cdot (\w'-\w) }\,  
\nn \\
 &
 \hspace{-2.5cm}
\times   
\bigg[\tilde Q(\w',\v',\v_*,\w_*)-\tilde d(\v_*,\w_*) \bigg]
\,    
\textrm{K}_0\left(\bar{Q}\, |\w'\!-\!\v'|\right) \;
 \left[
\textrm{K}_0\left(\bar{Q}\, |\w\!-\!\v|\right)
-\frac{\left(\bar{Q}^2\!-\!m^2\right)}{2\bar{Q}}\, |\w\!-\!\v|\, 
\textrm{K}_1\left(\bar{Q}\, |\w\!-\!\v|\right)
\right]
\, .
\label{Cross_Section_long_NEik_dyn_2} 
\end{align}
All in all, the NEik correction to the cross section for longitudinal photon beyond the generalized Eikonal contribution \eqref{Cross_Section_GEik_L_3} is given by the sum of the contributions \eqref{Cross_Section_long_NEik_dec_on_q_2},  \eqref{Cross_Section_long_NEik_dec_on_qbar_2} and \eqref{Cross_Section_long_NEik_dyn_2}.


\section{NEik DIS dijet production cross section via transverse  photon\label{sec:gammaT_cross_sec}}

Let us now consider the case of DIS dijet production initiated by a transverse photon. All the definitions and relations introduced in section \ref{sec:rel_Smat_ampl_cross_sec} in the longitudinal photon case are still valid in the transverse photon case, except that we consider the transverse photon cross section to be averaged over the two transverse polarizations.


\subsection{Generalized Eikonal contribution to the cross section for transverse photon}

In the transverse photon case, the dijet cross section at NEik accuracy can be written as a sum of a generalized eikonal contribution and a extra NEik correction, like in Eq.~\eqref{def_L_cross_sec} for for the longitudinal photon case. The generalized eikonal contribution is obtained as 
\begin{align}
\label{Cross_Section_GEik_T}
\frac{d\sigma_{\gamma^{*}_T\rightarrow q_1\bar q_2}}{d {\rm P.S.}}\Bigg|_{\rm Gen. \, Eik}
= 2q^+ \int d (\Delta b^-) e^{i\Delta b^-(k_1^++k_2^+ - q^+)} 
\frac{1}{2}\sum_{\lambda}\sum_{\rm hel. \,,\,   col. }
\Big\langle  \Big(\mathbf{M}_{q_1 \bar q_2 \leftarrow \gamma^*_T}^{\rm Gen.\,  Eik} \Big(- \frac{\Delta b^-}{2}\Big)\Big)^\dag \mathbf{M}_{q_1 \bar q_2 \leftarrow \gamma^*_T}^{\rm Gen.\,  Eik} \Big(\frac{\Delta b^-}{2} \Big) \Big\rangle
\, ,
\end{align}
%
which includes an averaging over the polarization $\lambda$ of the incoming transverse photon. The $b^-$ dependent amplitude involved in Eq.~\eqref{Cross_Section_GEik_T} is defined in the same way as in Eq.~\eqref{M_long_GEik_amp}, and can be read off from the generalized eikonal expression from the S-matrix \eqref{S-bef-GenEik_T}  in the transverse photon case. In such a way, one finds
\begin{align}
i \mathbf{M}_{q_1 \bar q_2 \leftarrow \gamma^*_T}^{\rm Gen.\,  Eik} (b^-)
 =&\, 
 \frac{e e_f}{2\pi} \, 
  \varepsilon_{\lambda}^i \; 
  \frac{1}{2q^+}\;
 \theta(q^+\!+\!k_1^+\!-\!k_2^+)\,
 \theta(q^+\!+\!k_2^+\!-\!k_1^+)\,
 \int d^2 \v\, e^{-i\v \cdot\k_1}\int d^2 \w\, e^{-i\w\cdot\k_2}\,
  \nn \\
&
\times
\bigg\{
-i\, \frac{(\w^j\!-\!\v^j)}{|\w\!-\!\v|}\, \hat{Q}\, \textrm{K}_1\left(\hat{Q}\, |\w\!-\!\v|\right)\, 
 \bar u(1) \gamma^+ \! \left[\frac{(k_2^+\!-\!k_1^+)}{q^+}\, \delta^{ij}
 +\frac{[\gamma^i,\gamma^j]}{2}\right] v(2)\, 
 \nn \\
&\,
+\textrm{K}_0\left(\hat{Q}\, |\w\!-\!\v|\right)\, 
m\, \bar u(1) \gamma^+  \gamma^i v(2)\, 
\bigg\}
\bigg[\mathcal{U}_F\Big(\v,b^- \Big)
\mathcal{U}_F^{\dag}\Big(\w,b^- \Big)
-1\bigg]
\label{Ampl-GenEik_T} 
\, .
\end{align}
The two Dirac structures in Eq.~\eqref{Ampl-GenEik_T} do not interfere at cross section level, as shown in Appendix~\ref{sec:Dirac_alg} (see Eqs.~\eqref{Dirac_trace_3T_res} and \eqref{Dirac_trace_4T_res}). Their squares, summed over $h_1$ and $h_2$ and averaged over $\lambda$ are obtained in Eqs.~\eqref{Dirac_trace_1T_res} and \eqref{Dirac_trace_2T_res} respectively.  
%
%
Using these results, the generalized eikonal cross section for transverse photon is found to be 
\begin{align}
\frac{d\sigma_{\gamma^{*}_T\rightarrow q_1\bar q_2}}{d {\rm P.S.}}\Bigg|_{\rm Gen. \, Eik}
=&\,
 2q^+ \int d (\Delta b^-) e^{i\Delta b^-(k_1^++k_2^+ - q^+)} 
 \left(\frac{e e_f}{2\pi}\right)^2 \, 
   \frac{1}{(2q^+)^2}\;
 \theta(q^+\!+\!k_1^+\!-\!k_2^+)\,
 \theta(q^+\!+\!k_2^+\!-\!k_1^+)\,
\nn \\
&
\times
\int_{\v, \v', \w, \w'}
e^{i \k_1 \cdot (\v'-\v) } \, e^{i\k_2 \cdot (\w'-\w) } 
\Bigg\{
8 k_1^+ k_2^+\, m^2\, 
\textrm{K}_0\left(\hat{Q}\, |\w'\!-\!\v'|\right)\, \textrm{K}_0\left(\hat{Q}\, |\w\!-\!\v|\right)\,
 \nn \\
&\;\;\;
+
4 k_1^+ k_2^+\, \left[1+\left(\frac{k_2^+\!-\!k_1^+}{q^+}\right)^2\right]\,
\hat{Q}^2\, 
 \frac{(\w'\!-\!\v')\!\cdot\!(\w\!-\!\v)}{|\w'\!-\!\v'||\w\!-\!\v|}\,
\textrm{K}_1\left(\hat{Q}\, |\w'\!-\!\v'|\right)\,\textrm{K}_1\left(\hat{Q}\, |\w\!-\!\v|\right)\,
\Bigg\}
 \nn \\
&
\times
\Bigg\langle
{\rm Tr} \Bigg[
\Big( \mathcal{U}_F \Big(\w',-\frac{\Delta b^-}{2}\Big) 
\mathcal{U}_F^\dag \Big(\v', -\frac{\Delta b^-}{2}\Big) - \mathbf{1} \Big) 
\Big( \mathcal{U}_F \Big(\v, \frac{\Delta b^-}{2}\Big) 
\mathcal{U}_F^\dag \Big(\w,  \frac{\Delta b^-}{2}\Big) - \mathbf{1} \Big) \Bigg] \Bigg\rangle
\, .
\label{Cross_Section_GEik_T_2}
\end{align}
%
Using the decomposition \eqref{GEik_color_op_expand} of the color operator, and the definitions \eqref{dipole} and \eqref{quadrupole}, one gets
\begin{align}
\frac{d\sigma_{\gamma^{*}_T\rightarrow q_1\bar q_2}}{d {\rm P.S.}}\Bigg|_{\rm Gen. \, Eik}
=&\,
N_c\,  \frac{\alpha_{\textrm{em}}}{\pi} \, e_f^2\,
   \frac{2k_1^+ k_2^+}{q^+}\;
 \theta(q^+\!+\!k_1^+\!-\!k_2^+)\,
 \theta(q^+\!+\!k_2^+\!-\!k_1^+)\,
 \int_{\v, \v', \w, \w'}
e^{i \k_1 \cdot (\v'-\v) } \, e^{i\k_2 \cdot (\w'-\w) }
\nn \\
&
\hspace{-3.5cm}\times
\Bigg\{
2\, m^2\, 
\textrm{K}_0\left(\hat{Q}\, |\w'\!-\!\v'|\right)\, \textrm{K}_0\left(\hat{Q}\, |\w\!-\!\v|\right)
+
\left[1+\left(\frac{k_2^+\!-\!k_1^+}{q^+}\right)^2\right]\,
\hat{Q}^2\,  \frac{(\w'\!-\!\v')\!\cdot\!(\w\!-\!\v)}{|\w'\!-\!\v'||\w\!-\!\v|}\, 
\textrm{K}_1\left(\hat{Q}\, |\w'\!-\!\v'|\right)\,\textrm{K}_1\left(\hat{Q}\, |\w\!-\!\v|\right)
\Bigg\}
 \nn \\
&
\times
\int d (\Delta b^-) e^{i\Delta b^-(k_1^++k_2^+ - q^+)} 
\bigg\{
Q\Big(\w',\v',\v,\w,\frac{\Delta b^-}{2}\Big)
-d\big(\w',\v'\big)-d\big(\v,\w\big)+1
\bigg\}
\, .
\label{Cross_Section_GEik_T_3}
\end{align}


\subsection{Explicit NEik correction to the cross section for transverse photon}

The NEik correction to the transverse photon cross section beyond the generalized eikonal contribution is calculated in the same way as in the longitudinal case, up to the averaging over $\lambda$, as
\begin{align}
 \frac{d\sigma_{\gamma^{*}_T\rightarrow q_1\bar q_2}}{d {\rm P.S.}}\Bigg|_{\rm NEik \, corr.}
&
=  (2q^+)\, 2\pi \delta(k_1^+\!+\!k_2^+\!-\!q^+)\, \frac{1}{2}\sum_{\lambda}\, \sum_{\rm hel. \,,\,  col. }
2 {\rm Re}  \Big\langle  \Big({\cal M}_{q_1 \bar q_2 \leftarrow \gamma^*_T}^{\rm strict~Eik}\Big)^\dag  {\cal M}_{q_1 \bar q_2 \leftarrow \gamma^*_T}^{\rm NEik \, corr.} \Big\rangle
\, .
\label{Cross_Section_T_NEik}
\end{align}
 It amounts to calculating the interference between the strict Eikonal amplitude and the NEik correction to the amplitude, normalized as in Eq.~\eqref{ampl_L_def}. From Eq.~\eqref{S-bef-StrictEik_T}, the strict eikonal amplitude for transverse photon is found to be
\begin{align}
& i {\cal M}_{q_1 \bar q_2 \leftarrow \gamma^*_T}^{\rm strict~Eik}
 = 
 \frac{e e_f}{2\pi} \, 
  \varepsilon_{\lambda}^i \, \frac{1}{2q^+}
 \int d^2 \v\, e^{-i\v \cdot\k_1}\int d^2 \w\, e^{-i\w\cdot\k_2}\,
 \Big[\mathcal{U}_F(\v )
\mathcal{U}_F^{\dag}(\w )
-1\Big]
  \nn \\
&
\times
\bigg\{
-i\, \frac{(\w^j\!-\!\v^j)}{|\w\!-\!\v|}\, \bar{Q}\, \textrm{K}_1\left(\bar{Q}\, |\w\!-\!\v|\right)\, 
 \bar u(1) \gamma^+ \! \left[\frac{(k_2^+\!-\!k_1^+)}{q^+}\, \delta^{ij}
 +\frac{[\gamma^i,\gamma^j]}{2}\right] v(2)\, 
+\textrm{K}_0\left(\bar{Q}\, |\w\!-\!\v|\right)\, 
m\, \bar u(1) \gamma^+  \gamma^i v(2)\, 
\bigg\}
\label{Ampl-StrictEik_T} 
\, .
\end{align}
By contrast, the NEik correction to the amplitude is the sum of five contributions 
\begin{align}
i {\cal M }_{q_1 \bar q_2 \leftarrow \gamma^*_T}^{\rm NEik \, corr.}= 
i{\cal M }_{q_1 \bar q_2 \leftarrow \gamma^*_T}^{\textrm{in}}
+
i{\cal M }_{q_1 \bar q_2 \leftarrow \gamma^*_T}^{\textrm{dec. on }q}
 +
i{\cal M }_{q_1 \bar q_2 \leftarrow \gamma^*_T}^{\textrm{dec. on }\bar{q}}
+
i{\cal M }_{q_1 \bar q_2 \leftarrow \gamma^*_T}^{L^+\textrm{ phase}}
+
i{\cal M }_{q_1 \bar q_2 \leftarrow \gamma^*_T}^{\textrm{dyn. target}}
\, ,
\label{NEik_corr_Ampl_T}
\end{align}   
corresponding to the contributions \eqref{S-mat-in_T}, \eqref{S-bef-q_dec_T}, \eqref{S-bef-qbar_dec_T}, \eqref{S-bef-Lplus_T} and \eqref{S-bef-dyn_T} to the S-matrix element respectively. We thus have
\begin{align}
i{\cal M }_{q_1 \bar q_2 \leftarrow \gamma^*_T}^{\textrm{in}} 
=&\,
  e e_f \, \varepsilon_{\lambda}^i\; \frac{1}{8k_1^+k_2^+}\;
\bar u(1) \gamma^+\!\left[\frac{(k_2^+\!-\!k_1^+)}{q^+}\, \delta^{i j}
+\frac{1}{2}\, [ \gamma^i,  \gamma^j] 
\right] v(2)
\nonumber\\
&\, \times \, 
\int d^2\z\,  e^{-i(\k_1+\k_2) \cdot \z}\: \int_{-L^+/2}^{L^+/2} dz^+\, 
\bigg[ \mathcal{U}_F\Big(\frac{L^+}{2},z^+;\z\Big) \overleftrightarrow{ {\cal D}_{\z^j}}  \mathcal{U}_F^{\dagger}\Big(\frac{L^+}{2},z^+;\z\Big)\bigg]
\label{ampl-in_T}
\, ,
\end{align}

\begin{align}
i{\cal M }_{q_1 \bar q_2 \leftarrow \gamma^*_T}^{\textrm{dec. on }q}
 = &\,
 \frac{e e_f}{2\pi} \,  \varepsilon_{\lambda}^i \, \frac{1}{2 q^+}\,\frac{1}{2 k_1^+}\,
\int d^2 \v\, e^{-i\v \cdot\k_1}\int d^2 \w\, e^{-i\w\cdot\k_2}  
 \nn \\
& \times
\bar u(1) \gamma^+   \bigg[
\left(\frac{[\gamma^l,\gamma^m]}{4}\,     \mathcal{U}^{(3)}_{F; lm} ( \v)
- i\,  \mathcal{U}^{(2)}_F ( \v) 
+  \mathcal{U}^{(1)}_{F;l} ( \v)\, \bigg( \frac{(\k_2^l\!-\!\k_1^l)}{2}  +\frac{i}{2}\, \overrightarrow{\partial_{\w^l}}\bigg)
\right)
\mathcal{U}_F^{\dag}(\w)
\bigg]  
\nn \\
& \times
\bigg\{
-i\, \frac{(\w^j\!-\!\v^j)}{|\w\!-\!\v|}\, \bar{Q}\, \textrm{K}_1\left(\bar{Q}\, |\w\!-\!\v|\right)\, 
 \left[\frac{(k_2^+\!-\!k_1^+)}{q^+}\, \delta^{ij}+\frac{[\gamma^i,\gamma^j]}{2}\right]
 +\textrm{K}_0\left(\bar{Q}\, |\w\!-\!\v|\right)\; m\, \gamma^i
 \bigg\}  v(2)
 \, ,
\label{Ampl-q_dec_T} 
\end{align}

\begin{align}
i{\cal M }_{q_1 \bar q_2 \leftarrow \gamma^*_T}^{\textrm{dec. on }\bar{q}}
= &\, 
 \frac{e e_f}{2\pi} \, \varepsilon_{\lambda}^i \, \frac{1}{2 q^+}\, \frac{1}{2 k_2^+}\, 
\int d^2 \v\, e^{-i\v \cdot\k_1}\int d^2 \w\, e^{-i\w\cdot\k_2}\,
 \nn \\
&
\times\;
\bar u(1)
 \gamma^+
\bigg\{
-i\, \frac{(\w^j\!-\!\v^j)}{|\w\!-\!\v|}\, \bar{Q}\, \textrm{K}_1\left(\bar{Q}\, |\w\!-\!\v|\right)\, 
 \left[\frac{(k_2^+\!-\!k_1^+)}{q^+}\, \delta^{ij}+\frac{[\gamma^i,\gamma^j]}{2}\right]
 +\textrm{K}_0\left(\bar{Q}\, |\w\!-\!\v|\right)\; m\, \gamma^i
 \bigg\}
 \nn \\
&
\times 
\bigg[
\mathcal{U}_F(\v)
\left(\frac{[\gamma^l,\gamma^m]}{4} \mathcal{U}^{(3)\dag}_{F; lm}(\w) 
\!-\!i\, \mathcal{U}^{(2)\dag}_F(\w)
+\bigg(\frac{i}{2}\,  \overleftarrow{\partial_{\v^l}}\!-\!\frac{(\k_2^l\!-\!\k_1^l)}{2}\bigg)\, \mathcal{U}^{(1)\dag}_{F;l} ( \w) 
\right)
\bigg]
 v(2) 
\label{Ampl-qbar_dec_T}
\, ,
\end{align}

\begin{align}
i{\cal M }_{q_1 \bar q_2 \leftarrow \gamma^*_T}^{L^+\textrm{ phase}}
 =  &\, 
  e e_f \, \varepsilon_{\lambda}^i \,
  \frac{(-1)L^+}{16k_1^+k_2^+}\;
  \bar u(1) \gamma^+\!
  \left[\frac{(k_2^+\!-\!k_1^+)}{q^+}\, \delta^{ij}+\frac{[\gamma^i,\gamma^j]}{2}\right]  
  v(2) \: 
 \int d^2 \z\, e^{-i\z \cdot(\k_1+\k_2)}\:
\Big[\mathcal{U}_F(\z)\overleftrightarrow{\partial_{\z^j}}
\mathcal{U}_F^{\dag}(\z)
\Big]
\label{Ampl-Lplus_T} 
\end{align}
and
\begin{align}
i{\cal M }_{q_1 \bar q_2 \leftarrow \gamma^*_T}^{\textrm{dyn. target}}
= &\,
 \frac{e e_f}{2\pi}\, \, \varepsilon_{\lambda}^i \, \frac{1}{2(q^+)^2}
\int d^2 \v\, e^{-i\v \cdot\k_1}\int d^2 \w\, e^{-i\w\cdot\k_2}\,
\bigg[\mathcal{U}_F\Big(\v,b^- \Big)\overleftrightarrow{\partial_{b^-}}
\mathcal{U}_F^{\dag}\Big(\w,b^- \Big)
\bigg]\bigg|_{b^-=0}\;  
 \nn \\
&
\times
\bar u(1) \gamma^+\Bigg\{ 
- \frac{(\w^i\!-\!\v^i)\, \bar{Q}}{|\w\!-\!\v|}\, \textrm{K}_1\left(\bar{Q}\, |\w\!-\!\v|\right)\,
-\frac{i(k_2^+\!-\!k_1^+)\, Q^2\, |\w\!-\!\v|}{4q^+\, \bar{Q}}\: 
 \textrm{K}_1\left(\bar{Q}\, |\w\!-\!\v|\right)\, m\, \gamma^i
 \nn\\
&\;\;\;\;
-\frac{(k_2^+\!-\!k_1^+)\, Q^2}{4q^+}\, (\w^j\!-\!\v^j) \textrm{K}_0\left(\bar{Q}\, |\w\!-\!\v|\right)  \left[\frac{(k_2^+\!-\!k_1^+)}{q^+}\, \delta^{ij}+\frac{[\gamma^i,\gamma^j]}{2}\right]
\Bigg\}v(2)
\label{Ampl-dyn_T} 
\,  .
\end{align}

The contributions  \eqref{ampl-in_T} and \eqref{Ampl-Lplus_T} to the amplitude contain the same two Dirac structure as the generalized Eikonal contribution \eqref{Ampl-GenEik_T}. Their contributions at cross section level can then be obtained thanks the relations \eqref{Dirac_trace_1T_res}, \eqref{Dirac_trace_2T_res}, \eqref{Dirac_trace_3T_res} and \eqref{Dirac_trace_4T_res}, as
\begin{align}
& \frac{d\sigma_{\gamma^{*}_T\rightarrow q_1\bar q_2}}{d {\rm P.S.}}\Bigg|_{\rm NEik \, corr.}^{\textrm{in}}
=\,
  2\pi \delta(k_1^+\!+\!k_2^+\!-\!q^+)\,  
N_c\, \alpha_{\textrm{em}}\,    e_f^2 \,  
  \left[1+\left(\frac{k_2^+\!-\!k_1^+}{q^+}\right)^2\right]\, 
2 {\rm Re}\;  (i)
 \int_{\z, \v', \w'}
e^{i \k_1 \cdot (\v'-\z) } \, e^{i\k_2 \cdot (\w'-\z) }\,
 \nn \\
&
\times
\frac{({\w'}^j\!-\!{\v'}^j)}{|\w'\!-\!\v'|}\, \bar{Q}\, \textrm{K}_1\left(\bar{Q}\, |\w'\!-\!\v'|\right)\, 
 \int_{-L^+/2}^{L^+/2} dz^+\, 
\Bigg\langle
\frac{1}{N_c}\, {\rm Tr} 
 \Big[\mathcal{U}_F(\w' )
\mathcal{U}_F^{\dag}(\v' )
-1\Big]
\bigg[ \mathcal{U}_F\Big(\frac{L^+}{2},z^+;\z\Big) \overleftrightarrow{ {\cal D}_{\z^j}}  \mathcal{U}_F^{\dagger}\Big(\frac{L^+}{2},z^+;\z\Big)\bigg]
 \Bigg\rangle
\, .
\label{Cross_Section_in_T}
\end{align}
and
\begin{align}
 \frac{d\sigma_{\gamma^{*}_T\rightarrow q_1\bar q_2}}{d {\rm P.S.}}\Bigg|_{\rm NEik \, corr.}^{L^+\textrm{ phase}}
=&\,
   2\pi \delta(k_1^+\!+\!k_2^+\!-\!q^+)\, N_c\, \alpha_{\textrm{em}}\,  
  e_f^2 \, 
 \left[1+\left(\frac{k_2^+\!-\!k_1^+}{q^+}\right)^2\right]\, 
2 {\rm Re}\;  (-i)\, \frac{L^+}{2}\;\int_{\z, \v', \w'}
e^{i \k_1 \cdot (\v'-\z) } \, e^{i\k_2 \cdot (\w'-\z) }\,
 \nn \\
&
\times
\frac{({\w'}^j\!-\!{\v'}^j)}{|\w'\!-\!\v'|}\, \bar{Q}\, \textrm{K}_1\left(\bar{Q}\, |\w'\!-\!\v'|\right)\, 
\Bigg\langle
\frac{1}{N_c}\, 
{\rm Tr} 
 \Big[\mathcal{U}_F(\w' )
\mathcal{U}_F^{\dag}(\v' )
-1\Big]
\Big[\mathcal{U}_F(\z)\overleftrightarrow{\partial_{\z^j}}
\mathcal{U}_F^{\dag}(\z)
\Big]
 \Bigg\rangle
\, .
\label{Cross_Section_Lplus_T}
\end{align}

The Dirac algebra for the contribution of \eqref{Ampl-q_dec_T} at cross section level can be performed as well using the relations \eqref{Dirac_trace_1T_res}, \eqref{Dirac_trace_2T_res}, \eqref{Dirac_trace_3T_res} and \eqref{Dirac_trace_4T_res}, except for the terms involving the longitudinal chromomagnetic background field $\mathcal{F}_{lm}$, which require instead the relations
\eqref{Dirac_trace_2primeT_calc_1}, \eqref{Dirac_trace_3primeT_calc_1}, \eqref{Dirac_trace_4primeT_calc_1}, \eqref{Dirac_trace_1primeT_calc_1}.
Then, one finds
\begin{align}
 &
 \frac{d\sigma_{\gamma^{*}_T\rightarrow q_1\bar q_2}}{d {\rm P.S.}}\Bigg|_{\rm NEik \, corr.}^{\textrm{dec. on }q}
= 
 2\pi \delta(k_1^+\!+\!k_2^+\!-\!q^+)\,  N_c\, \frac{\alpha_{\textrm{em}}}{\pi} \, e_f^2\,  
 \frac{2 k_2^+}{ q^+}\,
2 {\rm Re}\;   \int_{\v, \v', \w, \w'}
e^{i \k_1 \cdot (\v'-\v) } \, e^{i\k_2 \cdot (\w'-\w) }\;
 \nn \\
&\,
\times
\Bigg\{
\left[
 \left(\frac{ (\k_2^j\!-\!\k_1^j)}{2}\,
+\frac{i}{2}\,  \partial_{\w^j} \right)\!
\left(Q^{(1)}_j(\w',\v',\v_*,\w)\!-\!d^{(1)}_j(\v_*,\w)\right)
 -i \left(Q^{(2)}(\w',\v',\v_*,\w)\!-\!d^{(2)}(\v_*,\w)\right)
  \right]
\nn \\
&\,
\times
\left[
\frac{1}{2} \left(1+\left(\frac{k_2^+\!-\!k_1^+}{q^+}\right)^2\right)
 \frac{({\w'}\!-\!{\v'})\!\cdot\!(\w\!-\!\v)}{|\w'\!-\!\v'||\w\!-\!\v|}\, \bar{Q}^2\, \textrm{K}_1\left(\bar{Q}\, |\w'\!-\!\v'|\right)
 \textrm{K}_1\left(\bar{Q}\, |\w\!-\!\v|\right)\, 
+ m^2\, \textrm{K}_0\left(\bar{Q}\, |\w'\!-\!\v'|\right)\, \textrm{K}_0\left(\bar{Q}\, |\w\!-\!\v|\right)
\right]
\nn \\
&\,
+
 \frac{(k_1^+\!-\!k_2^+)}{q^+} \,   \frac{({\w'}^{i}\!-\!{\v'}^{i})(\w^j\!-\!\v^j)}{|\w'\!-\!\v'||\w\!-\!\v|}\, \bar{Q}^2\, \textrm{K}_1\left(\bar{Q}\, |\w'\!-\!\v'|\right)\, 
 \textrm{K}_1\left(\bar{Q}\, |\w\!-\!\v|\right)\, 
 \left(Q^{(3)}_{ij}(\w',\v',\v_*,\w)\!-\!d^{(3)}_{ij}(\v_*,\w)\right)
 \Bigg\}
\, ,
\label{Cross_Section_q_dec_T_2}
\end{align}
using the notations \eqref{dipole_1}, \eqref{dipole_2}, \eqref{quadrupole_1} and \eqref{quadrupole_2}, as well as
\begin{align}
d^{(3)}_{ij}(\v_*,\w) 
=&\,  
\Bigg\langle \frac{1}{N_c} {\rm Tr} 
\Big[  \mathcal{U}^{(3)}_{F; ij} ( \v) \mathcal{U}_F^\dag ( \w) \Big] \Bigg\rangle
\label{dipole_3}
\\
Q^{(3)}_{ij}(\w',\v',\v_*,\w)
=&\,
 \Bigg\langle \frac{1}{N_c} {\rm Tr} 
\Big[  \mathcal{U}_F ( \w') \mathcal{U}_F^\dag ( \v')  \mathcal{U}^{(3)}_{F; ij} ( \v) \mathcal{U}_F^\dag ( \w) \Big] \Bigg\rangle 
\label{quadrupole_3}
\, .
\end{align}

In the calculation of the contribution of \eqref{Ampl-qbar_dec_T} at cross section level, the Dirac algebra can be performed  using the relations  \eqref{Dirac_trace_1T_res}, \eqref{Dirac_trace_2T_res}, \eqref{Dirac_trace_3T_res} and \eqref{Dirac_trace_4T_res}, apart from the terms involving the longitudinal chromomagnetic background field $\mathcal{F}_{lm}$, which lead to the Dirac structures calculated in Eqs.~\eqref{Dirac_trace_2secT_calc_1}, \eqref{Dirac_trace_3secT_calc_1}, \eqref{Dirac_trace_4secT_calc_1}, \eqref{Dirac_trace_1secT_calc_1}. 
With all this, one finds 
\begin{align}
 \frac{d\sigma_{\gamma^{*}_T\rightarrow q_1\bar q_2}}{d {\rm P.S.}}\Bigg|_{\rm NEik \, corr.}^{\textrm{dec. on }\bar{q}}
&
=  2\pi \delta(k_1^+\!+\!k_2^+\!-\!q^+)\, N_c\, \frac{\alpha_{\textrm{em}}}{\pi} \, e_f^2\, 
  \frac{2 k_1^+}{ q^+}\,   
2 {\rm Re}
\int_{\v, \v', \w, \w'}
e^{i \k_1 \cdot (\v'-\v) } \, e^{i\k_2 \cdot (\w'-\w) }\;
\nn \\
&
\hspace{-3.4cm}
\times 
\Bigg\{
\Bigg[
\frac{1}{2}\, \left[1+\left(\frac{k_2^+\!-\!k_1^+}{q^+}\right)^2\right] 
\frac{({\w'}\!-\!{\v'})\!\cdot\!(\w\!-\!\v)}{|\w'\!-\!\v'||\w\!-\!\v|}\, \bar{Q}^2\, \textrm{K}_1\left(\bar{Q}\, |\w'\!-\!\v'|\right)\, 
  \textrm{K}_1\left(\bar{Q}\, |\w\!-\!\v|\right)\, 
+ m^2\, \textrm{K}_0\left(\bar{Q}\, |\w'\!-\!\v'|\right)\, \textrm{K}_0\left(\bar{Q}\, |\w\!-\!\v|\right)
\Bigg]
\nn \\
&
\hspace{-2.8cm}
\times 
 \bigg\{
 \Big[-\frac{(\k_2^j\!-\!\k_1^j)}{2} 
 +\frac{i}{2}\,\partial_{\v^j}
 \Big]
 \Big[{Q^{(1)}_j(\v',\w',\w_*,\v)}^{\dag}-{d^{(1)}_j(\w_*,\v)}^{\dag} \Big]
 -i  \Big[{Q^{(2)}(\v',\w',\w_*,\v)}^{\dag}-{d^{(2)}(\w_*,\v)}^{\dag}\Big]
   \bigg\}
\nn \\
&
\hspace{-2.8cm}
+
 \frac{(k_1^+\!-\!k_2^+)}{q^+}  
\frac{({\w'}^{i}\!-\!{\v'}^{i})(\w^j\!-\!\v^j)}{|\w'\!-\!\v'||\w\!-\!\v|}\, \bar{Q}^2\, \textrm{K}_1\left(\bar{Q}\, |\w'\!-\!\v'|\right)\, 
 \textrm{K}_1\left(\bar{Q}\, |\w\!-\!\v|\right)\, 
 \Big[{Q^{(3)}_{ij}(\v',\w',\w_*,\v)}^{\dag}-{d^{(3)}_{ij}(\w_*,\v)}^{\dag}\Big]
\Bigg\}
\, .
\label{Cross_Section_qbar_dec_T_3}
\end{align}

Finally, using the relations \eqref{Dirac_trace_1L_res}, \eqref{Dirac_trace_2L_res},  \eqref{Dirac_trace_1T_res}, \eqref{Dirac_trace_2T_res}, \eqref{Dirac_trace_3T_res}, \eqref{Dirac_trace_4T_res} and \eqref{Dirac_trace_3},
one gets the contribution corresponding to \eqref{Ampl-dyn_T} at cross section level
\begin{align}
& \frac{d\sigma_{\gamma^{*}_T\rightarrow q_1\bar q_2}}{d {\rm P.S.}}\Bigg|_{\rm NEik \, corr.}^{\textrm{dyn. target}}
=\,
   2\pi \delta(k_1^+\!+\!k_2^+\!-\!q^+)\, 
 N_c\, \frac{\alpha_{\textrm{em}}}{\pi} \, e_f^2\,  
\frac{k_1^+ k_2^+\,(k_2^+\!-\!k_1^+)}{(q^+)^3}\,
2 {\rm Re}\; (-i)
  \int_{\v, \v', \w, \w'}
e^{i \k_1 \cdot (\v'-\v) } \, e^{i\k_2 \cdot (\w'-\w) }\;
\nn \\
&
\times
\bigg[\tilde Q(\w',\v',\v_*,\w_*)-\tilde d(\v_*,\w_*) \bigg]
\Bigg\{ 
 \frac{1}{2}\, \left[1+\left(\frac{k_2^+\!-\!k_1^+}{q^+}\right)^2\right]\, 
\frac{({\w'}\!-\!{\v'})\!\cdot\!(\w\!-\!\v)}{|\w'\!-\!\v'|}\, \bar{Q}\, \textrm{K}_1\left(\bar{Q}\, |\w'\!-\!\v'|\right)\, 
 Q^2\, \textrm{K}_0\left(\bar{Q}\, |\w\!-\!\v|\right)
  \nn \\
&\;\;\;\;
+m^2\, Q^2\,  \textrm{K}_0\left(\bar{Q}\, |\w'\!-\!\v'|\right)\, 
\frac{|\w\!-\!\v|}{\bar{Q}}\: 
 \textrm{K}_1\left(\bar{Q}\, |\w\!-\!\v|\right)\,
+2 \, 
 \frac{({\w'}\!-\!{\v'})\!\cdot\!(\w\!-\!\v)}{|\w'\!-\!\v'||\w\!-\!\v|}\, \bar{Q}^2\, \textrm{K}_1\left(\bar{Q}\, |\w'\!-\!\v'|\right)\, 
  \textrm{K}_1\left(\bar{Q}\, |\w\!-\!\v|\right)\,
\Bigg\}
\, ,
\label{Cross_Section_dyn_T}
\end{align}
using the notations \eqref{dipole_dec} and \eqref{quadrupole_dec}.

All in all, the NEik correction to the cross section for transverse photon beyond the generalized Eikonal contribution \eqref{Cross_Section_GEik_T_3} is given by the sum of the contributions \eqref{Cross_Section_in_T}, \eqref{Cross_Section_Lplus_T}, \eqref{Cross_Section_q_dec_T_2}, \eqref{Cross_Section_qbar_dec_T_3} and \eqref{Cross_Section_dyn_T}.


\section{Summary and outlook\label{sec:conclu}}
In this paper, we computed the DIS dijet production cross section at next-to-eikonal accuracy in a dynamical gluon background field in the Color Glass Condensate framework. The dijet production cross section is calculated both for transverse and longitudinal polarization of the exchanged virtual photon and all possible sources of next-to-eikonal corrections are considered in a gluon background field. More specifically, we have accounted for the corrections that stem from (i) finite longitudinal width of the target, (ii) the interaction of the quark-antiquark pair with the subleading (transverse) component of the background field and (iii) the dynamics of the gluon background which is encoded in the $z^-$ dependence of the background field. 

The cross sections for both transverse and longitudinal photon are written as a generalized eikonal contribution and explicit NEik corrections. The generalized eikonal contribution includes the average $z^-$ dependence of the background gluon field at the amplitude level. Therefore, it 
goes beyond the strict eikonal approximation by including ``$+$''  - momentum exchange with the target. On the other hand, explicit NEik  contributions are independent of this effect since it brings further power suppression at high energy. 

Beyond the generalized eikonal approximation, the $z^-$ dependence of the background field provides a new type of explicit NEik correction (Eqs. \eqref{Cross_Section_long_NEik_dyn_2} for longitudinal and \eqref{Cross_Section_dyn_T} for transverse photon polarization) that encodes the relative $z^-$ dependence of the quark and antiquark at amplitude level. This correction gives a new type of decorated dipole \eqref{dipole_dec} and quadrupole \eqref{quadrupole_dec} operators that include a derivative of the Wilson lines along the ``$-$'' light-cone direction.  

An interesting observation regarding the NEik corrections is the following. Contrary to the other kind of NEik corrections, the one that involves the longitudinal chromomagnetic field $F_{ij}$ of the target drops from the longitudinal photon cross section identically. However, in the transverse photon cross section it can give a non-vanishing contribution. 

An immediate continuation of our work presented in this manuscript is to study the so-called correlation limit of the dijet production. In this limit, the produced jets fly almost back-to-back in momentum space and one can get accesses to various types of gluon transverse momentum dependent distributions (TMDs) for dijet production in different processes (see \cite{Petreska:2018cbf} for a review and references therein). At eikonal accuracy, DIS dijet production gives access to Weizs\"acker-Williams TMDs in the correlation limit. It would be very interesting to study this limit beyond eikonal accuracy in order to understand and establish how the CGC result together with its sub-eikonal corrections can be equivalent to the TMD factorization result together with its higher twist corrections.


Another interesting observable to study at NEik accuracy is inclusive photon+jet production at forward rapidity in proton-nucleus collisions. Since the produced photon does not rescatter on the target, this observable is expected to provide a clean environment to study the interaction of the quark probe with the dense target at a hadron collider. We are planing to study both the cross section and photon-jet angular correlations for this process at NEik accuracy in the future.

\acknowledgements{
TA is supported in part by the National Science Centre (Poland) under the research Grant No. 2018/31/D/ST2/00666 (SONATA 14). GB is supported in part by the National Science Centre (Poland) under the research Grant No. 2020/38/E/ST2/00122 (SONATA BIS 10). AC  is supported in part by the National Science Centre (Poland) under the research Grant No. 2021/43/D/ST2/01154 (SONATA 17). This work has been performed in the framework of MSCA RISE 823947 ``Heavy ion collisions: collectivity and precision in saturation physics'' (HIEIC) and has received funding from the European Union's Horizon 2020 research and innovation programme under grant agreement No. 824093.}


\appendix


\section{Dirac algebra for the cross section\label{sec:Dirac_alg}}


In this appendix, we calculate the Dirac structures which arise in the numerator at cross section level, in the calculation of DIS dijet at NEik accuracy, using the standard relations 
\begin{align}
\{\gamma^{\mu},\gamma^{\nu}\} =&\, 2 g^{\mu\nu} \label{fund_relation_Dirac}\\
(\gamma^{\mu})^{\dag} =&\, \gamma^0\gamma^{\mu}\gamma^0 \label{conj_gamma_matrix}\\
\sum_{h_1 = \pm\frac{1}{2}}  u(1)\, \bar u(1) =&\, {\slashed{\check{k}}}_1+m  \label{hel_sum_u_spinor}\\
\sum_{h_2 = \pm\frac{1}{2}}  v(2)\, \bar v(2) =&\, {\slashed{\check{k}}}_2-m  \label{hel_sum_v_spinor}
\, .
\end{align}
In particular, from Eq.~\eqref{fund_relation_Dirac}, one finds $\gamma^+\gamma^+=0$ and $\{\gamma^+,\gamma^j\}=0$, which will be used constantly. Moreover, since the Dirac matrices $\gamma^{\mu}$ are $4\times 4$ matrices, the trace of the corresponding identity matrix is ${\rm Tr}_{\rm D} [\mathbf 1]=4$.

In the calculation of the cross section induced by longitudinal photon in section \ref{sec:gammaL_cross_sec}, most terms (apart from the ones including $ \mathcal{U}^{(3)}_{F; ij}$) lead to the Dirac numerator
\begin{align}
{\cal N}_{1L} 
\equiv &\,
\sum_{h_1, h_2 = \pm\frac{1}{2}}
\left(\bar u(1) \gamma^+ v(2)\right)^{\dag}
\bar u(1) \gamma^+ v(2)
\, .
\label{Dirac_trace_1L_def}
\end{align}
Using the relations \eqref{fund_relation_Dirac}-\eqref{hel_sum_v_spinor}, one finds
\begin{align}
{\cal N}_{1L} 
= &\,
\sum_{h_1, h_2 = \pm\frac{1}{2}}
\bar v(2) \gamma^+ u(1)
\bar u(1) \gamma^+ v(2) 
=
{\rm Tr}_{\rm D}\left[
 \gamma^+({\slashed{\check{k}}}_1+m) \gamma^+({\slashed{\check{k}}}_2-m)
\right]
\nn\\
=&\,
{\rm Tr}_{\rm D}\left[
 \gamma^+\left\{{\slashed{\check{k}}}_1,\, \gamma^+\right\}{\slashed{\check{k}}}_2
\right]
=(2k^+_1)\; 
{\rm Tr}_{\rm D}\left[\gamma^+ {\slashed{\check{k}}}_2\right]
\label{Dirac_trace_1L_calc_1} 
\, .
\end{align}
Hence, using the cyclicity of the trace,
\begin{align}
{\cal N}_{1L} 
=&\,
(2k^+_1)\; 
{\rm Tr}_{\rm D}\left[\frac{\{\gamma^+,\, {\slashed{\check{k}}}_2\}}{2}\right]
=
(2k^+_1)\; \frac{(2k^+_2)}{2}
{\rm Tr}_{\rm D}\left[{\mathbf 1}\right]
=8k^+_1k^+_2
\label{Dirac_trace_1L_res} 
\, .
\end{align}
The other Dirac numerator encountered in the calculation of the longitudinal photon cross section, corresponding to the contribution containing $\mathcal{U}^{(3)}_{F; ij}$, is 
\begin{align}
{\cal N}_{2L} 
\equiv &\,
\sum_{h_1, h_2 = \pm\frac{1}{2}}
\left(\bar u(1) \gamma^+ v(2)\right)^{\dag}
\bar u(1) \gamma^+ [\gamma^i,\gamma^j]v(2)
\, .
\label{Dirac_trace_2L_def}
\end{align}
In the same way, one gets
\begin{align}
{\cal N}_{2L} 
= &\,
\sum_{h_1, h_2 = \pm\frac{1}{2}}
\bar v(2) \gamma^+ u(1)
\bar u(1) \gamma^+ [\gamma^i,\gamma^j]v(2) 
=
{\rm Tr}_{\rm D}\left[
 \gamma^+({\slashed{\check{k}}}_1+m) \gamma^+[\gamma^i,\gamma^j]({\slashed{\check{k}}}_2-m)
\right]
\nn\\
=&\,
{\rm Tr}_{\rm D}\left[
 \gamma^+\left\{{\slashed{\check{k}}}_1,\, \gamma^+\right\}[\gamma^i,\gamma^j]{\slashed{\check{k}}}_2
\right]
=(2k^+_1)\; 
{\rm Tr}_{\rm D}\left[\gamma^+[\gamma^i,\gamma^j] {\slashed{\check{k}}}_2\right]
\nn\\
=&\,
(2k^+_1)\; 
{\rm Tr}_{\rm D}\left[\frac{\{\gamma^+,\, {\slashed{\check{k}}}_2\}}{2}[\gamma^i,\gamma^j]\right]
=
(2k^+_1)\; \frac{(2k^+_2)}{2}
{\rm Tr}_{\rm D}\left[[\gamma^i,\gamma^j]\right]
\label{Dirac_trace_2L_calc_1} 
\, .
\end{align}
However, due to the cyclicity of the trace,
\begin{align}
{\rm Tr}_{\rm D}\left[\gamma^i\gamma^j\right]  
=&\,
{\rm Tr}_{\rm D}\left[\frac{\{\gamma^i,\, \gamma^j\}}{2}\right]  
=4 g^{ij} =-4 \delta^{ij} 
\, ,\label{Dirac_trace_gamma_i_gamma_j}
\end{align}
so that
\begin{align}
{\rm Tr}_{\rm D}\left[[\gamma^i,\gamma^j]\right]  
=&\,0 
\, ,\label{Dirac_trace_commutator}
\end{align}
and thus 
\begin{align}
{\cal N}_{2L} 
= &\,
0
\label{Dirac_trace_2L_res}
\, .
\end{align}

In the transverse photon case, the amplitude contains two types of Dirac structures in eikonal and NEik terms, excluding for the moment the contributions in $\mathcal{U}^{(3)}_{F; lm}$, and the NEik contribution from the target dynamics \eqref{Ampl-dyn_T}.
At cross section level, one has then to calculate the squares of these two structures as well as their interferences, defined as
\begin{align}
{\cal N}_{1T} 
\equiv &\, 
\frac{1}{2}\sum_{\lambda}\sum_{h_1, h_2 = \pm\frac{1}{2}}
 \varepsilon_{\lambda}^{i'*} \;  \varepsilon_{\lambda}^i \; 
\bigg(\bar u(1) \gamma^+\left[\frac{(k_2^+\!-\!k_1^+)}{q^+}\, \delta^{i'j'}
 +\frac{[\gamma^{i'},\gamma^{j'}]}{2}
\right]v(2)\bigg)^{\dag}\,
\bar u(1) \gamma^+\left[\frac{(k_2^+\!-\!k_1^+)}{q^+}\, \delta^{ij}
 +\frac{[\gamma^i,\gamma^j]}{2}
\right]v(2)
\label{Dirac_trace_1T_def} 
\\
{\cal N}_{2T} 
\equiv&\, 
\frac{1}{2}\sum_{\lambda}\sum_{h_1, h_2 = \pm\frac{1}{2}}
 \varepsilon_{\lambda}^{i'*} \;  \varepsilon_{\lambda}^i \; 
\Big(\bar u(1) \gamma^+\gamma^{i'}v(2)\Big)^{\dag}\,
\bar u(1) \gamma^+\gamma^i v(2)
\label{Dirac_trace_2T_def} 
\\
{\cal N}_{3T} 
\equiv&\, 
\frac{1}{2}\sum_{\lambda}\sum_{h_1, h_2 = \pm\frac{1}{2}}
 \varepsilon_{\lambda}^{i'*} \;  \varepsilon_{\lambda}^i \; 
\Big(\bar u(1) \gamma^+\gamma^{i'}v(2)\Big)^{\dag}\,
\bar u(1) \gamma^+
\left[\frac{(k_2^+\!-\!k_1^+)}{q^+}\, \delta^{ij}
 +\frac{[\gamma^i,\gamma^j]}{2}
\right]
v(2)
\label{Dirac_trace_3T_def}
\\ 
{\cal N}_{4T} 
\equiv &\, 
\frac{1}{2}\sum_{\lambda}\sum_{h_1, h_2 = \pm\frac{1}{2}}
 \varepsilon_{\lambda}^{i'*} \;  \varepsilon_{\lambda}^i \; 
\bigg(\bar u(1) \gamma^+\left[\frac{(k_2^+\!-\!k_1^+)}{q^+}\, \delta^{i'j'}
 +\frac{[\gamma^{i'},\gamma^{j'}]}{2}
\right]v(2)\bigg)^{\dag}\,
\bar u(1) \gamma^+\gamma^iv(2)
\label{Dirac_trace_4T_def} 
\, .
\end{align}
The transverse polarization vectors obey the completeness relation
\begin{align}
\sum_{\lambda} \varepsilon_{\lambda}^{i'*} \;  \varepsilon_{\lambda}^i
=&\, \delta^{i'i}
\label{2D_pol_vector_completeness}
\, .
\end{align}
Then, using the same method as before, one finds
\begin{align}
{\cal N}_{1T} 
= &\, 
\frac{\delta^{i'i}}{2}\sum_{h_1, h_2 = \pm\frac{1}{2}}
\bar v(2) 
\left[\frac{(k_2^+\!-\!k_1^+)}{q^+}\, \delta^{i'j'}
 +\frac{[\gamma^{j'},\gamma^{i'}]}{2}
\right]\gamma^+
u(1)\,
\bar u(1) \gamma^+\left[\frac{(k_2^+\!-\!k_1^+)}{q^+}\, \delta^{ij}
 +\frac{[\gamma^i,\gamma^j]}{2}
\right]v(2)
\nn\\
= &\, 
\frac{\delta^{i'i}}{2}
{\rm Tr}_{\rm D}\left[
\left[\frac{(k_2^+\!-\!k_1^+)}{q^+}\, \delta^{i'j'}
 +\frac{[\gamma^{j'},\gamma^{i'}]}{2}
\right]\gamma^+
({\slashed{\check{k}}}_1+m)
 \gamma^+\left[\frac{(k_2^+\!-\!k_1^+)}{q^+}\, \delta^{ij}
 +\frac{[\gamma^i,\gamma^j]}{2}
\right]({\slashed{\check{k}}}_2-m)
\right]
\nn\\
= &\, 
\frac{\delta^{i'i}}{2}\, (2k^+_1)\;
{\rm Tr}_{\rm D}\left[
\left[\frac{(q^+\!+\!k_2^+\!-\!k_1^+)}{q^+}\, \delta^{i'j'}
 +\gamma^{j'}\gamma^{i'}
\right]\gamma^+
\left[\frac{(q^+\!+\!k_2^+\!-\!k_1^+)}{q^+}\, \delta^{ij}
 +\gamma^i\gamma^j
\right]{\slashed{\check{k}}}_2
\right]
\nn\\
= &\, 
 k^+_1\; \frac{(2k^+_2)}{2}\,
{\rm Tr}_{\rm D}\left[
\delta^{i'i}\gamma^{j'}\gamma^{i'}\gamma^i\gamma^j
+2\frac{(q^+\!+\!k_2^+\!-\!k_1^+)}{q^+}\, \gamma^{j'}\gamma^{j}
+\frac{(q^+\!+\!k_2^+\!-\!k_1^+)^2}{(q^+)^2}\, \delta^{j'j}
\right]
\nn\\
= &\, 
 k^+_1k^+_2\;
{\rm Tr}_{\rm D}\left[
\delta^{i'i}\gamma^{j'}\frac{\{\gamma^{i'},\,\gamma^i\}}{2}\gamma^j
+2\frac{(q^+\!+\!k_2^+\!-\!k_1^+)}{q^+}\, \gamma^{j'}\gamma^{j}
+\frac{(q^+\!+\!k_2^+\!-\!k_1^+)^2}{(q^+)^2}\, \delta^{j'j}
\right]
\nn\\
= &\, 
 k^+_1k^+_2\;
{\rm Tr}_{\rm D}\left[
2\frac{(k_2^+\!-\!k_1^+)}{q^+}\, \frac{\{\gamma^{j'},\,\gamma^{j}\}}{2}
+\frac{(q^+\!+\!k_2^+\!-\!k_1^+)^2}{(q^+)^2}\, \delta^{j'j}
\right]
\nn\\
= &\, 
 4k^+_1k^+_2\, \delta^{j'j}\;
\left[
-2\frac{(k_2^+\!-\!k_1^+)}{q^+}
+\frac{(q^+\!+\!k_2^+\!-\!k_1^+)^2}{(q^+)^2}\, 
\right]
\, ,
\label{Dirac_trace_1T_calc_1} 
\end{align}
so that finally
\begin{align}
{\cal N}_{1T} 
= &\, 
 4k^+_1k^+_2\, \delta^{j'j}\;
\left[1
+\frac{(k_2^+\!-\!k_1^+)^2}{(q^+)^2}\, 
\right]
\, .
\label{Dirac_trace_1T_res} 
\end{align}  
Moreover
\begin{align}
{\cal N}_{2T} 
=&\, 
\frac{1}{2}\sum_{\lambda}\sum_{h_1, h_2 = \pm\frac{1}{2}}
 \varepsilon_{\lambda}^{i'*} \;  \varepsilon_{\lambda}^i \; 
\bar v(2)\gamma^{i'} \gamma^+u(1)\,
\bar u(1) \gamma^+\gamma^i v(2)
=
\frac{\delta^{i'i}}{2}\, 
{\rm Tr}_{\rm D}\left[
\gamma^{i'}\gamma^+({\slashed{\check{k}}}_1+m) \gamma^+\gamma^i({\slashed{\check{k}}}_2-m)
\right]  
\nn\\
=&\, 
\frac{\delta^{i'i}}{2}\, (2k^+_1)\; 
{\rm Tr}_{\rm D}\left[\gamma^{i'}\gamma^+ \gamma^i{\slashed{\check{k}}}_2\right]
=
\frac{\delta^{i'i}}{2}\, (2k^+_1)\; (-1)\, \frac{(2k^+_2)}{2}\, {\rm Tr}_{\rm D}\left[\gamma^{i'}\gamma^i\right]
=
-k^+_1k^+_2\delta^{i'i}\, (-4)\delta^{i'i}
\, ,
\label{Dirac_trace_2T_calc_1} 
\end{align}
using Eq.~\eqref{Dirac_trace_gamma_i_gamma_j}, so that
\begin{align}
{\cal N}_{2T} 
=&\, 
8k^+_1k^+_2
\, .
\label{Dirac_trace_2T_res} 
\end{align}
Concerning the interference contribution, one finds
\begin{align}
{\cal N}_{3T} 
=&\, 
\frac{1}{2}\sum_{\lambda}\sum_{h_1, h_2 = \pm\frac{1}{2}}
 \varepsilon_{\lambda}^{i'*} \;  \varepsilon_{\lambda}^i \; 
\bar v(2)\gamma^{i'} \gamma^+u(1)\,
\bar u(1) \gamma^+
\left[\frac{(k_2^+\!-\!k_1^+)}{q^+}\, \delta^{ij}
 +\frac{[\gamma^i,\gamma^j]}{2}
\right]
v(2)
\nn\\
=&\, 
\frac{\delta^{i'i}}{2}\, 
{\rm Tr}_{\rm D}\left[
\gamma^{i'}\gamma^+({\slashed{\check{k}}}_1+m) \gamma^+
\left[\frac{(k_2^+\!-\!k_1^+)}{q^+}\, \delta^{ij}
 +\frac{[\gamma^i,\gamma^j]}{2}
\right]
({\slashed{\check{k}}}_2-m)
\right]  
\nn\\
=&\, 
\frac{\delta^{i'i}}{2}\, 
{\rm Tr}_{\rm D}\left[
\gamma^{i'}\gamma^+{\slashed{\check{k}}}_1 \gamma^+
\left[\frac{(k_2^+\!-\!k_1^+)}{q^+}\, \delta^{ij}
 +\frac{[\gamma^i,\gamma^j]}{2}
\right]
{\slashed{\check{k}}}_2
\right]  
\, .
\label{Dirac_trace_3T_calc_1} 
\end{align}
This is the trace of an odd number of $\gamma^{\mu}$ matrices, so that
\begin{align}
{\cal N}_{3T} 
=&\, 
0
\, .
\label{Dirac_trace_3T_res} 
\end{align}
For the same reason, one has 
\begin{align}
{\cal N}_{4T} 
=&\, 
0
\, .
\label{Dirac_trace_4T_res} 
\end{align}
The two Dirac structures from the amplitude with transverse photon are thus not interfering after all.
Note that these calculations of ${\cal N}_{1T}$,  ${\cal N}_{2T}$, ${\cal N}_{3T}$ and  ${\cal N}_{4T}$ have been done without assuming $k_1^++k_2^+=q^+$, so that they are valid for the generalized eikonal contribution to the cross section, in which $k_1^++k_2^+$ can be different from $q^+$.

It remains now to evaluate the various Dirac structure induced at the cross section level by the contributions in $\mathcal{U}^{(3)}_{F; lm}$ in the transverse photon amplitude, see Eqs.~\eqref{Ampl-q_dec_T} and \eqref{Ampl-qbar_dec_T}. In the case of the $ \mathcal{F}_{lm}$ decoration inserted on the quark line, at $\v$, it is clear from Eq.~\eqref{Ampl-q_dec_T} that one obtains at the cross section level Dirac structures which are similar to ${\cal N}_{1T}$,  ${\cal N}_{2T}$, ${\cal N}_{3T}$ and  ${\cal N}_{4T}$,  but with $[\gamma^l,\gamma^m]$ inserted at the right of $\bar u(1) \gamma^+$, as
\begin{align}
{\cal N}_{1'T} 
\equiv &\, 
\frac{1}{2}\sum_{\lambda}\sum_{h_1, h_2 = \pm\frac{1}{2}}
 \varepsilon_{\lambda}^{i'*} \;  \varepsilon_{\lambda}^i \; 
\bigg(\bar u(1) \gamma^+\left[\frac{(k_2^+\!-\!k_1^+)}{q^+}\, \delta^{i'j'}
 +\frac{[\gamma^{i'},\gamma^{j'}]}{2}
\right]v(2)\bigg)^{\dag}\,
\bar u(1) \gamma^+[\gamma^l,\gamma^m]\left[\frac{(k_2^+\!-\!k_1^+)}{q^+}\, \delta^{ij}
 +\frac{[\gamma^i,\gamma^j]}{2}
\right]v(2)
\label{Dirac_trace_1primeT_def} 
\\
{\cal N}_{2'T} 
\equiv&\, 
\frac{1}{2}\sum_{\lambda}\sum_{h_1, h_2 = \pm\frac{1}{2}}
 \varepsilon_{\lambda}^{i'*} \;  \varepsilon_{\lambda}^i \; 
\Big(\bar u(1) \gamma^+\gamma^{i'}v(2)\Big)^{\dag}\,
\bar u(1) \gamma^+[\gamma^l,\gamma^m]\gamma^i v(2)
\label{Dirac_trace_2primeT_def} 
\\
{\cal N}_{3'T} 
\equiv&\, 
\frac{1}{2}\sum_{\lambda}\sum_{h_1, h_2 = \pm\frac{1}{2}}
 \varepsilon_{\lambda}^{i'*} \;  \varepsilon_{\lambda}^i \; 
\Big(\bar u(1) \gamma^+\gamma^{i'}v(2)\Big)^{\dag}\,
\bar u(1) \gamma^+[\gamma^l,\gamma^m]
\left[\frac{(k_2^+\!-\!k_1^+)}{q^+}\, \delta^{ij}
 +\frac{[\gamma^i,\gamma^j]}{2}
\right]
v(2)
\label{Dirac_trace_3primeT_def}
\\ 
{\cal N}_{4'T} 
\equiv &\, 
\frac{1}{2}\sum_{\lambda}\sum_{h_1, h_2 = \pm\frac{1}{2}}
 \varepsilon_{\lambda}^{i'*} \;  \varepsilon_{\lambda}^i \; 
\bigg(\bar u(1) \gamma^+\left[\frac{(k_2^+\!-\!k_1^+)}{q^+}\, \delta^{i'j'}
 +\frac{[\gamma^{i'},\gamma^{j'}]}{2}
\right]v(2)\bigg)^{\dag}\,
\bar u(1) \gamma^+[\gamma^l,\gamma^m]\gamma^iv(2)
\label{Dirac_trace_4primeT_def} 
\, .
\end{align}
Then,
\begin{align}
{\cal N}_{2'T} 
=&\, 
\frac{\delta^{i'i}}{2}\, 
{\rm Tr}_{\rm D}\left[
\gamma^{i'}\gamma^+({\slashed{\check{k}}}_1+m) \gamma^+[\gamma^l,\gamma^m]\gamma^i({\slashed{\check{k}}}_2-m)
\right]  
=
\frac{\delta^{i'i}}{2}\, (2k^+_1)\; 
{\rm Tr}_{\rm D}\left[\gamma^{i'}\gamma^+[\gamma^l,\gamma^m] \gamma^i{\slashed{\check{k}}}_2\right]
\nn\\
=&\, 
\frac{\delta^{i'i}}{2}\, (2k^+_1)\; (-1)\, \frac{(2k^+_2)}{2}\, {\rm Tr}_{\rm D}\left[\gamma^{i'}[\gamma^l,\gamma^m]\gamma^i\right]
=
-k^+_1k^+_2\, \delta^{i'i}\,  {\rm Tr}_{\rm D}\left[[\gamma^l,\gamma^m]\frac{\{\gamma^i,\, \gamma^{i'}\}}{2}\right]
\nn\\
=&\, 
2k^+_1k^+_2\, {\rm Tr}_{\rm D}\left[[\gamma^l,\gamma^m]\right]
=0
\, ,
\label{Dirac_trace_2primeT_calc_1} 
\end{align}
thanks to Eq.~\eqref{Dirac_trace_commutator}. Moreover,
\begin{align}
{\cal N}_{3'T} 
=&\, 0
\label{Dirac_trace_3primeT_calc_1} 
\\
{\cal N}_{4'T} 
=&\, 0
\, ,
\label{Dirac_trace_4primeT_calc_1} 
\end{align}
since they both correspond to the trace of an odd number of $\gamma^{\mu}$ matrices, like ${\cal N}_{3T}$ and ${\cal N}_{4T}$.
Concerning ${\cal N}_{1'T}$, one finds
\begin{align}
{\cal N}_{1'T} 
= &\, 
\frac{\delta^{i'i}}{2}
{\rm Tr}_{\rm D}\left[
\left[\frac{(k_2^+\!-\!k_1^+)}{q^+}\, \delta^{i'j'}
 +\frac{[\gamma^{j'},\gamma^{i'}]}{2}
\right]\gamma^+
({\slashed{\check{k}}}_1+m)
 \gamma^+[\gamma^l,\gamma^m]\left[\frac{(k_2^+\!-\!k_1^+)}{q^+}\, \delta^{ij}
 +\frac{[\gamma^i,\gamma^j]}{2}
\right]({\slashed{\check{k}}}_2-m)
\right]
\nn\\
= &\, 
\frac{\delta^{i'i}}{2}\, (2k^+_1)\;
{\rm Tr}_{\rm D}\left[
\left[\frac{(k_2^+\!-\!k_1^+\!-\!q^+)}{q^+}\, \delta^{i'j'}
 -\gamma^{i'}\gamma^{j'}
\right]
\gamma^+[\gamma^l,\gamma^m]
\left[\frac{(k_2^+\!-\!k_1^+\!-\!q^+)}{q^+}\, \delta^{ij}
 -\gamma^j\gamma^i
\right]{\slashed{\check{k}}}_2
\right]
\nn\\
= &\, 
\frac{\delta^{i'i}}{2}\, (2k^+_1)\; \frac{(2k^+_2)}{2}\,
{\rm Tr}_{\rm D}\left[
[\gamma^l,\gamma^m]
\left[\frac{(k_2^+\!-\!k_1^+\!-\!q^+)}{q^+}\, \delta^{ij}
 -\gamma^j\gamma^i
\right]
\left[\frac{(k_2^+\!-\!k_1^+\!-\!q^+)}{q^+}\, \delta^{i'j'}
 -\gamma^{i'}\gamma^{j'}
\right]
\right]
\nn\\
= &\, 
 k^+_1k^+_2\; 
{\rm Tr}_{\rm D}\left[[\gamma^l,\gamma^m]\left(
\delta^{i'i}\gamma^j\gamma^i\gamma^{i'}\gamma^{j'}
-2\frac{(k_2^+\!-\!k_1^+\!-\!q^+)}{q^+}\, \gamma^{j}\gamma^{j'}
+\frac{(k_2^+\!-\!k_1^+\!-\!q^+)^2}{(q^+)^2}\, \delta^{jj'}
\right)\right]
\nn\\
= &\, 
 k^+_1k^+_2\; 
{\rm Tr}_{\rm D}\left[[\gamma^l,\gamma^m]\left(
\delta^{i'i}\gamma^j\frac{\{\gamma^i,\, \gamma^{i'}\}}{2}\gamma^{j'}
-2\frac{(k_2^+\!-\!k_1^+\!-\!q^+)}{q^+}\, \gamma^{j}\gamma^{j'}
+\frac{(k_2^+\!-\!k_1^+\!-\!q^+)^2}{(q^+)^2}\, \delta^{jj'}
\right)\right]
\nn\\
= &\, 
 -2\frac{(k_2^+\!-\!k_1^+)}{q^+}\, k^+_1k^+_2\; 
{\rm Tr}_{\rm D}\left[[\gamma^l,\gamma^m] \gamma^{j}\gamma^{j'}
\right]
\nn\\
= &\, 
 \frac{16k^+_1k^+_2(k_1^+\!-\!k_2^+)}{q^+}\, 
\Big[\delta^{j'l}\, \delta^{jm}\!-\!\delta^{jl}\, \delta^{j'm}
\Big]
\, ,
\label{Dirac_trace_1primeT_calc_1} 
\end{align}
because, using the cyclicity of the trace,
\begin{align}
{\rm Tr}_{\rm D}\left[[\gamma^l,\gamma^m] \gamma^{j}\gamma^{j'}\right]
= &\, 
{\rm Tr}_{\rm D}\left[\gamma^l\{\gamma^m,\, \gamma^{j}\}\gamma^{j'}\right]
-{\rm Tr}_{\rm D}\left[\gamma^l \gamma^{j}\{\gamma^m,\,\gamma^{j'}\}\right]
\nn\\
= &\, 
-2\delta^{jm}\, {\rm Tr}_{\rm D}\left[\gamma^l\gamma^{j'}\right]
+2\delta^{j'm}\,{\rm Tr}_{\rm D}\left[\gamma^l \gamma^{j}\right]
\nn\\
= &\, 
8\Big[\delta^{j'l}\, \delta^{jm}\!-\!\delta^{jl}\, \delta^{j'm}
\Big]
\, .
\label{Dirac_trace_1primeT_calc_1} 
\end{align}

Finally, in the case of the $ \mathcal{F}_{lm}$ decoration inserted on the antiquark line, at $\w$, it is clear from Eq.~\eqref{Ampl-q_dec_T} that one obtains at the cross section level Dirac structures which are similar to ${\cal N}_{1T}$,  ${\cal N}_{2T}$, ${\cal N}_{3T}$ and  ${\cal N}_{4T}$,  but with $[\gamma^l,\gamma^m]$ inserted at the left of $v(2)$, as
\begin{align}
{\cal N}_{1''T} 
\equiv &\, 
\frac{1}{2}\sum_{\lambda}\sum_{h_1, h_2 = \pm\frac{1}{2}}
 \varepsilon_{\lambda}^{i'*} \;  \varepsilon_{\lambda}^i \; 
\bigg(\bar u(1) \gamma^+\left[\frac{(k_2^+\!-\!k_1^+)}{q^+}\, \delta^{i'j'}
 +\frac{[\gamma^{i'},\gamma^{j'}]}{2}
\right]v(2)\bigg)^{\dag}\,
\bar u(1) \gamma^+\!\!\left[\frac{(k_2^+\!-\!k_1^+)}{q^+}\, \delta^{ij}
 +\frac{[\gamma^i,\gamma^j]}{2}
\right]\!\![\gamma^l,\gamma^m]v(2)
\label{Dirac_trace_1secT_def} 
\\
{\cal N}_{2''T} 
\equiv&\, 
\frac{1}{2}\sum_{\lambda}\sum_{h_1, h_2 = \pm\frac{1}{2}}
 \varepsilon_{\lambda}^{i'*} \;  \varepsilon_{\lambda}^i \; 
\Big(\bar u(1) \gamma^+\gamma^{i'}v(2)\Big)^{\dag}\,
\bar u(1) \gamma^+\gamma^i[\gamma^l,\gamma^m]v(2)
\label{Dirac_trace_2secT_def} 
\\
{\cal N}_{3''T} 
\equiv&\, 
\frac{1}{2}\sum_{\lambda}\sum_{h_1, h_2 = \pm\frac{1}{2}}
 \varepsilon_{\lambda}^{i'*} \;  \varepsilon_{\lambda}^i \; 
\Big(\bar u(1) \gamma^+\gamma^{i'}v(2)\Big)^{\dag}\,
\bar u(1) \gamma^+
\left[\frac{(k_2^+\!-\!k_1^+)}{q^+}\, \delta^{ij}
 +\frac{[\gamma^i,\gamma^j]}{2}
\right]
[\gamma^l,\gamma^m]
v(2)
\label{Dirac_trace_3secT_def}
\\ 
{\cal N}_{4''T} 
\equiv &\, 
\frac{1}{2}\sum_{\lambda}\sum_{h_1, h_2 = \pm\frac{1}{2}}
 \varepsilon_{\lambda}^{i'*} \;  \varepsilon_{\lambda}^i \; 
\bigg(\bar u(1) \gamma^+\left[\frac{(k_2^+\!-\!k_1^+)}{q^+}\, \delta^{i'j'}
 +\frac{[\gamma^{i'},\gamma^{j'}]}{2}
\right]v(2)\bigg)^{\dag}\,
\bar u(1) \gamma^+\gamma^i[\gamma^l,\gamma^m]v(2)
\label{Dirac_trace_4secT_def} 
\, .
\end{align}
Using similar calculations or arguments than for ${\cal N}_{1'T}$,  ${\cal N}_{2'T}$, ${\cal N}_{3'T}$ and  ${\cal N}_{4'T}$, one can arrive at the results
\begin{align}
{\cal N}_{1''T} 
= &\, 
 \frac{16k^+_1k^+_2(k_1^+\!-\!k_2^+)}{q^+}\, 
\Big[\delta^{j'l}\, \delta^{jm}\!-\!\delta^{jl}\, \delta^{j'm}
\Big]
\label{Dirac_trace_1secT_calc_1}
\\
{\cal N}_{2''T} 
=&\, 0
\label{Dirac_trace_2secT_calc_1}  
\\
{\cal N}_{3''T} 
=&\, 0
\label{Dirac_trace_3secT_calc_1} 
\\
{\cal N}_{4''T} 
=&\, 0
\, .
\label{Dirac_trace_4secT_calc_1} 
\end{align}

Finally, the NEik correction at cross section level associated with the contribution \eqref{Ampl-dyn_T} of the dynamics of the target can be evaluated using relations calculated so far in this appendix, and
\begin{align}
\sum_{h_1, h_2 = \pm\frac{1}{2}} 
\Big(\bar u(1) \gamma^+\gamma^{i}v(2)\Big)^{\dag}\,
\bar u(1) \gamma^+ v(2)
=&\, 
{\rm Tr}_{\rm D}\left[
\gamma^{i}\gamma^+({\slashed{\check{k}}}_1+m) \gamma^+({\slashed{\check{k}}}_2-m)
\right]  
=
(2k_1^+)\,
 {\rm Tr}_{\rm D}\left[
\gamma^{i} \gamma^+{\slashed{\check{k}}}_2
\right]  
=0\, \, ,
\label{Dirac_trace_3} 
\end{align}
as a trace of three $\gamma^{\mu}$ matrices.

\section{Cross section on a $x^-$ dependent background field\label{sec:zminus_dep_cross_sec}}

In order to understand the relations between the S matrix, the amplitude and the cross section for scattering on a $x^-$ dependent background field, let us consider for simplicity a scalar model.

\subsection{Incoming wave packet}

An incoming state with one on-shell scalar particle of momentum $k_{\textrm{in}}^{\mu}$ is 
\begin{align}
a_{\textrm{in}}^{\dag}(\underline{k_{\textrm{in}}})\, |0> 
\, ,
\end{align}
with the creation operator in the incoming free Fock space normalized as
\begin{align}
\big[a_{\textrm{in}}(\underline{k}'),\, a_{\textrm{in}}^{\dag}(\underline{k})\big]
= & (2k^+)  (2\pi)^3 \delta^{(3)}(\underline{k}'\!-\!\underline{k})  
\, .
\end{align}

In order to avoid issues, one is led as usual to consider incoming wavepackets instead of pure momentum states, such as
\begin{align}
|\phi>  =& \int \frac{d^4k}{(2\pi)^4}\,  2\pi \delta(k^2\!-\!m^2)\, \theta(k^0)  \; \phi(\underline{k}) \; a_{\textrm{in}}^{\dag}(\underline{k})\, |0>
\nonumber\\
 =& \int \frac{dk^+}{2\pi}\,   \frac{\theta(k^+)}{2k^+} \int \frac{d^2\k}{(2\pi)^2}\; \phi(\underline{k}) \; a_{\textrm{in}}^{\dag}(\underline{k})\, |0>
\, ,
\end{align}
with the normalization
\begin{align}
<\phi|\phi>=& 1
\end{align}
or equivalently with the profile function $\phi(\underline{k})$ normalized as
\begin{align}
\int \frac{d^4k}{(2\pi)^4}\,  2\pi \delta(k^2\!-\!m^2)\, \theta(k^0)  \; |\phi(\underline{k})|^2 
= \int \frac{dk^+}{2\pi}\,   \frac{\theta(k^+)}{2k^+} \int \frac{d^2\k}{(2\pi)^2}\; |\phi(\underline{k})|^2 
=& 1
\end{align}

For simplicity, let us actually take a profile function which factorizes as 
\begin{align}
\phi(\underline{k}) =& \, \phi^{(+)}(k^+)\; \phi^{(\perp)}(\k)
\, ,
\end{align}
with the normalizations
\begin{align}
\int \frac{dk^+}{2\pi}\,   \frac{\theta(k^+)}{2k^+}\; |\phi^{(+)}(k^+)|^2 
=& 1
\label{norm_plus_wavep}
\\
 \int \frac{d^2\k}{(2\pi)^2}\; |\phi^{(\perp)}(\k)|^2 
=& 1
\label{norm_perp_wavep}
\, .
\end{align}

Let us assume that the wavepacket is centered at position $\x=0$ in the transverse plane. Then, we can introduce translations of this wavepacket in the transverse direction, in order to gain control on the impact parameter of the collision that we will consider. The translation of the wavepacket in the transverse plane by a vector $\B$ is performed thanks to the transverse momentum operator as
\begin{align}
|\phi_\B> \equiv &\,  e^{-i\B \cdot \hat{\P}} |\phi>
= \int \frac{dk^+}{2\pi}\,   \frac{\theta(k^+)}{2k^+} \int \frac{d^2\k}{(2\pi)^2}\; \phi(\underline{k}) \;
e^{-i\B \cdot {\k}}\;  a_{\textrm{in}}^{\dag}(\underline{k})\, |0>
\end{align}


\subsection{S matrix}

Let us consider the scattering of this wavepacket on a classical background field. For the final state, we will simply take a Fock state ${\cal F}$ in the outgoing asymptotic Fock space 
\begin{align}
<{\cal F}_{\textrm{out}}| \equiv &\, <0| \Big\{ \prod_{f\in {\cal F}}   a_{\textrm{out}}(\underline{p_f})   \Big\}
\, .
\end{align}

Then, the S matrix element for the scattering of the incoming wavepacket on the classical background field into ${\cal F}$ and with an impact parameter $\B$ is
\begin{align}
{\cal S}_{\phi \rightarrow {\cal F}}(\B)
 \equiv &\, <{\cal F}_{\textrm{out}}|\phi_\B>  
=  \int \frac{dk^+}{2\pi}\,   \frac{\theta(k^+)}{2k^+} \int \frac{d^2\k}{(2\pi)^2}\; \phi(\underline{k}) \;
e^{-i\B \cdot {\k}}\;  <{\cal F}_{\textrm{out}}|a_{\textrm{in}}^{\dag}(\underline{k})\, |0>
\, .
\end{align}
The background field appears only implicitly, in the mapping between the $\textrm{in}$ and $\textrm{out}$ Fock spaces.


\subsection{Cross section: general formula}

The fully differential cross section for this scattering is the integral over the  impact parameter $\B$ of the square of the S matrix element. Hence
\begin{align}
\frac{d\sigma_{\phi \rightarrow {\cal F}}}{d \textrm{P.S.}({\cal F})} 
\equiv &\,  \int d^2 \B\;  \left| {\cal S}_{\phi \rightarrow {\cal F}}(\B)\right|^2
\nonumber\\
=&\, 
 \int \frac{d{k'}^+}{2\pi}\,   \frac{\theta({k'}^+)}{2{k'}^+} \int \frac{d^2\k'}{(2\pi)^2}\; \phi(\underline{k'})^* \;
 \int \frac{dk^+}{2\pi}\,   \frac{\theta(k^+)}{2k^+} \int \frac{d^2\k}{(2\pi)^2}\; \phi(\underline{k}) \;
 \int d^2 \B\; e^{i\B \cdot({\k'}- {\k})}\;
\\
&\hspace{2cm} \times \;
<0|a_{\textrm{in}}(\underline{k'})\, |{\cal F}_{\textrm{out}}>\:
 <{\cal F}_{\textrm{out}}|a_{\textrm{in}}^{\dag}(\underline{k})\, |0>
\nonumber\\
=&\, 
 \int \frac{d{k'}^+}{2\pi}\,   \frac{\theta({k'}^+)}{2{k'}^+} \; \phi^{(+)}({k'}^+)^* \;
 \int \frac{dk^+}{2\pi}\,   \frac{\theta(k^+)}{2k^+}\; \phi^{(+)}(k^+)
 \int \frac{d^2\k}{(2\pi)^2}\;  |\phi^{(\perp)}(\k)|^2 \;
\\
&\hspace{2cm} \times \;
<0|a_{\textrm{in}}({k'}^+, \k)\, |{\cal F}_{\textrm{out}}>\:
 <{\cal F}_{\textrm{out}}|a_{\textrm{in}}^{\dag}(k^+,\k)\, |0> 
\, .\label{xsec_1}
\end{align}
Assuming that the profile function $\phi^{(\perp)}(\k)$ is sharply peaked at some $\k=\k_{\textrm{in}}$ and using its normalization relation \eqref{norm_perp_wavep}, one finds
\begin{align}
\frac{d\sigma_{\phi \rightarrow {\cal F}}}{d \textrm{P.S.}({\cal F})} 
=&\, 
 \int \frac{d{k'}^+}{2\pi}\,   \frac{\theta({k'}^+)}{2{k'}^+} \, \phi^{(+)}({k'}^+)^* 
 \int \frac{dk^+}{2\pi}\,   \frac{\theta(k^+)}{2k^+}\, \phi^{(+)}(k^+)\,
<0|a_{\textrm{in}}({k'}^+, \k_{\textrm{in}})\, |{\cal F}_{\textrm{out}}>\,
 <{\cal F}_{\textrm{out}}|a_{\textrm{in}}^{\dag}(k^+,\k_{\textrm{in}})\, |0> 
\, .
\label{xsec_2}
\end{align}


\subsection{Cross section: case of $x^-$ independent background field\label{sec:xmin_indep_cross_sec}}

If the background field is independent of $x^-$, the whole problem is invariant under translations along $x^-$, and the $p^+$ component of the momentum is conserved by the scattering. In that case, $<{\cal F}_{\textrm{out}}|a_{\textrm{in}}^{\dag}(k^+,\k_{\textrm{in}})\, |0> $ should be proportional to $\delta(k^+\!-\!p^+_{\cal F})$, where $p^+_{\cal F}$ is the total light-cone momentum of the particles belonging to the final Fock state ${\cal F}$. 
By convention, we can then define the scattering amplitude  ${\cal M}_{\cal F}(k^+,\k_{\textrm{in}})$ via
\begin{align}
<{\cal F}_{\textrm{out}}|a_{\textrm{in}}^{\dag}(k^+,\k_{\textrm{in}})\, |0> 
=& \, (2k^+)\,  2\pi\, \delta(k^+\!-\!p^+_{\cal F})\, i{\cal M}_{\cal F}(k^+,\k_{\textrm{in}})
\, ,\label{ampl_1}
\end{align}
if ${\cal F}$ is not a one scalar Fock state like the initial state. Instead, if it is a one scalar Fock state as well, we should subtract the "no scattering" contribution from \eqref{ampl_1}, as usual.
Inserting that expression into Eq.~\eqref{xsec_2}, one finds
\begin{align}
\frac{d\sigma_{\phi \rightarrow {\cal F}}}{d \textrm{P.S.}({\cal F})} 
=&\, 
 \int \frac{d{k'}^+}{2\pi}\,   \frac{\theta({k'}^+)}{2{k'}^+} \, \phi^{(+)}({k'}^+)^* 
 \int \frac{dk^+}{2\pi}\,   \frac{\theta(k^+)}{2k^+}\, \phi^{(+)}(k^+)\,
\, (2{k'}^+)\, 2\pi\, \delta({k'}^+\!-\!p^+_{\cal F})\, (-i){\cal M}_{\cal F}({k'}^+,\k_{\textrm{in}})^{\dag} 
\nonumber\\
& \hspace{2cm} \times\;
 \,(2k^+)\, 2\pi\, \delta(k^+\!-\!p^+_{\cal F})\, i{\cal M}_{\cal F}(k^+,\k_{\textrm{in}})
\nonumber\\
=&\, 
\int \frac{dk^+}{2\pi}\, \theta(k^+)\, \phi^{(+)}(k^+)\,
 \int d{k'}^+\,   \theta({k'}^+) \, \phi^{(+)}({k'}^+)^* 
 \delta({k'}^+\!-\!k^+)\,  2\pi\, \delta(k^+\!-\!p^+_{\cal F})\,
{\cal M}_{\cal F}({k'}^+,\k_{\textrm{in}})^{\dag}\,  {\cal M}_{\cal F}(k^+,\k_{\textrm{in}})
\nonumber\\
=&\, 
\int \frac{dk^+}{2\pi}\,   \frac{\theta(k^+)}{2k^+}\, |\phi^{(+)}(k^+)|^2 
\;  (2k^+)\, 2\pi\, \delta(k^+\!-\!p^+_{\cal F})\,
 | {\cal M}_{\cal F}(k^+,\k_{\textrm{in}})|^2 
\, .
\label{xsec_3} 
\end{align}
Assuming that the profile function $\phi^{(+)}(k^+)$ is sharply peaked at some $k^+=k^+_{\textrm{in}}$ and using its normalization relation \eqref{norm_plus_wavep}, one finds
\begin{align}
\frac{d\sigma_{\phi \rightarrow {\cal F}}}{d \textrm{P.S.}({\cal F})} 
=&
\, (2k^+_{\textrm{in}})\, 2\pi\, \delta(k^+_{\textrm{in}}\!-\!p^+_{\cal F})\,
 | {\cal M}_{\cal F}(k^+_{\textrm{in}},\k_{\textrm{in}})|^2 
\, .
\label{xsec_4}
\end{align}
This is the standard expression used in CGC-type calculations, together with Eq.~\eqref{ampl_1}.


\subsection{Cross section: case of background field with $x^-$ dependence
\label{App:X_section_x_minus}
}
If the background field depends on $x^-$ and thus breaks invariance under translations along $x^-$, one has in general $k^+\neq p^+_{\cal F}$, so that the relation \eqref{ampl_1} is not valid anymore. In that case, it is convenient to Fourier transform 
 with respects to $p^+_{\cal F}-k^+$ as
\begin{align}
<{\cal F}_{\textrm{out}}|a_{\textrm{in}}^{\dag}(k^+,\k_{\textrm{in}})\, |0> 
=& 2k^+
\int dz^-\, e^{i\, z^- (p^+_{\cal F}-k^+)} \; 
i{\mathbb M}_{\cal F}(z^-,k^+,\k_{\textrm{in}})
\, ,\label{ampl_2}
\end{align}
where ${\mathbb M}_{\cal F}(z^-,k^+,\k_{\textrm{in}})$ now does not depend on $p^+_{\cal F}$ (but other choices of variables should be possible). 

As a remark, in the limit of no $x^-$ dependence of the background field, the $z^-$ dependence of  ${\mathbb M}_{\cal F}(z^-,k^+,\k_{\textrm{in}})$ would disappear and we would obtain 
\begin{align}
<{\cal F}_{\textrm{out}}|a_{\textrm{in}}^{\dag}(k^+,\k_{\textrm{in}})\, |0> 
\rightarrow & 
(2k^+)\, 2\pi \delta(p^+_{\cal F}-k^+) \; 
i{\mathbb M}_{\cal F}(0,k^+,\k_{\textrm{in}})
\, ,\label{ampl_3}
\end{align}
and thus 
\begin{align}
{\mathbb M}_{\cal F}(0,k^+,\k_{\textrm{in}})
\rightarrow & \, {\cal M}_{\cal F}(k^+,\k_{\textrm{in}})
\, ,\label{ampl_4}
\end{align}
by comparison with Eq.~\eqref{ampl_1}.

Coming back to the case of a $x^-$ dependent background field, one can insert the relation \eqref{ampl_2} into \eqref{xsec_2}, and obtain
\begin{align}
\frac{d\sigma_{\phi \rightarrow {\cal F}}}{d \textrm{P.S.}({\cal F})} 
=&\, 
 \int \frac{d{k'}^+}{2\pi}\,   \frac{\theta({k'}^+)}{2{k'}^+} \, \phi^{(+)}({k'}^+)^* 
 \int \frac{dk^+}{2\pi}\,   \frac{\theta(k^+)}{2k^+}\, \phi^{(+)}(k^+)\,
 (2k^+)
\int dz^-\, e^{i\, z^- (p^+_{\cal F}-k^+)} \; 
\nonumber\\
& \hspace{2cm} \times\;
(2{k'}^+)\int d{z'}^-\, e^{-i\, {z'}^- (p^+_{\cal F}-{k'}^+)} \; 
{\mathbb M}_{\cal F}({z'}^-,{k'}^+,\k_{\textrm{in}})^{\dag}\; 
{\mathbb M}_{\cal F}(z^-,k^+,\k_{\textrm{in}})
\, .
\label{xsec_5}
\end{align}
With the change of variables $(z^-,{z'}^-)\mapsto (x^-,r^-)$, defined as
\begin{align}
x^- =&\, \frac{z^-+{z'}^-}{2}\\
r^- =&\, z^--{z'}^-
\, ,
\end{align}
the cross section becomes
\begin{align}
\frac{d\sigma_{\phi \rightarrow {\cal F}}}{d \textrm{P.S.}({\cal F})} 
=&\, 
 \int \frac{d{k'}^+}{2\pi}\,   \theta({k'}^+) \, \phi^{(+)}({k'}^+)^* 
 \int \frac{dk^+}{2\pi}\,   \frac{\theta(k^+)}{2k^+}\, \phi^{(+)}(k^+)\,
(2k^+) \int dr^-\, e^{i\, r^- (p^+_{\cal F}-\frac{(k^++{k'}^+)}{2})} \; 
\nonumber\\
& \hspace{2cm} \times\;
\int dx^-\, e^{i\, x^- ({k'}^+-k^+)} \; 
{\mathbb M}_{\cal F}\Big(x^--\frac{ r^-}{2},{k'}^+,\k_{\textrm{in}}\Big)^{\dag}\; 
{\mathbb M}_{\cal F}\Big(x^-+\frac{ r^-}{2},k^+,\k_{\textrm{in}}\Big)
\, .
\label{xsec_6}
\end{align}
At this stage, at first, it does not seem possible to get rid of the wave packet profile function without further approximations. Instead, we have to remember that here, we were considering the cross section for scattering on a given configuration of the background field. In CGC-type calculations, it is necessary to take a statistical average over the configurations of the classical background field, which is supposed to be proportional to the quantum expectation value in the "one target" state of the same quantity, now considered to be a quantum operator. 

Let us focus on the inclusive version of the studied scattering. It means that we are not measuring anything in the target fragmentation region, so that for example we include both events in which the target breaks up and events in which it stays intact. For such observable, the target average is done at cross section level, as
\begin{align}
\left<\frac{d\sigma_{\phi \rightarrow {\cal F}}}{d \textrm{P.S.}({\cal F})} \right>
=&\, 
 \int \frac{d{k'}^+}{2\pi}\,   \theta({k'}^+)\, \phi^{(+)}({k'}^+)^* 
 \int \frac{dk^+}{2\pi}\,   \frac{\theta(k^+)}{2k^+}\, \phi^{(+)}(k^+)\,
(2k^+) \int dr^-\, e^{i\, r^- (p^+_{\cal F}-\frac{(k^++{k'}^+)}{2})} \; 
\nonumber\\
& \hspace{2cm} \times\;
\int dx^-\, e^{i\, x^- ({k'}^+-k^+)} \; 
\left<{\mathbb M}_{\cal F}\Big(x^--\frac{ r^-}{2},{k'}^+,\k_{\textrm{in}}\Big)^{\dag}\; 
{\mathbb M}_{\cal F}\Big(x^-+\frac{ r^-}{2},k^+,\k_{\textrm{in}}\Big)\right>
\, .
\label{xsec_7}
\end{align}
 In this averaging over field configurations, we are including configurations which can in particular have support anywhere along $z^-$. In such a way, there is an overall invariance under translations along the $-$ direction which is restored by the target average. In particular, we have 
\begin{align}
\left<{\mathbb M}_{\cal F}\Big(x^--\frac{ r^-}{2},{k'}^+,\k_{\textrm{in}}\Big)^{\dag}\; 
{\mathbb M}_{\cal F}\Big(x^-+\frac{ r^-}{2},k^+,\k_{\textrm{in}}\Big)\right>
=&\, 
\left<{\mathbb M}_{\cal F}\Big(-\frac{ r^-}{2},{k'}^+,\k_{\textrm{in}}\Big)^{\dag}\; 
{\mathbb M}_{\cal F}\Big(\frac{ r^-}{2},k^+,\k_{\textrm{in}}\Big)\right>
\, .
\end{align}
Using this property, the integral over $x^-$ becomes trivial in Eq.~\eqref{xsec_7} and forces ${k'}^+={k}^+$. Then,
\begin{align}
\left<\frac{d\sigma_{\phi \rightarrow {\cal F}}}{d \textrm{P.S.}({\cal F})} \right>
=&\, 
 \int \frac{dk^+}{2\pi}\,   \frac{\theta(k^+)}{2k^+}\, |\phi^{(+)}(k^+)|^2\, (2k^+) \,
\int dr^-\, e^{i\, r^- (p^+_{\cal F}-k^+)} \; 
\left<{\mathbb M}_{\cal F}\Big(-\frac{ r^-}{2},k^+,\k_{\textrm{in}}\Big)^{\dag}\; 
{\mathbb M}_{\cal F}\Big(\frac{ r^-}{2},k^+,\k_{\textrm{in}}\Big)\right>
\, .
\label{xsec_8}
\end{align}
We can now assume that the profile function $\phi^{(+)}(k^+)$ is sharply peaked at some $k^+=k^+_{\textrm{in}}$ and using its normalization relation \eqref{norm_plus_wavep}, one obtain
\begin{align}
\left<\frac{d\sigma_{\phi \rightarrow {\cal F}}}{d \textrm{P.S.}({\cal F})} \right>
=&\, 
 2k^+_{\textrm{in}} \,
\int dr^-\, e^{i\, r^- (p^+_{\cal F}-k^+_{\textrm{in}})} \; 
\left<{\mathbb M}_{\cal F}\Big(-\frac{ r^-}{2},k^+_{\textrm{in}},\k_{\textrm{in}}\Big)^{\dag}\; 
{\mathbb M}_{\cal F}\Big(\frac{ r^-}{2},k^+_{\textrm{in}},\k_{\textrm{in}}\Big)\right>
\, .
\label{xsec_9} 
\end{align}
This expression is then the generalization of Eq.~\eqref{xsec_4} in the case of an arbitrary $x^-$ dependence of the background field. Interestingly, it is valid only at the level of the target-averaged cross section, not for each configuration of the  background field like Eq.~\eqref{xsec_4}.

In the case of a slow $x^-$ dependence of the background field we can then Taylor expand ${\mathbb M}_{\cal F}$ and ${\mathbb M}_{\cal F}^{\dag}$ in Eq.~\eqref{xsec_9} around $r^-=0$, and perform the integration over $r^-$ explicitly. At zeroth order, we would recover Eq.~\eqref{xsec_4}, and at every following orders, we would get terms with derivatives of $\delta(k^+_{\textrm{in}}\!-\!p^+_{\cal F})$.


\end{document}